\documentclass[aps,prx,preprint,onecolumn,citeautoscript,superscriptaddress,footinbib,
eqsecnum]{revtex4-1}  
\usepackage{amsmath,amssymb,bm,bbm} 
\usepackage{graphicx}  
\usepackage{color} 
\usepackage[dvipsnames]{xcolor}
\usepackage[papersize={8.5in,11in}]{geometry}
\usepackage[colorlinks=true]{hyperref}  
\usepackage{ulem}
\hypersetup{
    bookmarks=true,         
    unicode=false,          
    pdftoolbar=true,        
    pdfmenubar=true,        
    pdffitwindow=false,     
    pdfstartview={FitH},    
    pdftitle={Confinement transition of the orthogonal semi-metal},    
    pdfauthor={Snir Gazit, Fakher Assaad,  Subir Sachdev, Ashvin Vishwanath, Chong Wang},     
    pdfsubject={},   
    pdfcreator={},   
    pdfproducer={}, 
    pdfkeywords={} {} {}, 
    pdfnewwindow=true,      
    colorlinks=true,       
    linkcolor=magenta, 
    citecolor=blue,        
    filecolor=magenta,      
    urlcolor=blue           
} 

\usepackage{amsfonts}
\usepackage{upgreek}
\usepackage{slashed}
\usepackage{latexsym}

\newcommand{\beq}{\begin{equation}}
\newcommand{\eeq}{\end{equation}}
\def\bea{\begin{eqnarray}}
\def\eea{\end{eqnarray}}

\usepackage{braket}
\usepackage{comment}
\usepackage{slashed}
\usepackage{subcaption}
\newcommand{\ttext}[1]{\mbox{\tiny #1}}

\begin{document}

\preprint{arXiv:1804.01095}

\title{Confinement transition of $\mathbb{Z}_2$ gauge theories coupled to massless fermions:
emergent QCD$_3$ and $SO(5)$ symmetry}

\author{Snir Gazit}
\affiliation{Department of Physics, University of California, Berkeley, CA 94720, USA}

\author{Fakher F. Assaad}
\affiliation{Institut f\"ur Theoretische Physik und Astrophysik, Universit\"at W\"urzburg, 97074 W\"urzburg, Germany}

\author{Subir Sachdev}
\affiliation{Department of Physics, Harvard University, Cambridge MA 02138, USA}
\affiliation{Perimeter Institute for Theoretical Physics, Waterloo, Ontario, Canada N2L 2Y5}

\author{Ashvin Vishwanath}
\affiliation{Department of Physics, Harvard University, Cambridge MA 02138, USA}

\author{Chong Wang}
\affiliation{Department of Physics, Harvard University, Cambridge MA 02138, USA}

\date{\today
\\
\vspace{0.4in}}


\begin{abstract}
We study a model of fermions on the square lattice at half-filling  coupled to an Ising gauge theory, that was recently shown in Monte Carlo simulations to exhibit  $\mathbb{Z}_2$ topological order and massless Dirac fermion excitations. On tuning parameters, a confining phase with broken symmetry (an antiferromagnet in one choice of Hamiltonian) was also established, and the transition between these phases was found to be continuous, with co-incident onset of symmetry breaking and confinement. While the confinement transition in pure gauge theories is well understood in terms of condensing magnetic flux excitations, the same transition in the presence of gapless fermions is a challenging problem owing to the statistical interactions between fermions and the condensing flux excitations. The conventional scenario then proceeds via a two step transition, involving a symmetry breaking transition leading to gapped fermions followed by confinement. In contrast, here, using large scale quantum Monte Carlo simulations, we provide further evidence for a direct, continuous transition and also find numerical evidence for an enlarged $SO(5)$ symmetry rotating between antiferromagnetism and valence bond solid orders proximate to criticality. Guided by our numerical finding, we develop a field theory description of the direct transition involving an emergent non-abelian ($SU(2)$) gauge theory and a matrix Higgs field. We contrast our results with the conventional Gross--Neveu--Yukawa transition.

\end{abstract}

\maketitle

\section{Introduction}

Classical and quantum phase transitions have traditionally been studied in the framework of the Landau-Ginzburg-Wilson paradigm. Phases are distinguished on the basis of whether they preserve or break global symmetries of the Hamiltonian. Two distinct phases can be separated by a continuous phase transition only when one of them breaks a single symmetry which is preserved in the other.

More recently, studies of correlated quantum systems have led to many examples of important physical models displaying phase transitions which do not fit this familiar paradigm. We can have continuous quantum phase transitions between phases which break distinct symmetries \cite{Senthil_DC}. And upon allowing for topological order, several new types
of quantum phase transitions become possible.
We can have continuous phase transitions between a phase with topological order to a phase without topological order, both of which preserve all symmetries: the earliest example of this is the phase transition in the Ising gauge theory in 2+1 dimensions described by Wegner \cite{WegnerILGT}. We can also have continuous 
quantum transitions between phases with distinct types of topological order, and many examples are known in fractional quantum Hall systems \cite{WW93,CFW93}. We can have a continuous transition from a phase with topological order to one with a broken symmetry \cite{RJSS91,CSS94,SSMV99}. Finally, we can have phase transitions between a Dirac semimetal and various gapped states, including a symmetric gapped state without topological order dubbed symmetric mass generation \cite{BenTov2015,  Chandrasekharan2016, Catterall2016, Meng2016, You_2018,You_2018a} 

Theories of these novel transitions all involve quantum field theories 
with deconfined emergent gauge fields. The presence of the gauge fields reflects the long-range quantum entanglement near the critical point: this entanglement is not easily captured by the symmetry-breaking degrees of freedom, or their fluctuations. 
The gauge theories have varieties of Higgs and confining phases, and transitions between these phases allow for the transitions described above. 

In the present paper, we will present a novel example of a deconfined critical point, between a deconfined phase with topological order and a confining phase with broken symmetry. The deconfined phase has Z$_2$ topological order, but in contrast to conventional topologically ordered states which are gapped, it also features gapless fermionic excitations, whose gaplessness is protected by the symmetries of the underlying Hamiltonian.   This is an example of a `nodal' Z$_2$ topological order, that has been invoked in the context of the square lattice antiferromagnet \cite{Senthil_2000}, and in the Kitaev model on the honeycomb lattice \cite{Kitaev06}. Here, we will augment the model to include both spin and charge conservation leading to a larger number of Dirac fermions (four flavors of complex two component fields)  which can be simulated without a sign problem. Thus, we will be studying how the confinement transition in a gauge theory is modified by the presence of gapless charged fermions. 

The confining phase, in the formulation used in our paper,  is an insulator
on the square lattice at half-filling 
with two-sublattice antiferromagnetic (AFM) order, similar to that found in a cuprate compound like La$_2$CuO$_4$. 
The topologically ordered phase is also at half-filling. 
This phase can be considered as a toy model of a `pseudogap' phase relevant to the doped cuprate superconductors, although our analysis will be restricted to half-filling. 
We will argue
that there is a direct transition between the insulating antiferromagnet and 
the phase with topological order, and present a critical field theory with an emergent non-abelian $SU(2)$ gauge field.

Emergent symmetries will also play an in important role in the analysis of our deconfined critical point. These are symmetries which are not present in the underlying Hamiltonian, or in the non-critical phases, but which become asymptotically exact at long distances and times in the critical regime.
An emergent $SO(5)$ symmetry was proposed \cite{Tanaka_2005,Senthil_2006,Wang_2017} for the deconfined critical point between the insulating AFM and valence bond solid (VBS) states on the square lattice: numerical computations on lattice models have observed such enhanced symmetries \cite{Nahum_2015,Suwa_2016,Karthik_2017,Sato_2017,Powell_2018}.
We will present numerical evidence for the same $SO(5)$ symmetry between the AFM and VBS order parameters in our model. This emergent symmetry is intriguing, because neither of the two phases near the transition (OSM or AFM) involves actual long-range VBS order, and yet VBS fluctuations become as strong as AFM fluctuations at the critical point. As we shall see later, this feature arises naturally in our formulation of the critical theory. 

We shall study a model introduced in
recent quantum Monte Carlo (QMC) studies \cite{Assaad2016,Gazit2017}. The model can be considered as an effective theory of electrons ($c$) on the square lattice.
The model is expressed as
an Ising lattice gauge theory (ILGT) coupled to `orthogonal' fermions ($f$). 
The QMC studies showed that this model exhibits a topological ordered `orthogonal
semi-metal' (OSM) phase: this phase hase a $\mathbb{Z}_2$ topological order, and massless Dirac fermion excitations which carry $\mathbb{Z}_2$ electric charges and also the
spin and electromagnetic charges of the underlying electrons ($c$). The charges carried by these fermions are identical to the `orthogonal fermions' introduced in Ref.~\cite{Nandkishore_2012}, and so we have adopted their terminology. The previous studies also presented evidence for a confining AFM phase, along with a possible direct and continuous phase transition between the OSM and AFM phases. However the underlying mechanism of this transition was not understood.

Our  QMC simulation finds numerical evidence for an emergent $SO(5)$ symmetry, rotating between AFM and VBS orders, at criticality. We contrast this finding with the more standard Gross--Neveu--Yukawa (GNY) universality class, where such an enlarged symmetry is absent. Guided by the numerical results, we conjecture that the critical theory describing the confinement transition is given by a two-color ($SU(2)$) quantum chromodynamics (QCD) with $N_f=2$ flavors of Dirac fermions coupled to a near critical matrix Higgs field. The Higgs mechanism is shown to naturally allow access to both the confined and deconfined phases, using a single tuning parameter. The mechanism described here resembles the theory of symmetric mass generation (SMG) \cite{You_2018,You_2018a} in some respects, but differs in the representation of the Higgs field under the gauge group. While SMG required Higgs fields transforming in the fundamental representation, so the Higgs phase was free of topological order, here we will employ a Higgs field in the adjoint representation, and the Higgs phase will therefore inherit a $\mathbb{Z}_2$ topological order along with gapless fermions. 

\section{Model, symmetries and phase diagram}

\subsection{Model} 
\begin{figure}
        \includegraphics[scale=0.6]{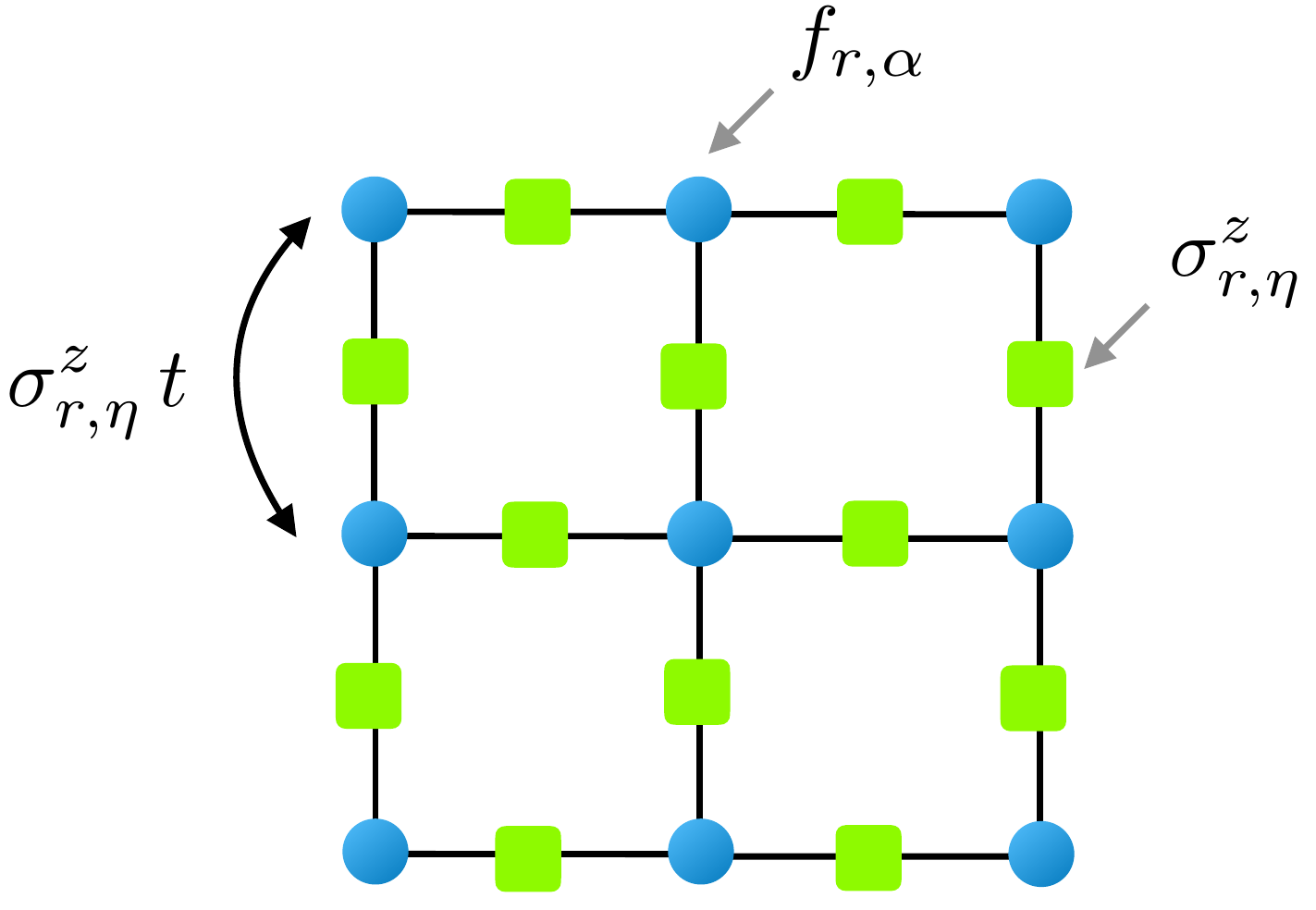}
        \caption{Lattice model -- the fermions $f_{r,\alpha}$ reside on the square lattice sites (blue circles) and the Ising gauge fields, $\sigma^z_{r,\eta}$ reside on the bonds (green squares). The Ising gauge field determines the sign of the hopping amplitude along the corresponding bond.}
        \label{fig:ILGT}
\end{figure}

We consider the Hamiltonian $\mathcal{H}=\mathcal{H}_{\mathbb{Z}_2}+\mathcal{H}_f$, illustrated in Fig.~\ref{fig:ILGT}. 
The ILGT part of the Hamiltonian \cite{WegnerILGT} reads,
\begin{equation}
\begin{gathered}
    \mathcal{H}_{\mathbb{Z}_2}=-J\sum_{\square}\prod_{b\in\square}\sigma^z_b-h\sum_b \sigma_b^x.
\end{gathered}
\label{eq:ILGT}
\end{equation}
Here, $\sigma^z_b$ and $\sigma^x_b$ are the conventional Pauli matrices, $\square$ labels the square lattice elementary plaquettes and $b=\{r,\hat{\eta}\}$ denotes the square lattice bonds with $r=\{r_x,r_y\}$ being the lattice site and $\hat{\eta}=\hat{x} / \hat{y}$.
The fermionic part of the Hamiltonian is given by,
\begin{equation}
    \mathcal{H}_f=-t\sum_{r,\hat{\eta},\alpha}\sigma_{r,\hat{\eta}}^z f_{r+\hat{\eta},\alpha}^\dagger f_{r,\alpha}+\text{h.c.}+U\sum_r\left(n^\uparrow_r-\frac 1 2\right)\left(n^\downarrow_r-\frac 1 2\right),
\label{eq:Hf}
\end{equation}
where, the operator $f_{r,\alpha}^\dagger$ creates an `orthogonal fermion' \cite{Nandkishore_2012} at site $r$ with spin polarization $\alpha$ and $n_r=\sum_{\alpha}f^\dagger_\alpha f_\alpha=n^{\uparrow}_r+n^\downarrow_r$ is the fermion density. 

By itself, this model cannot be a complete representation of the spin and charge excitations of a lattice electron model, like the Hubbard model. This is because it is not possible to write down a gauge-invariant electron operator, $c_r$, in terms of the $f_r$ and the $\sigma_{r, \hat\eta}$. We need another bosonic degree of freedom which carries a $\mathbb{Z}_2$ electric charge. We will introduce such a degree of freedom later in the Section~\ref{sec:rotate}; but for now, we assume that this boson is gapped in all the phases we study below, and we will not include it in our numerical study of $\mathcal{H}$.

As we demonstrate below, by varying the strength of the on-site Hubbard interaction term, in Eq.~\ref{eq:Hf}, we map a more generic phase diagram, compared to the ones obtained in Refs.~\cite{Assaad2016,Gazit2017}. Furthermore, this extension allows us to test the stability of the OSM confinement transition, and to compare its critical properties with the more standard GNY and three dimensional classical Ising universality classes. 

\subsection{Symmetries}  
The global and local symmetries of the Hamiltonian will play an important role in our analysis. First, the Hamiltonian is invariant under global $SU_s(2)$ rotations corresponding to spin rotation symmetry. Second, because we restrict ourselves to half-filling, 
our model is also invariant under the particle-hole (PH) transformation $f_\alpha \to (-1)^{r_x+r_y}f_\alpha^\dagger$. Finally, combining PH symmetry with the $U_c (1)$ symmetry corresponding to particle number conservation forms an enlarged $SU_c(2)$ pseudo-spin symmetry rotating between charge density wave (CDW) and superconducting order parameters \cite{Zhang_PseudoSpin}. 

Partial particle-hole (PH) symmetry, acting only on one of the spin species, maps between the charge, $n_r$, and the spin, $S^z_r=n^{\uparrow}_r - n_r^\downarrow$, operators.  Consequently, partial PH symmetry interchanges between the symmetries $SU_s(2)$ and  $SU_c(2)$ and, when these symmetries are broken, between AFM and BCS/CDW orders respectively. The Hubbard term in Eq.~(\ref{eq:Hf}) explicitly breaks partial PH symmetry, since under the symmetry action repulsive interaction is mapped to attractive interaction, $U\to-U$ \cite{auerbach_book}.

The correspondence of our model to lattice gauge theories (LGT) is manifest in the extensive number of {\it local} Ising symmetries generated by the operators $G_r=(-1)^{n_r}\prod_{b\in+_r}\sigma_b^x$, with $+_r$ denoting the set of bonds emanating from the site $r$. The eigenvalues,  $Q_r=\pm1$, of $G_r$ are conserved quantities and within the Hamiltonian formalism of LGT \cite{Kogut_RMP} are identified with the {\it static} background  $\mathbb{Z}_2$ charge. 

The Hilbert space then decomposes into a direct sum of subspaces labeled by the $\mathbb{Z}_2$ charge configuration $Q_r$, and comprises quantum states that obey an Ising variant of Gauss' law $G_r=Q_r$. For a uniform charge configuration, we can distinguish between two possibilities: an even LGT, $Q=1$, with no background charge and an odd LGT, $Q=-1$ with a single $\mathbb{Z}_2$ background charge at each site. We note that partial PH symmetry maps $Q\to-Q$. 

Gauss's law can be either explicitly enforced \cite{Gazit2017}, or alternatively, it is generated dynamically at sufficiently low temperatures \cite{Assaad2016}. In the numerical computation below, we will consider both options depending on numerical convenience. The zero temperature universal properties of our model, which are the focus of this study, do not depend on the above choice. 

\subsection{Phase diagram} 
\begin{figure*}[t!p]
    \centering
    \begin{subfigure}[b]{0.34\textwidth}
        \includegraphics[width=\textwidth]{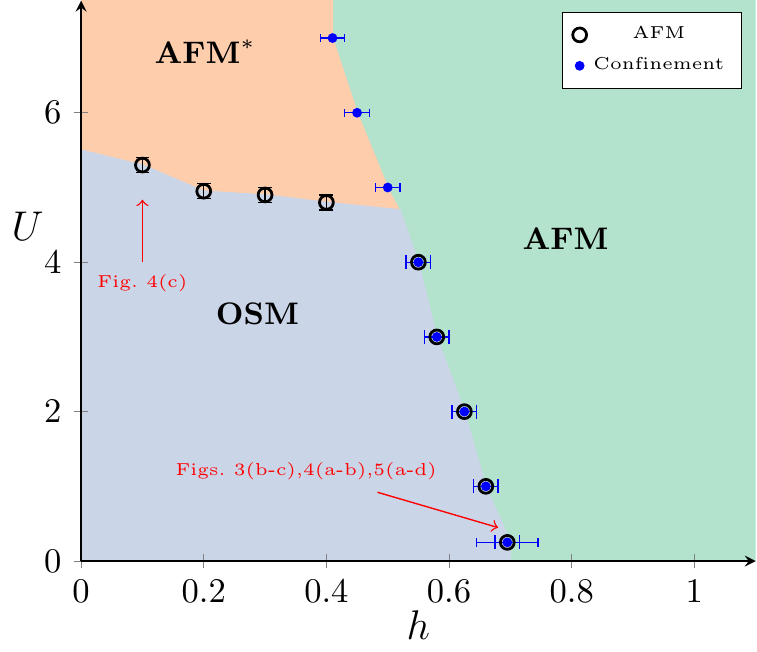}
        \caption{}
        \label{fig:phase_diag}
    \end{subfigure}
	\begin{subfigure}[b]{0.3\textwidth}
        \includegraphics[width=\textwidth]{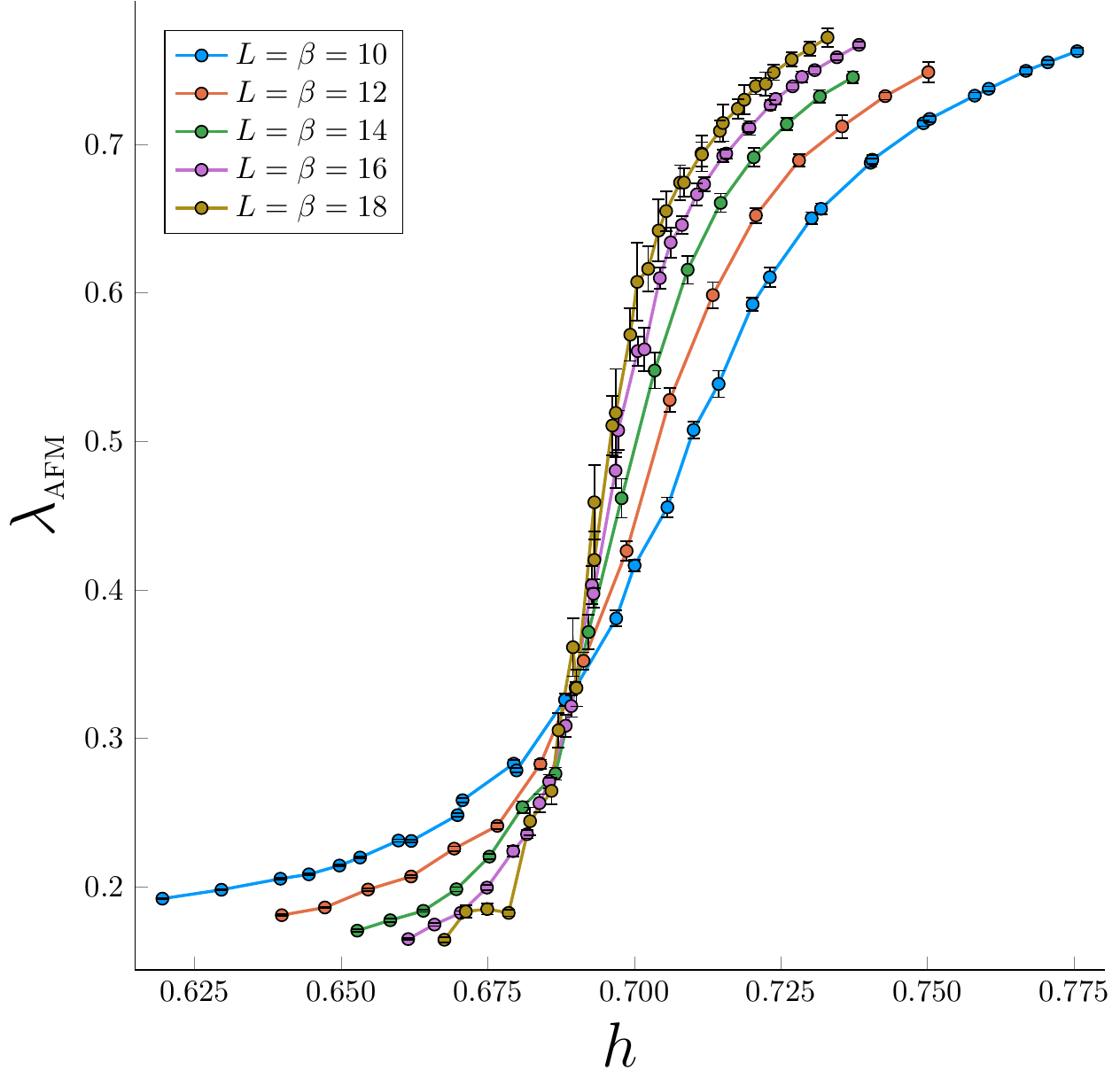}
        \caption{}
        \label{fig:lambda_ratio}
    \end{subfigure}
    \begin{subfigure}[b]{0.3\textwidth}
        \includegraphics[width=\textwidth]{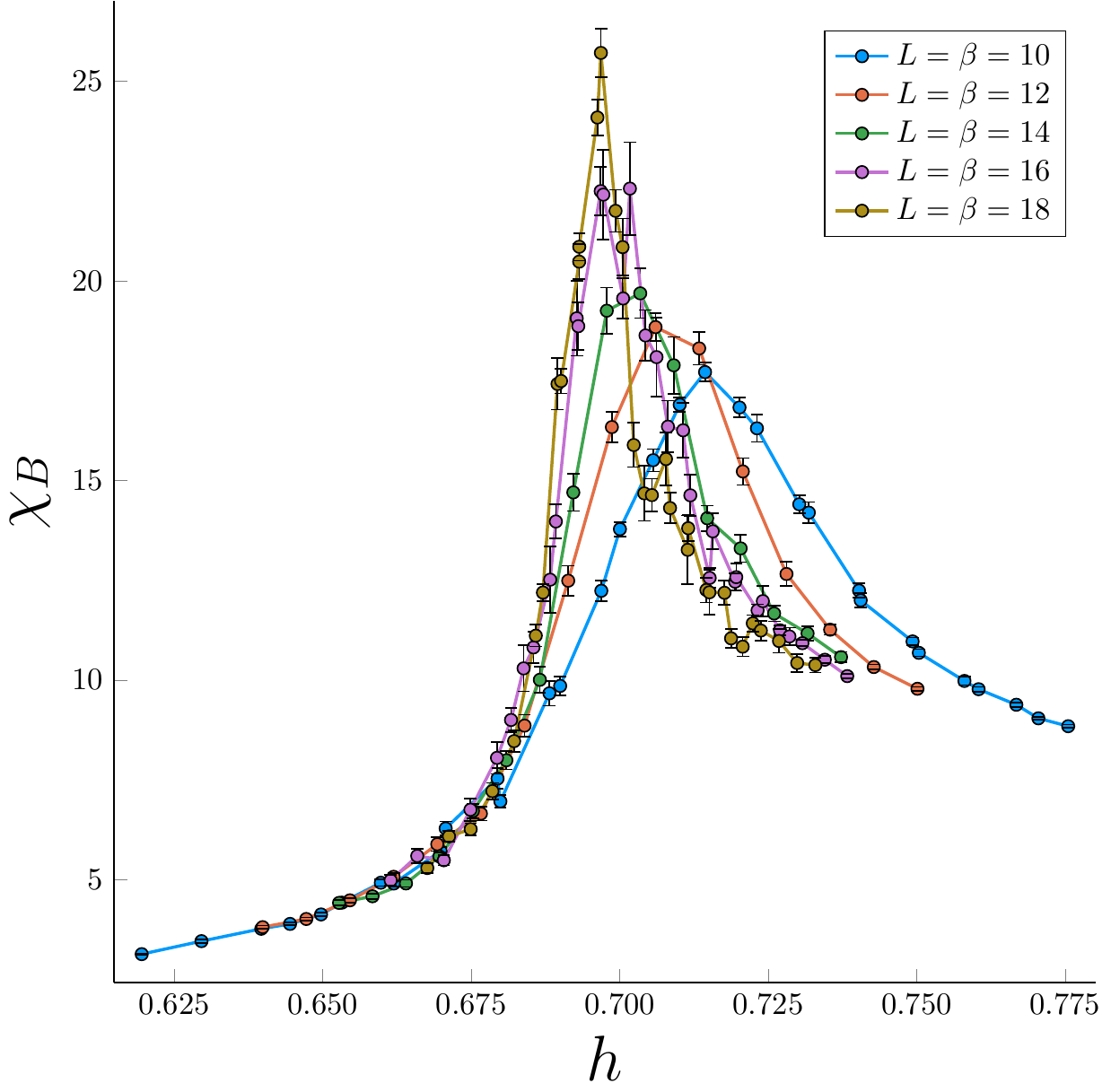}
        \caption{}
        \label{fig:chi_B}
    \end{subfigure}
    \caption{ (a) Phase diagram of the ILGT coupled to fermions. Red arrows point to parameter cuts studied in the figures below. (b-c) Simulation of the OSM confinement transition, carried out at $U=0.25$ as a function of $h$. (b) The onset of AFM order is located by a curve crossing analysis of the susceptibility ratio $\lambda_{\ttext{AFM}}$ and (c) the confinement transition is located from the divergence of the Ising flux susceptibility, $\chi_B$. }
\end{figure*}

We now determine  the general structure of the zero temperature phase diagram (see Fig.~\ref{fig:phase_diag}) by studying several limiting cases. 
For concreteness, we consider negative values of $J$ (the case $J>0$ is discussed in Ref.~\cite{Gazit2017}) and set $-t=|J|$. All other energy scales are measured in units of $|J|$. We will only consider the odd LGT case, which, as we explain in the following, is compatible with repulsive Hubbard interactions, $U>0$. The corresponding results for the even LGT can be easily obtained by applying a partial PH transformation with the appropriate identification of symmetries and order parameters, as discussed above. 

We first consider the strong coupling limit $h\gg t,U,|J|$. In the extreme limit $h\to\infty$, the ILGT ground state is given by the product state $\left|\Psi^{\sigma}_{\text{conf}}\right\rangle=\prod_b\left|\sigma_b^x=1\right\rangle$, as follows directly from minimizing the transverse field term in Eq.~(\ref{eq:ILGT}). In the above limit, we can safely neglect quantum fluctuations and substitute $\sigma_b^x=1$ in the Ising Gauss's law. This yields the relation $Q_r=(-1)^{n_r}$ such that the local fermion parity, $(-1)^{n_f}$, becomes a conserved quantity, identified with the background Ising charge, $Q_r$.

Following the standard LGT analysis \cite{Kogut_RMP}, we now establish the effective interaction between a pair of Ising charge excitations in the strong coupling limit. To comply with Ising Gauss's law, a string of flipped Ising gauge field, $\sigma_b^x=-1$, must connect any pair of Ising charges. The energy cost associated with each spin flip is proportional to $h$, and thus the interaction potential grows linearly with the separation giving rise to confinement.  

The repulsive Hubbard interaction favors single on-site occupancy and consequently gaps even parity (doublons and holons) states. The resulting low energy sector is an odd LGT with a emergent Gauss law constraint $G_r=-1$. This leaves the on-site fermion spin as the only remaining dynamical degree of freedom. Reintroducing quantum fluctuations, at large but finite transverse field $h$, allows for virtual hopping processes. Similarly to the super-exchange mechanism, such fluctuations induce an effective AFM Heisenberg interaction proportional to $t^2/h$ . The zero temperature ground state will then spontaneously break the spin rotational symmetry, $SU_s(2)$, by forming a N\'{e}el AFM state. 

Next, we examine the weak coupling limit $J\gg t,h,U$. Here, minimizing the Ising flux term in Eq.~(\ref{eq:ILGT}) (for negative $J$) realizes a uniform $\pi$--flux state, $\left|\Psi^\sigma_{\text{de-conf}}\right\rangle=\prod_\square\left|\Phi_\square=-1\right\rangle$, where $\Phi_\square=\prod_{b\in\square}\sigma^z_b$ is the Ising flux threading the elementary plaquette, $\square$. Crucially, the single-particle spectrum of the $\pi$--flux lattice hosts a pair of gapless Dirac fermions \cite{Affleck_1988}. In the resulting phase, the matter fields are deconfined, since, in contrast to the confining phase, gauge field fluctuations mediate only short-range a attractive interaction with a vanishing string tension. The deconfined phase hosts fractionalized excitations carrying long-range entanglement \cite{KITAEV_2003}. We note that a $\pi$--flux phase can be stabilized even if $J$ is positive by taking the large hopping amplitude $t$ limit \cite{Gazit2017,Assaad2016}.

The gapless deconfined phase resembles the well-known gapless $\mathbb{Z}_2$ spin liquid, using the condensed matter theory (CMT) parlance \cite{Senthil_2000,Wen_2002}. However, there is one crucial difference: in our case the fermionic matter fields carry in addition to the SU(2) spin charge (similarly to conventional spinons) an $U(1)$ electromagnetic charge. This makes our model more closely related to an orthogonal-metal construction \cite{Nandkishore_2012}, where the fractionalization pattern involves decomposing the physical fermion into a product of a fermion carrying both spin and charge and an Ising spin. Both slave particles carry an Ising gauge charge. We, therefore, dub this phase by the name orthogonal semi-metal (OSM).

Due to the vanishing density of states at half-filling, the Dirac phase is stable against AFM order for weak Hubbard interactions, $U\ll t$. However, a transition to an AFM$^*$ phase is expected at sufficiently large coupling. Here, the asterisk expresses the fact that the gauge theory remains deconfined in the AFM$^*$ phase. This situation should be contrasted with the confined phase, where along with AFM symmetry breaking order, the gauge sector is confined.

\subsection{Phase transitions} 

The different phases of our model are classified according to the presence or absence of topological order and conventional symmetry breaking AFM order. Thus, the associated phase transitions are expected to involve either confinement or symmetry breaking or both. 

More specifically, the phase transition between the deconfined Dirac phase and the AFM$^*$ phase is solely marked by the rise of AFM order, while the Ising gauge field sector remains deconfined. Therefore, the transition belongs to the conventional chiral GNY universality class \cite{Herbut_GN,Assaad13,Toldin14,Sorella_GN}.  On the other hand, across the transition between the confined AFM and AFM$^*$ phases the gapped fermions are only spectators and the transition is signaled by the emergence of topological order in the AFM$^*$ phase. Thus, the phase transition corresponds to the standard confinement transition of the pure Ising gauge theory, which belongs to the three dimensional classical Ising model universality class (the spin-wave (Goldstone) modes are not expected to modify the universality class of this transition, as can be seen by the methods of Ref.~\cite{SSTM02}).

The most interesting phase transition, which is the subject of this study, is between the  deconfined Dirac phase and the confined AFM. Previous numerical simulations \cite{Gazit2017,Assaad2016} and new results shown below have found evidence for a single and continuous phase transition involving both symmetry breaking and confinement. Gaining a better analytic and numerical understanding of this transition is the main subject of the remainder of this paper.

\begin{figure*}[t]
    \centering
    \begin{subfigure}[b]{0.3\textwidth}
        \includegraphics[width=\textwidth]{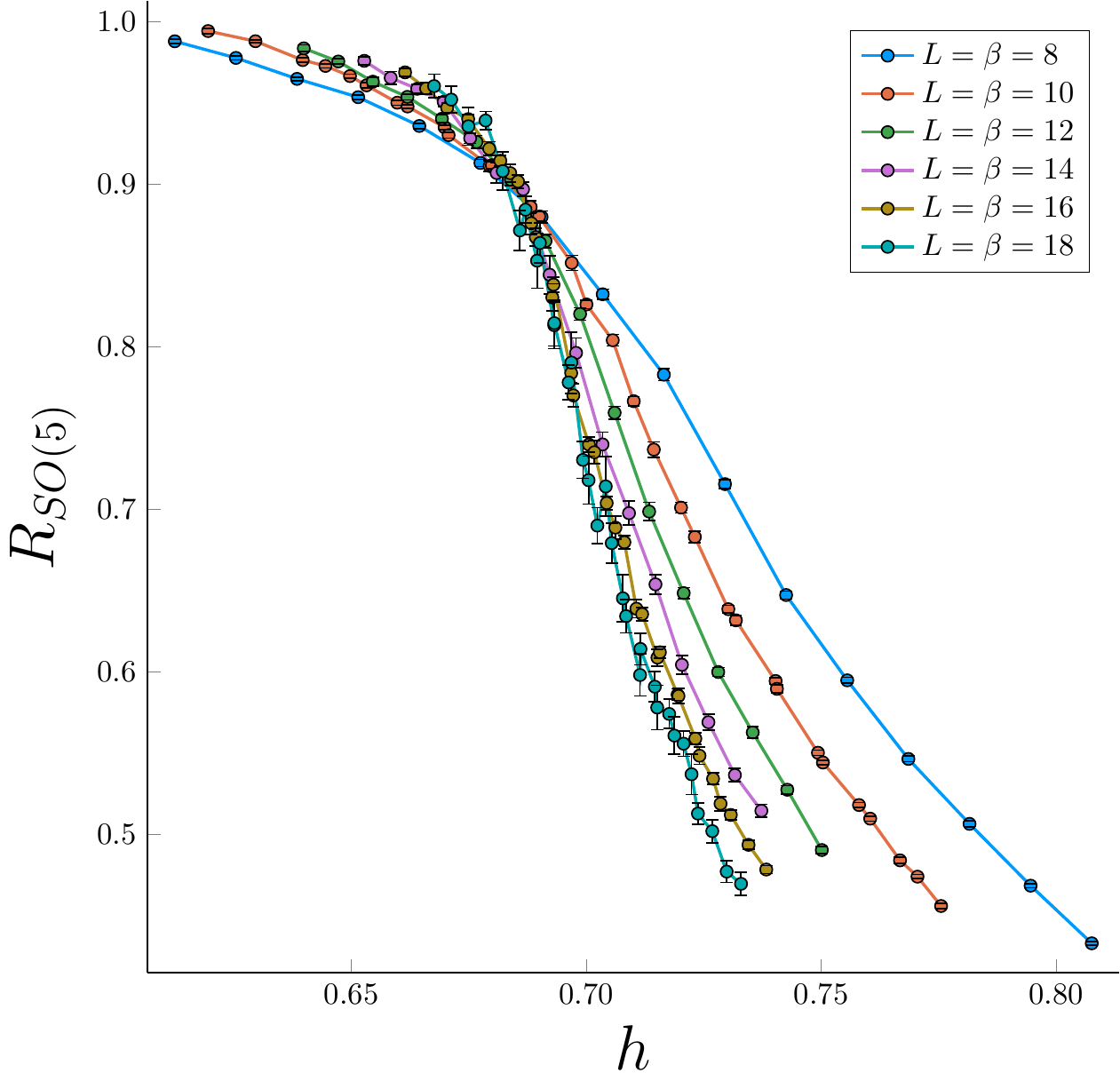}
        \caption{}
        \label{fig:ratio_QCD}
    \end{subfigure}
    \begin{subfigure}[b]{0.37\textwidth}
        \includegraphics[width=\textwidth]{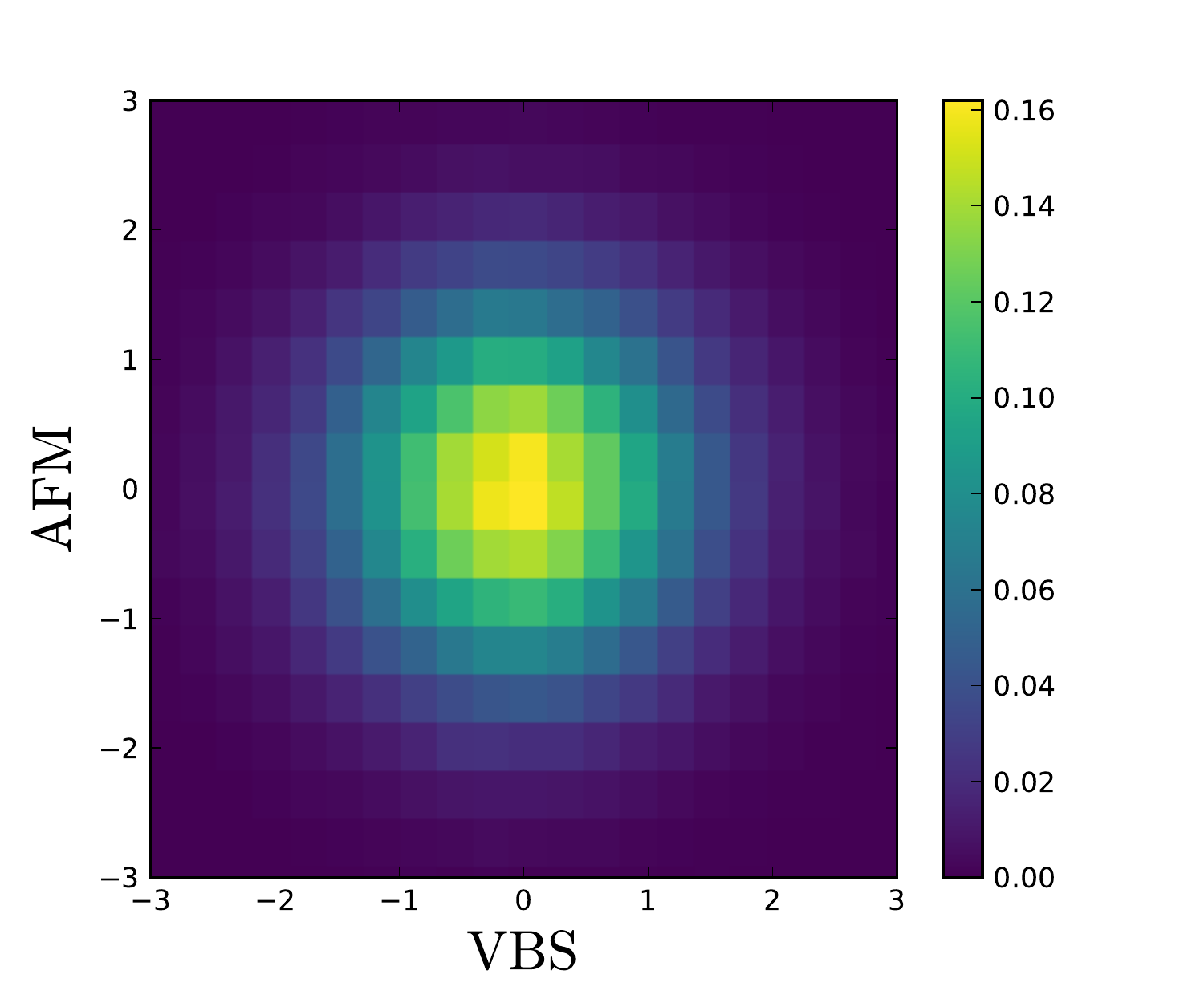}
        \caption{}
        \label{fig:hist_VBS_AFM}
    \end{subfigure}
     \begin{subfigure}[b]{0.3\textwidth}
        \includegraphics[width=\textwidth]{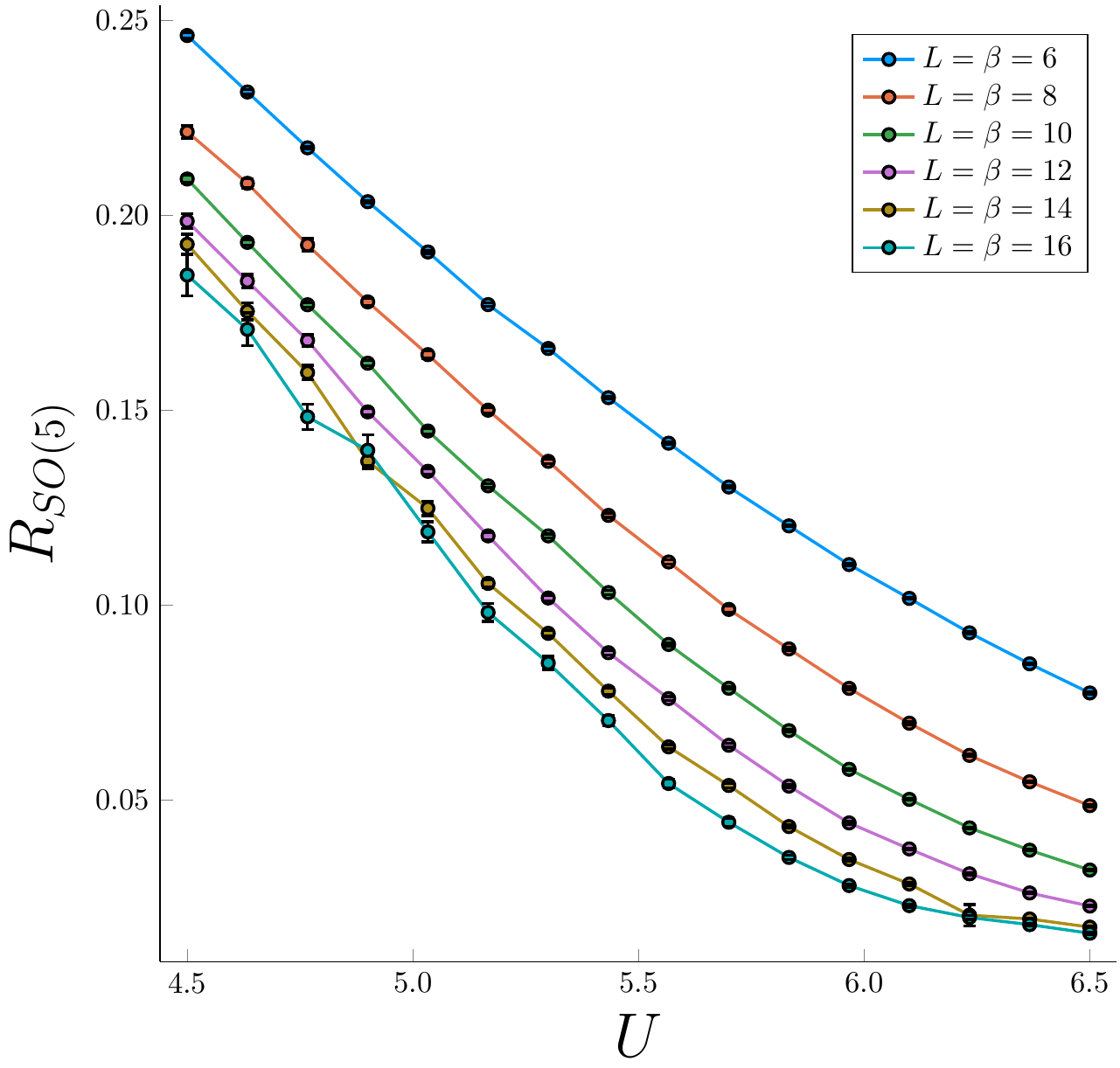}
        \caption{}
        \label{fig:ratio_GN}
    \end{subfigure}
    \caption{ Signature of an $SO(5)$ symmetry. (a) A clear curve crossing is observed in the  susceptibility ratio $R_{SO(5)}$ across the OSM confinement transition for $U=0.25$ as a function of $h$. (b) Joint probability distribution $\mathbf{P}(D^x,\mathbf{S}^z)$ of the VBS and AFM order parameters at criticality. $\mathbf{P}(D^x,\mathbf{S}^z)$ exhibits a circular symmetry (c) Susceptibility ratio $R_{SO(5)}$ across the AFM transition for $h=0.1$ as a function of $U$. The absence  of curves crossing rules out the emergence of an $SO(5)$ symmetry at the GNY transition.} \label{fig:ratio_rg}
\end{figure*} 
\section{Quantum Monte Carlo}

\subsection{Methods} 

The ILGT coupled to fermions is free of the numerical sign-problem for arbitrary fermion density (here we are interested only in the half-filled case) \cite{Gazit2017,Assaad2016}. This allows us to study our model using an unbiased and a numerically exact (up to statistical errors) QMC simulations. We employ the standard auxiliary-field QMC algorithm \cite{Assaad2008,ALF_v1} using both single spin-flip updates and global moves inspired by the worm algorithm \cite{Gazit2017}. In all cases, we set the imaginary time Trotter step to be $|t|\Delta\tau =1/12$, a value for which the discretization errors are controlled. In what follows, we set $t=J=-1$ and explore the phase diagram as a function of $h$ and $U$. Unless otherwise stated, we also explicitly impose Gauss's law constraint. Further technical details of our numerical scheme as well as additional numerical data can be found in Appendices~\ref{appA}, \ref{appB} and \ref{appC}.

\subsection{Observables} 

We probe the VBS and AFM order parameters using the bond kinetic energy, $\mathbf{D}^{x/y}$, and the fermion spin, $\mathbf{S}^{\gamma}$, operators, respectively. Their corresponding lattice definitions at finite wave vector, $q$, are given by,
\begin{equation}
\begin{aligned}
\mathbf{D}^{\eta}(q)&=\sum_{r,\alpha} e^{i q\cdot r}\left(\sigma^z_{r,\eta}f^\dagger_{r+\eta,\alpha}f_{r,\alpha}+\text{h.c}\right) \\
\mathbf{S}^{\gamma}(q)&=\sum_{r,\alpha,\beta}e^{i q\cdot r}f^\dagger_{r,\alpha}\mathbf{\tau}^\gamma_{\alpha\beta}f_{r,\beta}
\end{aligned}
\label{eq:BSdef}
\end{equation}
where, $\mathbf{\tau}_{\alpha\beta}^{\gamma}$ are the usual Pauli matrices. 

On the $\pi$-flux square lattice, the set of fermion bilinears appearing in Eq.~\ref{eq:BSdef} form a five component super-vector that transforms as a fundamental under $SO(5)$ rotations. Within this formalism, the competition between AFM and VBS fluctuations is explicitly manifest \cite{Tanaka_2005,Senthil_2006}.

To study fluctuations, we use the imaginary time static susceptibility, which for a generic operator, $\mathcal{O}$, is defined by $\chi_{\mathcal{O}}(q)=\frac {1} {\beta L^2} \left\langle \left(\int_0^\beta d\tau \,\mathcal{O} (q,\tau) \right)^2\right\rangle$. Here, expectation values are defined with respect to the thermal density matrix, $\beta=1/T$ is the inverse temperature $T$ ,and $L$ is the linear system size. The ordering wave vector associated with AFM (VBS) order (along the $\hat{x}/\hat{y}$ bonds) equals $G_{\ttext{AFM(VBS)}}=\{\pi,\pi\}(\{\pi,0\}/\{0,\pi\})$. 

To locate the onset of AFM order, we use the renormalization group (RG) invariant ratio $\lambda_{\ttext{AFM}}=1-{\chi_{\mathbf{S}}\left(G_{\ttext{AFM}}\right)}/{\chi_{\mathbf{S}}\left(G_{\ttext{AFM}}-\Delta q\right)}$, with $|\Delta q|=2\pi/L$ being the smallest wave vector on our finite lattice. $\lambda_{\ttext{AFM}}$ approaches unity deep in an AFM phase and vanishes when the symmetry is restored \cite{Pujari_2016}. For a continuous transition, curves of $\lambda_{\ttext{AFM}}$ corresponding to different Euclidean space-time volumes are expected to cross at the critical coupling. Anticipating the emergence of strong VBS fluctuations at criticality, we also define the analogous RG ratio, $\lambda_{\ttext{VBS}}=1-{\chi_{\mathbf{D}}\left(G_{\ttext{VBS}}\right)}/{\chi_{\mathbf{D}}\left(G_{\ttext{VBS}}-\Delta q\right)}$.

For pure lattice gauge theories, it is standard to probe confinement via the Polyakov loop \cite{Polyakov_1978}. In the presence of matter fields, the Polyakov loop no longer sharply defines confinement due to charge screening. In principle, one can detect the rise of topological order by extracting the topological contribution to the entanglement entropy \cite{Isakov2011,GroverEE_2013} or by measuring the Fredenhagen-Marcu \cite{Fredenhagen_1986,Gregor_2010} order parameter. However, such probes are difficult to reliably scale with system size in fermionic QMC simulations. In our analysis, we detect the thermodynamic singularity associated with the confinement transition by probing the expected divergence of the Ising flux susceptibility, $\chi_B=\partial\langle \Phi \rangle/\partial J$, with $\Phi$ being the Ising flux density defined above \cite{Gazit2017}.

\subsection{Numerical Results} 

Our first task is to determine numerically the phase diagram shown in Fig.~\ref{fig:phase_diag}. We exemplify our analysis by studying the OSM confinement transition. For concreteness, we fix $U=0.25$, and drive the transition by increasing the strength of the transverse field, $h$. In our finite size scaling analysis, we consider linear system sizes up to $L=18$. We further assume relativistic scaling and accordingly consider inverse temperatures that grow linearly with the system size, $\beta=L$.

In Fig.~\ref{fig:lambda_ratio}, we track the evolution of $\lambda_{\ttext{ AFM}}$ as a function of $h$. We observe a clear curve crossing that varies very little with system size and strongly indicates a continuous transition. The crossing point marks the rise of AFM order and allows us to estimate the critical coupling, $h^{\ttext{ AFM}}_c(U=0.25)=0.69(2)$. Moving to the IGLT sector, in Fig.~\ref{fig:chi_B}, we depict the Ising flux susceptibility, $\chi_B$, across the confinement transition. With increase in the system size, $\chi_B$ displays a progressively diverging and narrowing peak. We use the peak position to estimate the critical coupling of the confinement transition to be $h^{\text{conf}}_c=0.69(2)$. This value coincides, within the error bars, with the emergence of AFM order, found above, suggesting that symmetry breaking and confinement occur simultaneously.

We employ a similar analysis to determine the rest of the phase boundaries appearing in Fig.~\ref{fig:phase_diag}. We find that the critical confinement line separating the AFM and AFM$^*$ phases meets with the AFM transition line separating the OSM and the AFM$^*$ phases at a tricritical point. The two critical lines then merge into a single line corresponding to the OSM confinement transition. 

We now test the emergence of enlarged symmetries in the OSM confinement transition. In the presence of an $SO(5)$ symmetry, the scaling dimension of the VBS and AFM order parameters must coincide \cite{Nahum_2015}. As a direct consequence, similarly to $\lambda_{\ttext{AFM}}$, the susceptibilities ratio, $R_{SO(5)}=\chi_{\ttext{AFM}}(G_{\ttext{AFM}})/\chi_{\ttext{VBS}}(G_{\ttext{VBS}})$ becomes a renormalization group (RG) invariant.

In Fig.~\ref{fig:ratio_QCD}, we depict the susceptibility ratio, $R_{SO(5)}$, as a function of $h$, across the confinement transition, for different system sizes. Indeed, we find that all curves cross at a single point, independent of the space-time volume. We use the crossing point to pin down the critical coupling, $h^{SO(5)}_c=0.69(2)$, in excellent agreement with the above calculations, using other observables. We note that this result is a necessary but not a sufficient condition for the emergence of an $SO(5)$ symmetry. Nevertheless, it serves as a non-trivial test for this effect.

\begin{figure}[t]
    \centering
    \begin{subfigure}[b]{0.24\textwidth}
        \includegraphics[width=\textwidth]{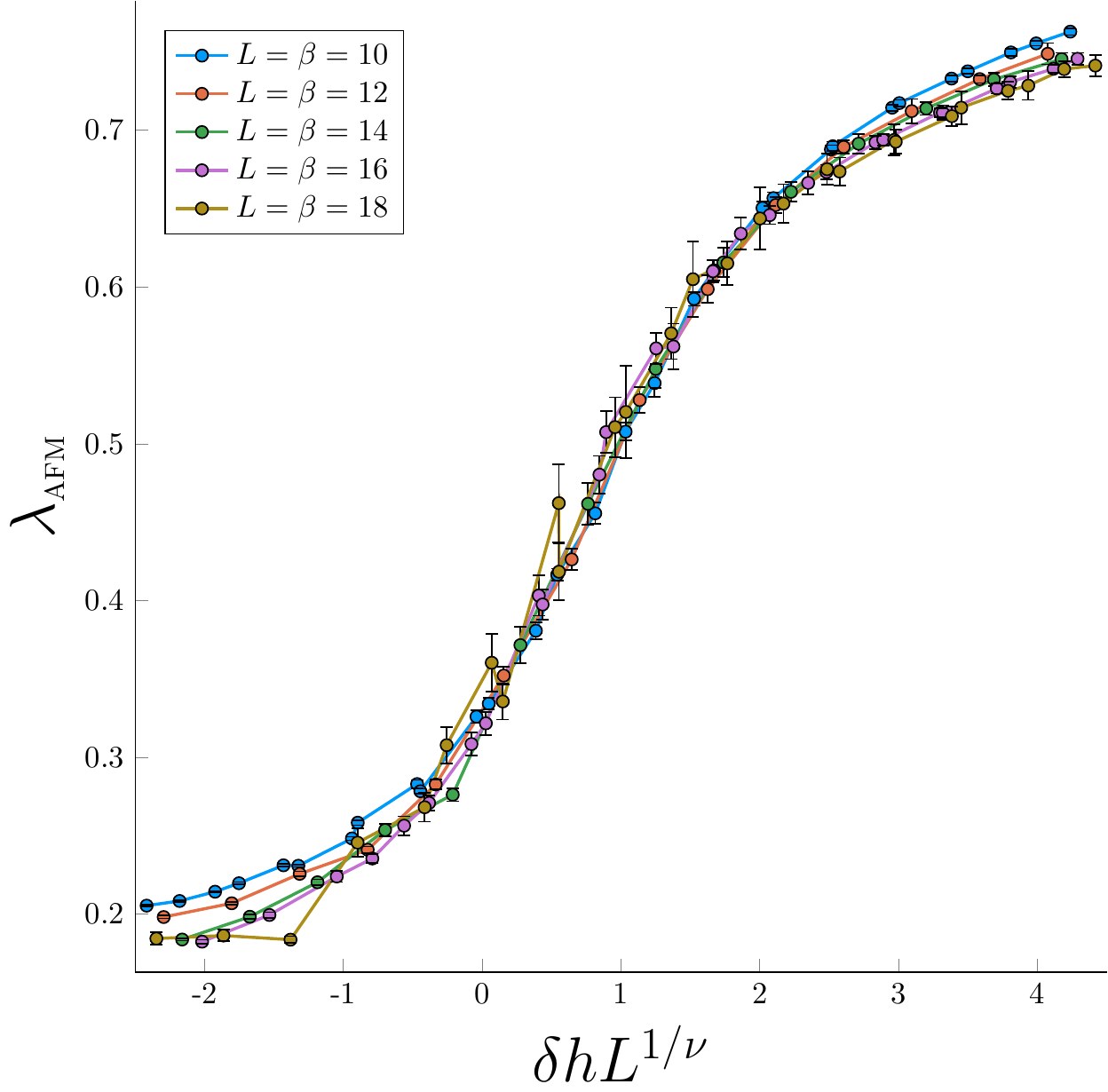}
        \caption{}
        \label{fig:afm_lambda_scale}
    \end{subfigure}
    \begin{subfigure}[b]{0.24\textwidth}
        \includegraphics[width=\textwidth]{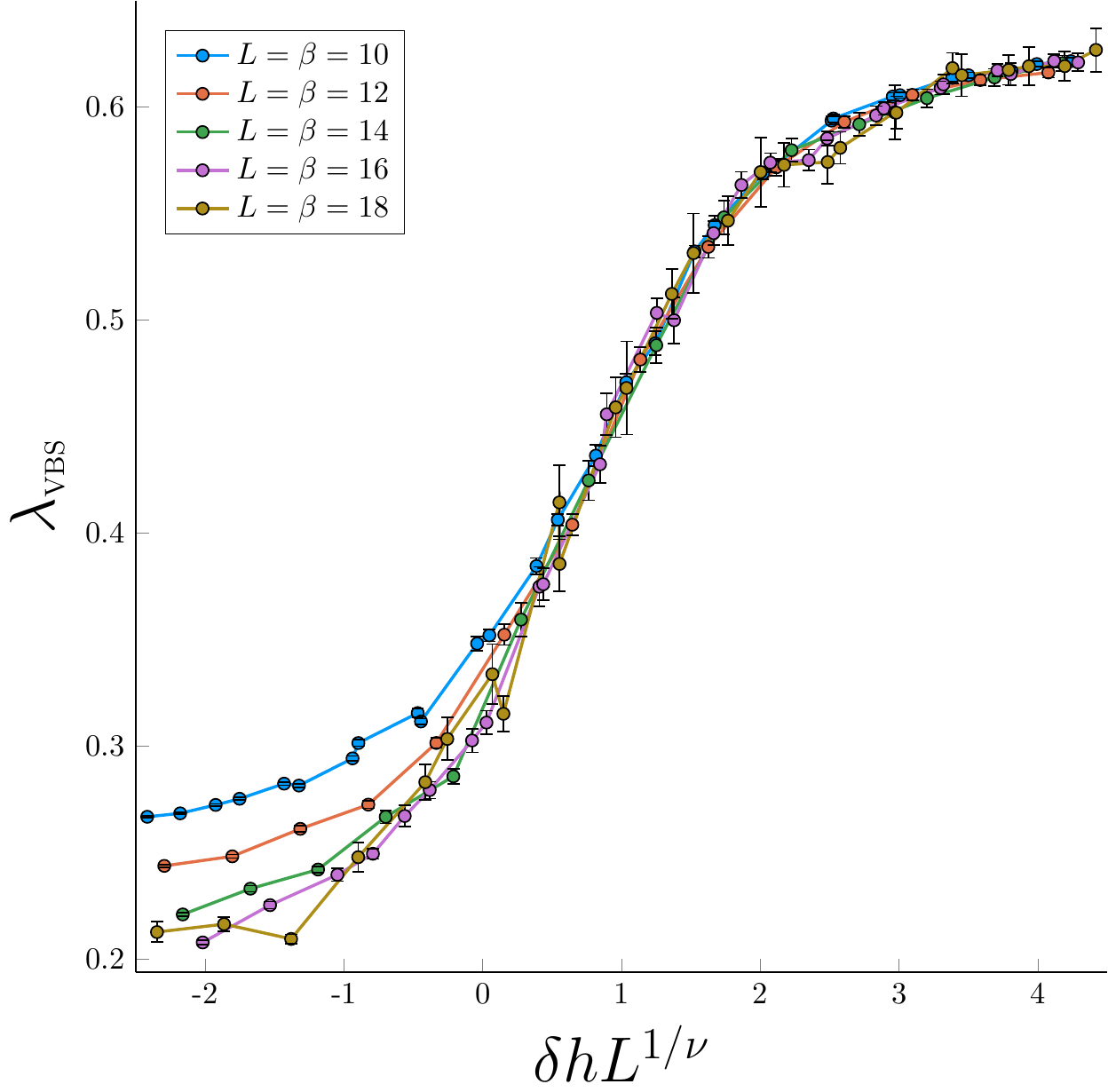}
        \caption{}
        \label{fig:vbs_lambda_scale}
    \end{subfigure}\\
    \begin{subfigure}[b]{0.24\textwidth}
        \includegraphics[width=\textwidth]{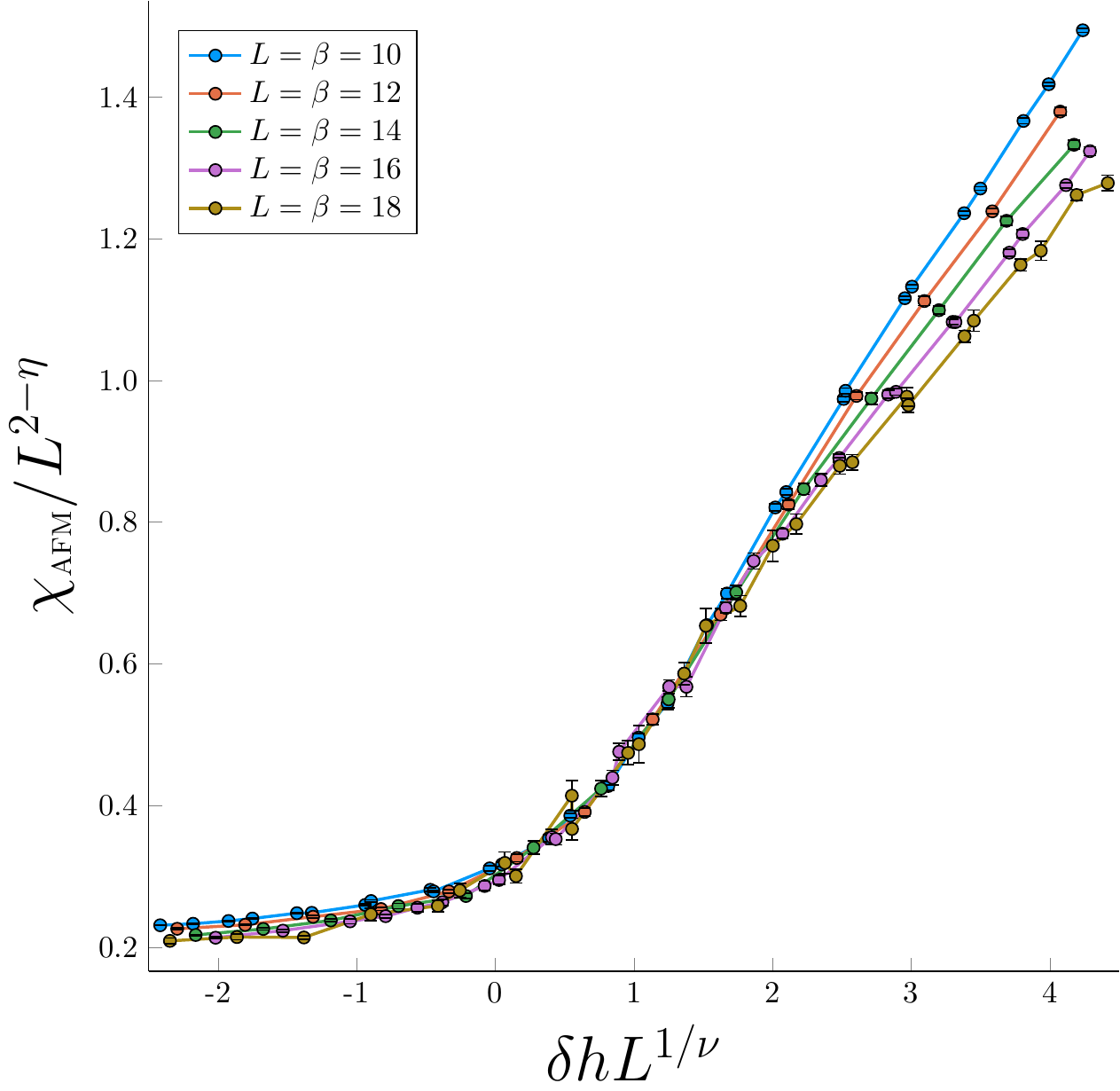}
        \caption{}
        \label{fig:scaling_afm}
    \end{subfigure}
    \begin{subfigure}[b]{0.24\textwidth}
        \includegraphics[width=\textwidth]{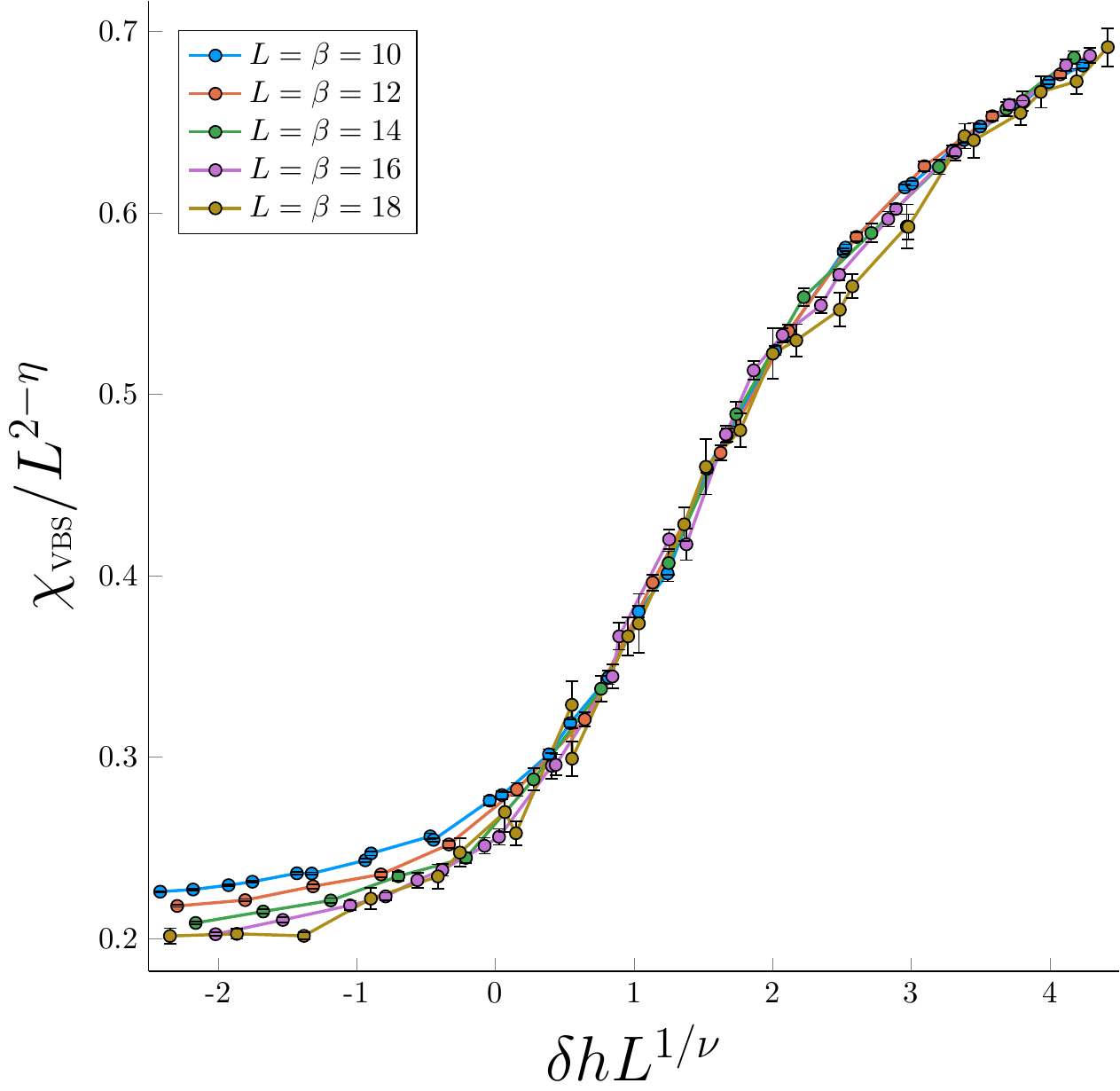}
        \caption{}
        \label{fig:scaling_vbs}
    \end{subfigure}
    \caption{ Finite size scaling analysis of (a) $\lambda_{\ttext{AFM}}$ (b) $\lambda_{\ttext{VBS}}$ (c) $\chi_{\ttext{AFM}}$ and (d) $\chi_{\ttext{VBS}}$. In all cases, curve collapse is obtained using the critical coupling $h_c=0.69$, the correlation length exponent $\nu=0.58$. The {\it same} anomalous exponent $\eta=1.4$ is used to scale both the AFM and VBS fluctuations.}\label{fig:scaling}
\end{figure}
To further illustrate the emergence of an $SO(5)$ symmetry, in Fig.~\ref{fig:hist_VBS_AFM}, we depict a two dimensional histogram approximating the joint probability distribution of the VBS and AFM order parameters at criticality.  We note that due to algorithmic limitations, in computing the AFM histogram, one must simulate the constraint-free model. Doing so, only slightly shifts the  critical coupling and as explained above, it does not affect critical properties. Remarkably, the joint distribution exhibits a circular form, which provides further indication for the emergence of an $SO(5)$ symmetry. We have also verified, using a similar analysis, that the joint probability distribution of the VBS order along the x and y directions affords an emergent, $SO(2)$, rotational symmetry at criticality, see Appendix~\ref{appB}.

To better appreciate the above result, it is instructive to apply the susceptibility ratios analysis on the more conventional GNY transition. To that end, we investigate the transition between the OSM phase and the AFM$^*$phase. We fix $h=0.1$, and cross the AFM transition by increasing $U$. The results of this analysis are shown in Fig.~\ref{fig:ratio_GN}. In stark contrast to the confinement transition, we find no evidence for a curve crossing. Thus, we can deduce that the putatively continuous OSM confinement transition must belong to a universality class that is {\it distinct} from the conventional GNY transition. This conclusion is one of our main results.

Motivated by the above results, we now extract the critical properties of the OSM confinement transition from the numerical data. The dimensionless susceptibility ratios are expected to follow a simple scaling form $\lambda_{\ttext{AFM/VBS}}(h,L)=\tilde{\lambda}_{\ttext{AFM/VBS}}(\delta h L^{1/\nu})$, where $\delta h=h-h_c$ defines the quantum detuning parameter from the critical coupling $h_c$, and $\nu$ is the correlation length exponent. In Figs.~\ref{fig:afm_lambda_scale} and \ref{fig:vbs_lambda_scale} we present the universal scaling functions $\tilde{\lambda}_{\ttext{AFM/VBS}}$ obtained from a curve collapse analysis using $h_c=0.69(2)$ and $\nu=0.58(1)$.

In the presence of $SO(5)$ symmetry, the AFM and VBS order parameters are expected to share the same anomalous exponent $\eta$. We assume the standard scaling form $\chi_{\ttext{VBS/AFM}}=L^{2-\eta}\tilde{\chi}_{\ttext{VBS/AFM}}(\delta h L^{1/\nu})$, where  $\tilde{\chi}_{\ttext{VBS/AFM}}$ are the universal scaling functions of the VBS and AFM order parameters. In Figs.~\ref{fig:scaling_afm} and \ref{fig:scaling_vbs} we depict the universal scaling functions $\tilde{\chi}_{\ttext{VBS/AFM}}$ using our previous estimates for $h_c$ and $\nu$ and the {\it same} anomalous exponent $\eta=1.4(1)$. The increased system size and improved methodology used is this work allowed for a more reliable determination of critical exponents, compared to the ones appearing in Ref.~\cite{Gazit2017}. 

In the above scaling analysis, we found that curves corresponding to the smallest system sizes deviate from the expected universal curve. These scaling violations are most likely attributed to non-universal corrections to scaling that may be sizable at small system sizes. Nevertheless, we note that the critical regime over which we obtain a nearly perfect curve collapse systematically increases with the systems size.

We note that although the AFM and VBS exponents co-incide, the scaling functions, $\tilde{\chi}_{\ttext{VBS/AFM}}$,
in Figs.~\ref{fig:scaling_afm} and \ref{fig:scaling_vbs} do not appear to be the same. The theory to be presented
in Section~\ref{sec:crit} requires these functions to be the same at leading order, with differences only appearing upon considering corrections to scaling. This feature needs to be understood better in future work.

\section{Critical theory of the confinement transition}
\label{sec:crit}

\subsection{Previous work}

It is useful to first recall other theories of confinement transitions out of a state with $\mathbb{Z}_2$ topological order \cite{SSreview}. 
The confinement transition of the even ILGT without dynamical matter was already described by Wegner \cite{WegnerILGT}, 
which he showed was in the (inverted) Ising universality
class.
The odd ILGT without dynamical matter has a confinement transition to a state with VBS order, and the square lattice critical point is described by a deconfined $U (1)$ gauge theory \cite{RJSS91,SSMV99,Senthil_DC2}. This can be understood by viewing the $\mathbb{Z}_2$ gauge theory of the topological state
as a compact $U(1)$ gauge theory in which a charge 2 Higgs field has condensed \cite{FradkinShenker}. Then the uncondensing of the Higgs
field leads to a confining phase of the $U (1)$ gauge theory, across a critical point where the U(1) gauge fields are deconfined: the background $\mathbb{Z}_2$ electric charges of the odd ILGT suppress the $U(1)$ monopoles at the critical point, leading to deconfinement. This furnishes an example of an enlarged gauge group appearing at the confinement-deconfinement critical point of a $\mathbb{Z}_2$ gauge theory. Analogously, we will see that for our problem of confinement of ILGT coupled to massless fermions, enlarging the gauge group can account for this transition as well. However, here we will need to introduce an SU(2) gauge symmetry as described below.

\subsection{Fractionalization and Higgs field: parton construction}
\label{sec:parton}

The $f$ fermions that appear in the Ising gauge theory can be constructed via the following `parton' construction by fractionalizing the physical, gauge invariant degrees of freedom. Notice, the gauge invariant operators in that model are purely bosonic, and include the spin ${\bf S}$ and psuedospin ${\bf I}$ generators. The latter include the U(1) charge operators $I^z$, and $I^\pm$ that create/destroy charged bosons. These can be decomposed into partons as follows. First define:
\begin{equation}
    X_r = \left ( \begin{array}{cc}
       f_{r\uparrow}  & -f^\dagger_{r\downarrow} \\
        f_{r\downarrow} & f^\dagger_{r\uparrow}
    \end{array}\right )
\end{equation}
The spin and psuedospin rotations act via multiplication of SU(2) matrices to the right or left: $X \rightarrow U^s X [U^{ps}]^\dagger$. Then the physical operators are:
\begin{eqnarray}
{\bf S}_r = \frac14 {\rm Tr} \{  X_r^\dagger {\bm \tau} X_r\}; \,\,\,&& \,\,\,{\bf I}_r = \frac14 {\rm Tr} \{  X_r {\bm \mu} X^\dagger_r\} 
\label{eq:X}
\end{eqnarray}
Here we are using the convention for spin/pseudospin Pauli matrices ${\bm \tau}/{\bm \mu}$ from Eq.~(\ref{eq:BSdef}).
Clearly there is a $\mathbb{Z}_2$ gauge redundancy in this definition corresponding to changing the sign of the fermion operators. Thus a minimal parton Hamiltonian will have hopping of $f$ fermions mediated by an Ising ($\mathbb{Z}_2$) gauge field, as in to our starting model. However, in order to accomplish the observed transition we will need a different set of variables. To this end, define a fermion matrix field $Y_r$ which is superficially similar to the $X_r$ above, however which only carries the spin quantum number. The psuedospin is assumed to be carried by a triad of  bosonic matrix fields $\hat{H}_a$, $a=1,\,2,\,3$ each of which is a $2\times2$ matrix. This can also be written as $\hat{H}_a = \sum_{b=1}^3 {H}_{ab} {\mu}^b=\vec{H}_a \cdot\vec{\mu}$. In terms of these fields we can decompose the physical operators as:
\begin{eqnarray}
{\bf S}_r = \frac14 {\rm Tr} \{  Y_r^\dagger {\bm \tau} Y_r\} ;\,\,\,&& \,\,\, {I}_{ar} = \frac14 {\rm Tr} \{  Y_r \hat{H}_{ar} Y^\dagger_r\} 
\label{eq:Y}
\end{eqnarray}
While spin rotations are implemented as before $Y\rightarrow U^sY$, psuedospin rotations only act on $\hat{H}^a$ which transforms  as a vector. This decomposition though has additional gauge freedom, for instance we can simultaneously rotate:  
\begin{equation}
  Y_r \rightarrow Y_r \left [ U^g_r \right ]^\dagger ;\;\;\; \hat{H}_{ar} \rightarrow U^g_r \hat{H}_{ar} \left [ U^g_r \right ]^\dagger 
\end{equation}
which leaves the physical operators invariant. Therefore this decomposition has an SU(2) gauge redundancy. Therefore the effective theory will now involve $Y$ fermions coupled to an SU(2) gauge field. We can readily recover the $\mathbb{Z}_2$ Dirac  phase as follows. Consider a Higgs transition in which the fields $H_{ab}$ acquire an expectation value: 
\begin{equation}
\langle H_{ab} \rangle =H_0  \delta_{ab}\,.  \label{H0}
\end{equation}
Then, $\hat{H}_a  = H_0 \mu^a$ and Eq.~(\ref{eq:Y}) reduces to Eq.~(\ref{eq:X}).  
We will later see that the dynamics at the transition will naturally favor such a Higgs condensate.

\subsection{Fractionalization and Higgs field: Rotating reference frame construction}
\label{sec:rotate}

An alternate derivation of the fractionalized degrees of freedom can be obtained by first expanding the Hilbert space of the model to include electron excitations $c_\alpha$.  We can then show that the AFM and VBS order parameters of the possible confining phases,
and the orthogonal fermions $f_\alpha$ of the $\mathbb{Z}_2$ deconfined phase, all emerge
by transforming the underlying gauge-invariant electrons, $c_\alpha$, to a rotating reference frame under $SU_c(2)$. 

A similar approach was adopted in Refs.\cite{SSNambu,CSS17} which considered phases with $\mathbb{Z}_2$ topological order in which there are dynamical fermions carrying $\mathbb{Z}_2$ gauge charges and the global $U_c(1)$ charge ($U_c(1)$ is a subgroup of $SU_c(2)$), but these fermions 
are spinless under $SU_s (2)$. The transition of these phases to confining Fermi liquids (which can be unstable to superconductivity) was described by embedding the $\mathbb{Z}_2$ gauge theory in a $SU(2)$ gauge group. This larger gauge group was needed for a proper description of the confining phase in terms of composites of  the fractionalized degrees of freedom \cite{SS09}. It was introduced by transforming to a `rotating reference frame' under $SU_s (2)$. In the topological phase, the $SU(2)$ gauge invariance was broken down to $\mathbb{Z}_2$ by condensing a $SO(3)$ Higgs field which was neutral under $U_c(1)$ and $SU_s (2)$.

In our case, we transform to a rotating reference frame under $SU_c (2)$ by writing \cite{LeeWenRMP,XS10}
\begin{equation}
\left( \begin{array}{c} c_{r,\uparrow} \\ c_{r,\downarrow}^\dagger \end{array} \right) = R_r \left( \begin{array}{c} f_{r,\uparrow} \\ f_{r,\downarrow}^\dagger \end{array} \right) 
\label{defR}
\end{equation}
where $R_r$ is a position and time dependent $SU(2)$ matrix which performs the transformation to a $SU_c (2)$ rotating reference frame. This definition immediately
introduces a $SU_g(2)$ gauge invariance because the r.h.s. is invariant under 
\begin{equation}
R_r \rightarrow R_r \,  U^g_r  \quad,\quad      \left( \begin{array}{c} f_{r,\uparrow} \\ f_{r,\downarrow}^\dagger \end{array} \right) \rightarrow \left [ U^g_r \right ]^\dagger \left( \begin{array}{c} f_{r,\uparrow} \\ f_{r,\downarrow}^\dagger \end{array} \right) \,,
\end{equation}
where $U^g_r$ is an arbitrary spacetime-dependent $SU_g(2)$ matrix, as in Eq.~(\ref{eq:Y}).
The definition in Eq.~(\ref{defR}) shows that $R_r$ transforms as a $SU_c(2)$ fundamental
under left multiplication, and a $SU_g (2)$ fundamental under right multiplication. Note that in this $SU_g(2)$ gauge theory formulation,
and unlike the $\mathbb{Z}_2$ gauge theory in Eq.~(\ref{eq:Hf}), at this point the $f$ fermions do not carry a $SU_c(2)$ charge; they only carry
a $SU_g(2)$ charge, and the $SU_c(2)$ charge has been transferred from the $f$ to the $R$.

We now want to obtain an OSM state, proximate to confining AFM/VBS states, from the $SU_g(2)$ gauge theory defined by Eq.~(\ref{defR}).
Condensing the $R$ boson would completely Higgs $SU_g (2)$, and so we assume that $R$ remains gapped across the transition.
But we can break $SU(2)$ down to $\mathbb{Z}_2$ by condensing a matrix Higgs field, $H_{ab}$, which is composed of a pair of $R$ bosons:
\begin{equation}
H_{ab} \sim \mbox{Tr} \left( \mu^a R\, \mu^b R^\dagger \right)\,, \label{defH}
\end{equation}
where $a,b=1,2,3$.
This is an alternative interpretation of the Higgs field $H_{ab}$ introduced in the Section~\ref{sec:parton}.
Eq.~(\ref{defH}) is the
analog of the paired condensate of `slave' bosons carrying
$U(1)$ gauge charges in the OM construction of Ref.~\cite{Nandkishore_2012}.
$H_{ab}$ transforms as spin-one under the $SU_g(2)$ gauge and $SU_c(2)$ pseudo-spin symmetries via a left and right multiplications, respectively. 

Now introducing a Higgs condensate as in Eq.~(\ref{H0})
breaks the gauge $SU_g(2)$ down to $\mathbb{Z}_2$. It also ties together the global $SU_g (2) \times SU_c (2)$ transformations
to a diagonal subgroup, so that the $f$ fermions effectively acquire a $SU_c (2)$ index. These are precisely the characteristics of the 
observed OSM phase. 

We note that the Higgs field in Eq.~(\ref{defH}) is the only possible $R$ pair without spatial gradients. Other possibilities for $R$ pair Higgs fields
are either trivial ($\mbox{Tr} \left( R R^\dagger \right) = 2$) or vanish identically 
($\mbox{Tr} \left( \mu^a R R^\dagger \right) = \mbox{Tr} \left( R\, \mu^b R^\dagger \right) = 0$).
We can also make Higgs fields from pairs of the $f$ fermions, as was done recently in Ref.~\cite{Thomson_2018}. Such Higgs fields carry only $SU_g (2)$ charges, and their condensation leads to topologically ordered phases with fermionic excitations with 
global $SU_s (2)$ charges only: these are not orthogonal fermions, and so condensation of the $f$ pair Higgs field does not lead to an OSM.

\subsection{Critical theory}

We can now write down a continuum theory for a phase transition out of the OSM phase by assembling the degrees
of freedom described above in a $SU_g(2)$ gauge theory. First we take the continuum limit of the $(f_\uparrow, f_\downarrow^\dagger)$ fermions moving in a $\pi$ flux background to a obtain two-components Dirac spinors, $\psi_v$, which carry a valley index $v=1,2$
and a fundamental $SU_g (2)$ gauge charge (index not explicitly displayed). The fermions also carry a $SU_s(2)$ charge, but its action
is clearer in a Majorana fermion representation \cite{Wang_2017,Thomson_2018}. Minimally coupling these fermions to a $SU_g(2)$ gauge
field, we obtain two-color QCD coupled to $N_f=2$ flavors of Dirac fermions in three space-time dimensions. This theory 
was examined recently by Wang {\it et al.} \cite{Wang_2017}, and following them we dub it QCD$_3(N_f=2)$.

Wang {\it et al.} noted that QCD$_3(N_f=2)$ has a global $SO(5)$ symmetry, and that a gauge-invariant fermion bilinear transforms
as an $SO(5)$ vector. Tracing this fermion bilinear back to the lattice fermions, $f_\alpha$, they noted that this 
$SO(5)$ order parameter is precisely the composite of the $3$-component AFM order parameter and the $2$-component VBS order parameter.
A confining phase of QCD$_3$ is expected to break the $SO(5)$ symmetry, and so we have achieved our aim of writing down a theory
which is proximate to confining phases with AFM or VBS order. We have also obtained an understanding of the evidence for $SO(5)$ symmetry in our numerics.

Finally, we combine QCD$_3(N_f=2)$ with a phenomenological action for $H$ to obtain our theory for the transition between
the OSM and AFM phases.
	\begin{equation}
    \begin{aligned}
	\mathcal{S}&= \int d^3x\,\sum_{v=1}^{N_f}\bar{\psi}_{v}\slashed{D}_a\psi_v-\frac{1}{2}\mbox{Tr}\left[\left(D^H_a H\right)^T\left(D^H_a H\right)\right] 
    \\&+\frac{1}{2}m^2\, \mbox{Tr}[H^TH] +\kappa\det{H} +\frac{1}{4}\lambda \, \mbox{Tr}[H^TH]^2 \\ &+
    \frac{1}{4}\lambda' \, \mbox{Tr}[(H^TH)^2]+\frac 1 4 f^2_{\mu\nu}.
	\label{eq:qcd_higgs}
    \end{aligned}
	\end{equation}	
Here $a^c_\mu$ represents the $SU(2)$ gauge field, and the covariant derivative of the Dirac fermions is defined as , $\slashed{D}_a=\gamma_\mu\left(i\partial_\mu+a_\mu^c \mathbf{\tau}^c\right) $, where $\mathbf{\tau}^c$ are the Pauli matrices. Similarly, the covariant derivative of the Higgs field reads, ${D}^H_a=\left(\partial_\mu+a_\mu^c O^c\right) $, where $O^c$ are the generators of $SO(3)$ rotations. Finally, the last term is the standard Maxwell term, with   $f^c_{\mu\nu}$ being the non-abelian field strength. Note that all terms in Eq.~(\ref{eq:qcd_higgs}) respect the global $SO(5)$ symmetry.

 \begin{figure}
  \centering
    \includegraphics[scale=0.35]{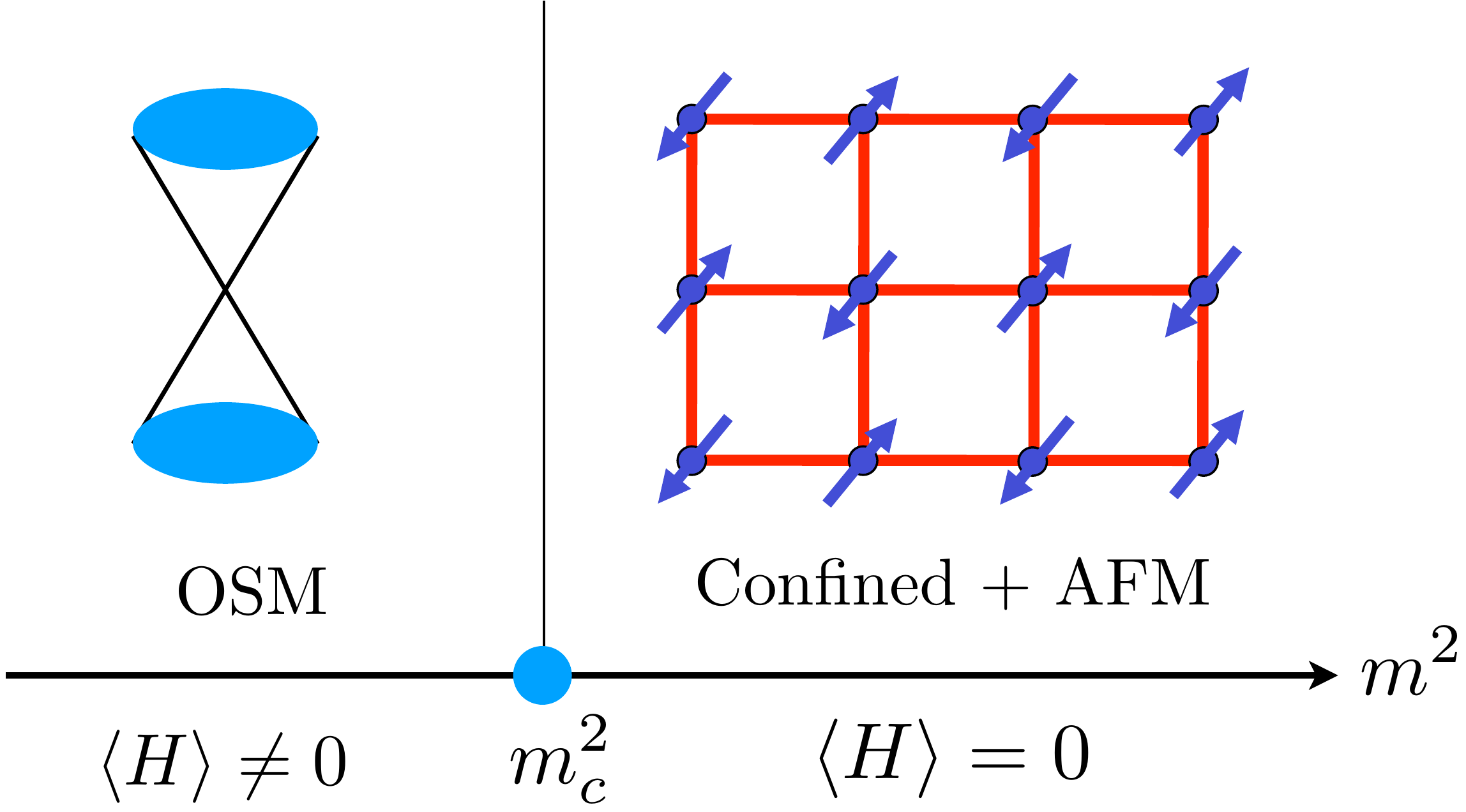}
    \caption{Higgs mediated confinement transition. For positive Higgs mass, $m^2 >0$, the Higgs field is gapped. The effective field theory is then QCD$_3$($N_f=2$), which confines and spontaneously breaks chiral symmetry, leading to an insulator with AFM order. Conversely, for $m^2 <0$, the Higss field condenses and reduces the $SU_g(2)$ gauge symmetry down to $\mathbb{Z}_2$ giving rise to the OSM.    }
    \label{fig:QCD}
 \end{figure}
The transition between OSM and AFM phases is described by tuning the Higgs mass, $m^2$, as shown in Fig.~\ref{fig:QCD}. 
For negative $m^2$, the Higgs field is condensed
as in Eq.~(\ref{H0}), and we obtain an OSM phase as described above. Note that the $\psi_v$ fermions remain massless even when the Higgs field
is condensed. This is because there is no allowed tri-linear Yukawa term between the Higgs boson and the fermions; such a Yukawa term is forbidden by 
$SU_c (2)$ symmetry, as the matrix Higgs field $H$ carries a $SU_c(2)$ charge, while the fermions $\psi$ do not. This feature is in contrast to earlier
theories of phases with $\mathbb{Z}_2$ topological order \cite{SSNambu,CSS17,Thomson_2018}, where the Yukawa term was symmetry allowed, and led to a gap in the fermion spectrum
when the Higgs field was condensed.

For positive $m^2$, we can neglect the massive Higgs field, and then Eq.~(\ref{eq:qcd_higgs}) reduces to QCD$_3(N_f=2)$.
For sufficiently large $N_f$, QCD$_3(N_f)$ defines a deconfined conformal field theory, with non-trivial scaling dimensions that can be computed in a 
$1/N_f$ expansion. However, it is expected that there is
a critical $N^c_f$ such that for $N_f < N_c$ the theory is confining. 
The most recent lattice QMC calculation \cite{Rajamani_2018} estimates $N_f^c=4-6$. For $N_f=2$, which is relevant to our case, a clear `chiral' symmetry breaking was observed, corresponding to a breaking of $SO(5)$ symmetry in our language. \footnote{Strictly speaking, the simulated QCD$_3$ at $N_f=2$ does not have the full $SO(5)$ symmetry on the lattice scale, because the full symmetry is anomalous. In principle, there is a more exotic scenario\cite{Wang_2017}, in which the QCD theory with full $SO(5)$ symmetry flows to the continuous Neel to VBS transition (the deconfined quantum critical point), and chiral symmetry breaking happens only when the full $SO(5)$ is explicitly broken (for example to $SO(3)\times SO(2)$). Our theory holds even if this scenario is correct, since the full $SO(5)$ is already broken in our microscopic model.}
Therefore, in Eq.~(\ref{eq:qcd_higgs}), the Higgs transition provides a means to simultaneously drive confinement and symmetry breaking using a {\it single} tuning parameter corresponding to the mass-squared of the Higgs field. Once we are in the SO(5)-broken regime, other irrelevant operators (not shown in Eq.~(\ref{eq:qcd_higgs})) will become important, and we assume these select the AFM order observed, rather than the VBS order.

Finally we turn to the critical point between the OSM and AFM phases. We assume that this is described by the SO(5)-symmetric deconfined critical theory 
in Eq.~(\ref{eq:qcd_higgs}) after the Higgs mass mass $m^2$ has been tuned to its critical value. The idea is that the additional contributions of the critical
Higgs modes, when combined with the gapless fermions, are sufficient to suppress the confining effects of the $SU_g (2)$ gauge field. The continuous transition observed in our numerics, along with the evidence for global $SO(5)$ symmetry is evidence in support of our proposal.

We note also the cubic term, proportional to $\kappa$ in Eq.~(\ref{eq:qcd_higgs}). In purely scalar field theories, this would be sufficient to imply a first-order
phase transition. However, when combined with strong gauge fluctuations and massless fermions, it is 
not clear whether estimates which expand about the upper-critical dimension can be reliable. 
In the large-$N_f$ expansion of such a Higgs critical theory, the $\kappa$ determinant term involves of order
$N_f$ powers of the Higgs field, 
and is clearly irrelevant at the critical point. Our evidence for a
continuous transition is evidence that this is also likely the case at $N_f = 2$.

Even if irrelevant at criticality, on moving into the Higgs phase, the $\kappa$ determinant term will dictate the nature of the Higgs condensate. Note, that since multiple Higgs fields are present due to the global symmetry, different patters of Higgs condensates are possible depending on how many $\hat{H}_a$ we condense. These are all degenerate to quadratic order, but are differentiated by the determinant term that selects a simultaneous condensate as in Eq.~(\ref{H0}) independent of the sign of $\kappa$. This form of the Higgs condensate is crucial to obtaining the OSM phase. 
	
\section{Discussion and summary}

We have carried out a detailed numerical analysis of the confinement transition
of the orthogonal semi-metal (OSM) in a model with a repulsive on-site Hubbard interaction. This serves as a model of a confinement transition in a $\mathbb{Z}_2$ gauge theory coupled to gapless Dirac fermions that carry gauge charge, which is also free of the fermion sign problem. Our key numerical finding is an emergent $SO(5)$ symmetry at criticality that enlarges the microscopic $SO(3)\times C_4$ symmetry associated with spin rotations and the discrete square lattice point group symmetry. Crucially, we demonstrate that this result is a qualitatively unique feature of the OSM confinement transition that fundamentally distinguishes it from the more conventional Gross-Neveu-Yukawa (GNY) and Ising criticality. In addition, our refined numerical calculations allowed us to improve previous estimates of critical data, and further support the scenario of deconfined criticality (DC) with a second order phase transition.

We note that, even more than a decade after the initial theoretical proposal, the ultimate thermodynamic fate of DC for insulating square lattice antiferromagnets remains in debate. Numerical studies of lattice models show conflicting results, where estimates of certain universal quantities exhibit a significant drift with system size, and in certain models even an indication for a first order transition.  On the other hand, several numerical studies indicate an enlarged $SO(5)$ symmetry that is hard to reconcile with a first order transition (see Ref.\cite{Wang_2017} for a recent discussion).

As our model involves fermionic degrees of freedom, its computational cost using standard QMC methodology does not scale favorably with systems size, compared to models of non-LGW transitions consisting of bosonic degrees of freedom. It is therefore more challenging to assert a strong statement on the thermodynamic limit of our model. Nevertheless, up to the largest length scale studied, we did not observe any sign of deviation from critical scaling and critical properties seem to remain robust for a wide range of microscopic parameters without any degree of fine tuning. Most relevant for this work, it is difficult to imagine a scenario, in which an enlarged symmetry could generically arise at a first order phase transition. 

We used the numerical results as a guide for constructing a field theory description of the OSM confinement transition, which is linked to recent studies of descended phase of QCD$_3(N_f=2)$ \cite{Wang_2017,You_2018,You_2018a,Thomson_2018}. We introduced a {\it matrix\/} Higgs mechanism, which is distinct from the vector Higgs approach presented in Ref.~\cite{Thomson_2018}. In the latter case, the Higgs fields were bilinears of the fermions $f_\alpha$, in contrast to the boson bilinears we employed in Eq.~(\ref{defH}), and their condensation led to spin liquids with fermionic spinons which do not carry the electromagnetic charge. In contrast, condensation of our matrix Higgs field led to an orthogonal metal, in which the fermions carry both spin and electromagnetic charge. At the same time the fermions carry $\mathbb{Z}_2$ gauge charge, unlike in the symmetric mass generation scenario of Refs. \cite{You_2018,You_2018a}, where a Higgs field in the fundamental representation condenses giving rise to gapless fermions, without gauge charge. 

Looking to the future, it would be interesting to explore some extension of our Higgs mechanism.  Our QCD$_3$ mechanism has a natural prediction when time-reversal symmetry is explicitly broken, in which case the Dirac fermions obtain a mass term with total Chern number $C=2$. Deep in the deconfined phase this leads to a Semion$\times$Semion topological order ($\nu=4$ in Kitaev's $16$-fold classification \cite{Kitaev06}). However, near the critical point (when the Chern mass scale is greater than the Higgs mass scale), we obtain an $SU(2)_1$ Chern-Simons theory which is simply the Semion chiral spin liquid. The two topological orders can in principle be distinguished by their ground state degeneracy on a torus or infinite cylinder, perhaps through DMRG calculation. This Semion topological order, if observed, would be a strong signature of the enhanced gauge symmetry near the critical point.

Another extension, which may be implemented in quantum Monte Carlo, is to consider similar transitions described by QED$_3$, namely a $U(1)$ (instead of $SU(2)$) gauge theory coupled to $N_f=4$ Dirac fermions. There are two scenarios in which this would be natural. First, one could consider explicitly breaking the pseudo-spin $SU_c(2)$ symmetry down to $U(1)$, say by breaking the particle-hole symmetry. Alternatively, one can study a similar system but with $\mathbb{Z}_4$ gauge field on the lattice -- in fact in this scenario we can have more controlled arguments about the ultimate IR fate of the phases and phase transition, as we briefly outline in Appendix~\ref{appD}. In both cases the gauge symmetry can be naturally enlarged to $U(1)$ but not $SU(2)$. At the critical point of such QED$_3$-Higgs transition we expect an enlarged $SO(2)\times(SO(6)\times U(1))/\mathbb{Z}_2$ symmetry, instead of $SO(3)\times SO(5)$ in the QCD$_3$-Higgs transition (the Neel-VBS $SO(5)$ observed in this work is a subgroup of both symmetries).

On the numerical front, we see several exciting future directions. First, identifying observables that can probe the emergent $SU(2)$ gauge fields and matrix Higgs field, $H$, would allow for direct confirmation of the critical theory in the numerical simulations. Second, the emergence of an $SO(5)$ symmetry at criticality can be further tested by studying certain high order correlation functions that are required to vanish by symmetry \cite{Nahum_2015}. Finally, eliminating the observed non-universal corrections to scaling requires simulations on larger lattices, beyond standard methodologies. In that regard, one promising approach is the Hamiltonian variant of the fermion bag algorithm \cite{Huffman_2016,Huffman17}.

Lastly, we note that since the theory we simulated, $\mathcal{H}$,  does not contain any gauge neutral fermion it can be thought of arising from an underlying bosonic theory. It is tempting to conjecture that the associated bosonic description will also be free of the numerical sign problem. Identifying such bosonic lattice models would allow access to significantly larger system sizes and an accurate study of critical properties. 

\subsection*{Acknowledgements}

We thank Shubhayu Chatterjee, Tarun Grover and Mathias Scheurer for valuable discussions: SC and MS pointed out that the $\det{H}$ term in
Eq.~(\ref{eq:qcd_higgs}) was allowed. SG and AV thank Mohit Randeria for an earlier collaboration on a related topic. The authors gratefully acknowledge the Gauss Centre for Supercomputing e.V. (www.gauss-centre.eu) for funding this project by providing computing time on the GCS Supercomputer SuperMUC at Leibniz Supercomputing Centre (www.lrz.de). FFA thanks the DFG through SFB 1170 ToCoTronics for financial support. This research was supported by the National Science Foundation under Grant No. DMR-1360789 (SS). Research at Perimeter Institute is supported by the Government of Canada through Industry Canada and by the Province of Ontario through the Ministry of Research and Innovation. SS also acknowledges support from Cenovus Energy at Perimeter Institute. SG was supported by the ARO (W911NF-17-1-0606) and the ERC Synergy grant UQUAM. This work was partially performed at the Aspen Center for Physics (NSF grant PHY-1607611) and the Kavli Institute for Theoretical Physics (NSF grant PHY-1125915). AV was supported by a Simons Investigator Grant, and AV and SG were  supported by NSF DMR- 1411343. CW was supported by the Harvard Society of Fellows. This research used the Lawrencium computational cluster resource provided by the IT Division at the Lawrence Berkeley National Laboratory (Supported by the Director, Office of Science, Office of Basic Energy Sciences, of the U.S. Department of Energy under Contract No. DE-AC02-05CH11231)

\appendix


\section{Circumventing the zero mode problem at finite Hubbard interactions}
\label{appA}

To enforce Gauss's law in the numerical simulation \cite{Gazit2017}, we introduce a set of discrete Lagrange multipliers, $\lambda_r$, at each lattice site, $r$, which are identified with the temporal component of the Ising gauge field. Explicitly, for an even LGT $G_r=1$, we project to the physical Hilbert space using the projector $\hat{P}=\prod_r \hat{P}_r$, where,
\begin{equation}
\begin{aligned}
\hat{P}_r&=\frac{1}{2}\left(1+\prod_{b\in +_r}\sigma^x_{r,b}(-1)^{n_r^f}\right)\\
&=\sum_{\lambda_r=\pm 1}e^{i\pi\left(\frac{1-\lambda_r}{2}\right)\left(\sum_{b\in +_r}\left(\frac{1-\sigma^x_{r,b}}{2}\right)+n_r^f\right)}.
\end{aligned}
\end{equation}
Substituting the above expression in the path integral representation yields the following fermionic weight (see \cite{Gazit2017} for a complete derivation),
\begin{equation}
W_f(\lambda,\sigma^z)=\text{Tr}\left[e^{i\pi\sum_r\left(\frac{1-\lambda_r}{2}\right)n_r^f}\prod_{\tau}e^{f^\dagger K(\sigma^z(\tau)) f}\right].
\label{eq:weight_PH}
\end{equation}
In the above, $K(\sigma^z(\tau))$ is the infinitesimal $\sigma^z(\tau)$-dependent hopping kernel. Remarkably, the constraint does not introduce a sign problem for an arbitrary fermion density. However, at half-filling, configurations satisfying $\prod_r \lambda_r=-1$ sustain an exact zero mode, as can be verified by applying a partial PH symmetry to Eq.~(\ref{eq:weight_PH}). Such configurations have a vanishing Boltzmann weight and are not sampled in the Monte Carlo simulation. 

This introduces a systematic bias in expectation values of observables that are not symmetric under partial PH transformation. In Ref.~\cite{Gazit2017}, a method that compensates on the missing weight was presented. 

In the present work, we consider a simpler solution, explained below. The Hubbard term is decoupled using an auxiliary-field $s_r$. For attractive interactions the density channel decoupling is given by,
\begin{equation}
e^{\epsilon U\left(n^\uparrow_r-\frac 1 2\right)\left(n^\downarrow_r-\frac 1 2\right)}=\frac{1}{2}\sum_{s=\pm1}e^{\gamma s_r (n^\downarrow_r+n^\uparrow_r-1)}
\end{equation}
where $\gamma=\cosh^{-1}\left(\exp(\epsilon U/2)\right)$. It is clear from the above equation that the associated weight is not symmetric under partial PH transformation for a {\it generic} auxiliary-field configuration. Therefore for finite Hubbard interactions, the zero mode is lifted allowing for an accurate sampling. We have verified this fact explicitly by benchmarking the QMC data with exact diagonalization on small system sizes.
    

\section{Additional QMC data}
\label{appB}
\begin{figure}[t]
    \centering
    \begin{subfigure}[b]{0.33\textwidth}
        \includegraphics[width=\textwidth]{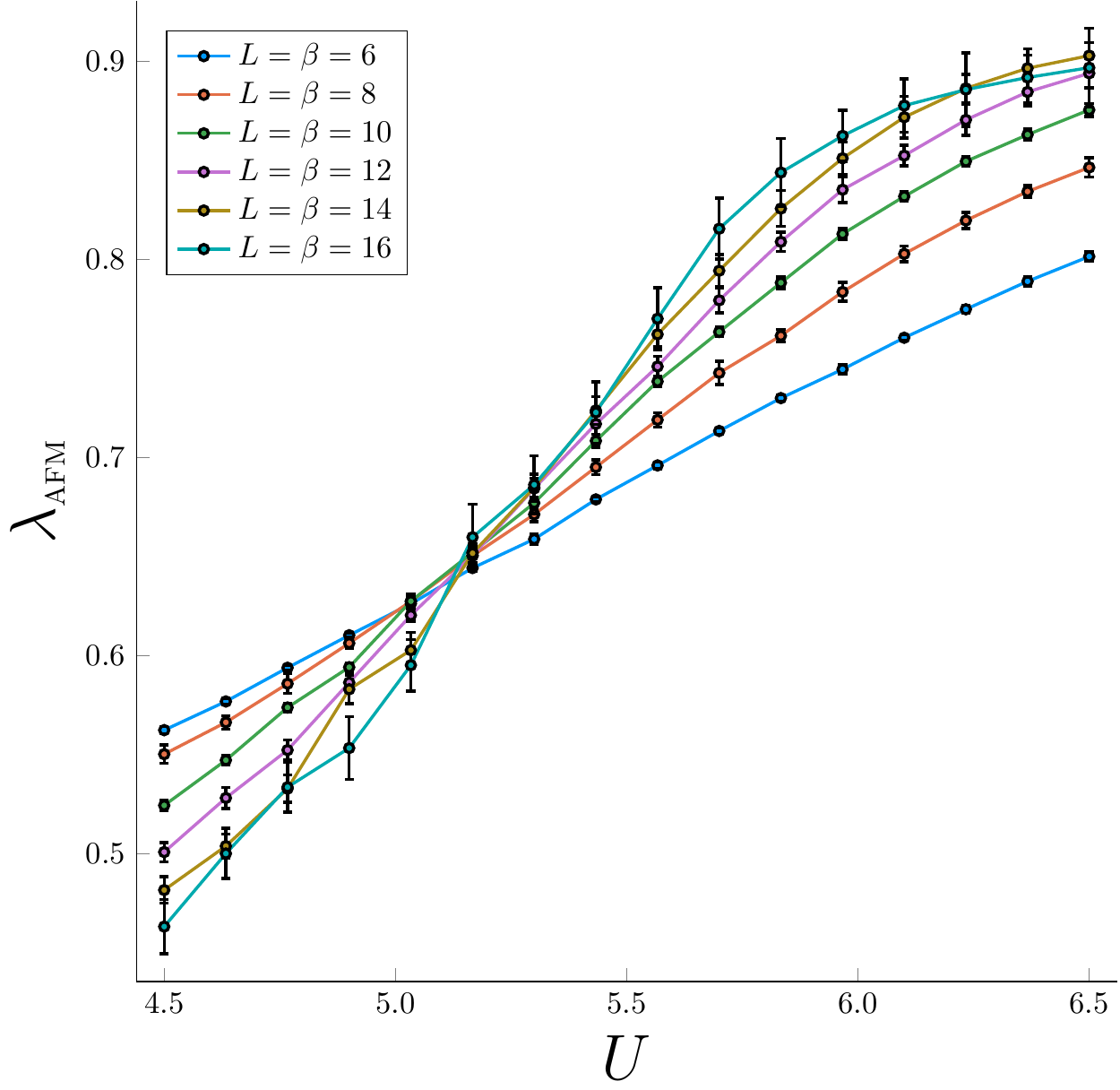}
        \caption{}
        \label{fig:gn_afm_lambda}
    \end{subfigure}
    \begin{subfigure}[b]{0.33\textwidth}
        \includegraphics[width=\textwidth]{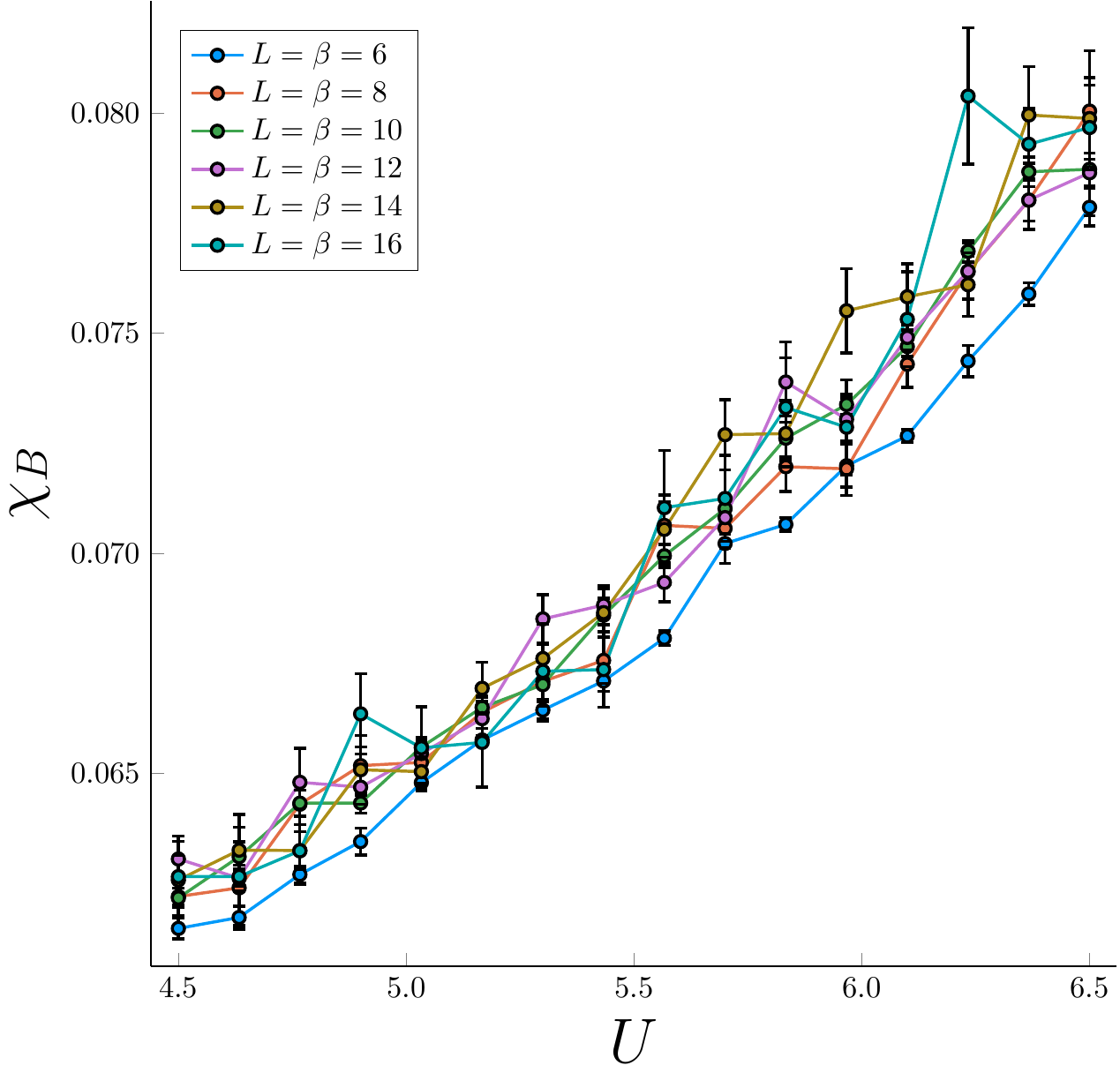}
        \caption{}
        \label{fig:gn_chi_B}
    \end{subfigure}\\
    \begin{subfigure}[b]{0.33\textwidth}
        \includegraphics[width=\textwidth]{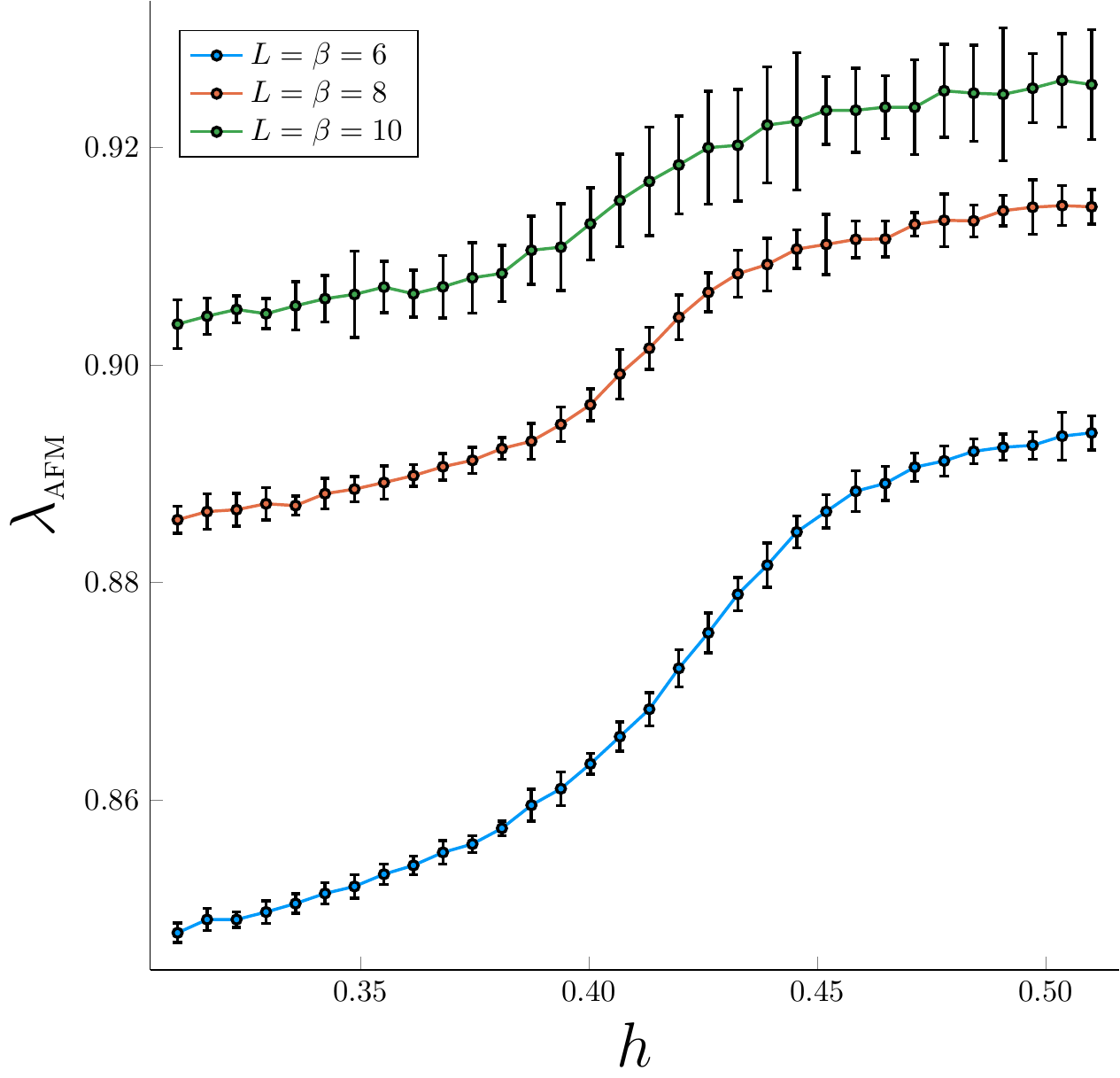}
        \caption{}
        \label{fig:isi_afm_lambda}
    \end{subfigure}
    \begin{subfigure}[b]{0.33\textwidth}
        \includegraphics[width=\textwidth]{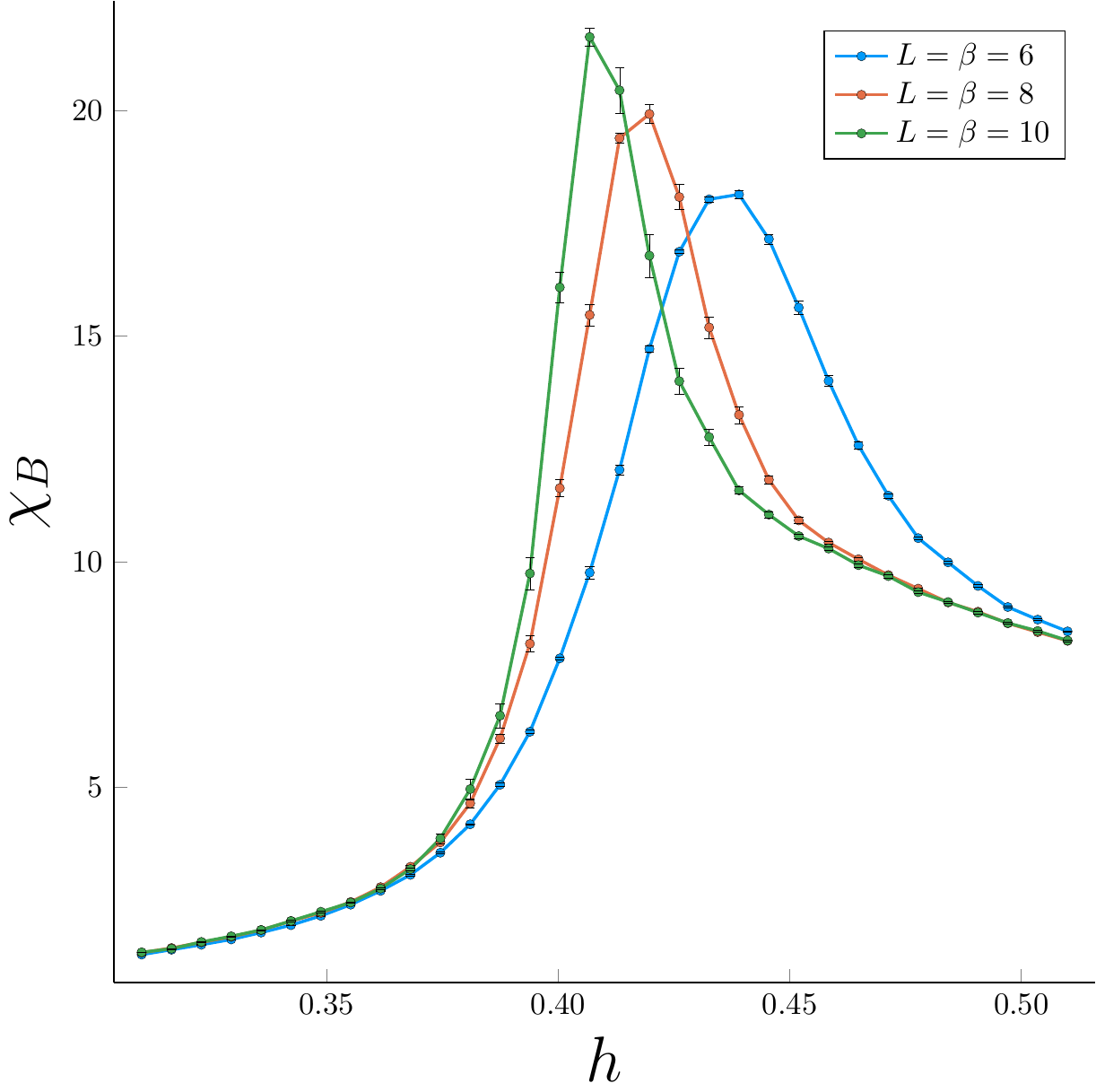}
        \caption{}
        \label{fig:isi_chi_B}
    \end{subfigure}
    \caption{(a-b) AFM ordering transition separating the AFM$^*$ and OFM phases. QMC data is calculated at $h=0.1$ and as a function of $U$. (a) $\lambda_{\ttext{AFM}}$ exhibits a clear curve crossing (b) $\chi_B$ crosses the transition smoothly. (c-d) Ising confinement transition separating the AFM$^*$ and AFM phases. QMC data is calculated at $U=7$ and as a function of $h$. (c) $\lambda_{\ttext{AFM}}$, indicates that the AFM order remains finite across the transition. (d) The Ising flux susceptibility, $\chi_B$, diverges at the confinement transition with increase in the system size. } \label{fig:ILGT_GN_cross}
\end{figure}

In this section, we present additional QMC data supporting our finding in the main text. In Fig.~\ref{fig:ILGT_GN_cross}, we consider two parameter cuts corresponding to the AFM ordering transition between the OSM and AFM$^*$ phases and the Ising confinement transition between the AFM$^*$ and AFM phases. As expected, we find that the former involves only AFM ordering as seen in a curve crossing analysis of $\lambda_{\ttext{AFM}}$, while the latter in marked solely by a singularity in flux susceptibility, $\chi_B$, indicating confinement. 

\begin{figure}[t]
    \centering
    \includegraphics[width=0.53\textwidth]{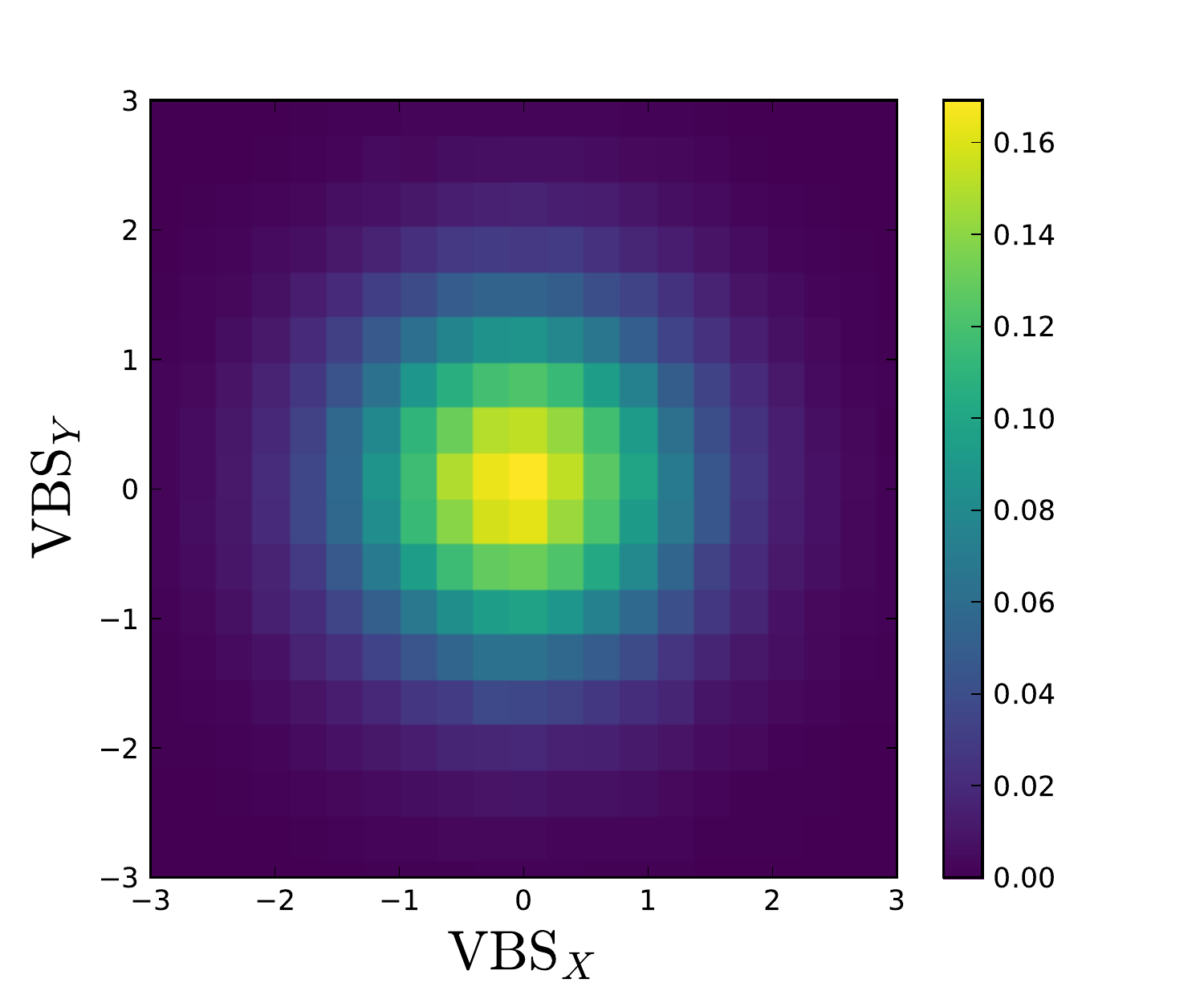}
   \caption{Joint probability distribution $\mathbf{P}(\mathbf{B}^x,\mathbf{B}^y)$ at the OSM confinement transition. The apparent circular symmetry indicates that the $C_4$ square lattice symmetry is enlarged to an $SO(2)$ rotational symmetry.}
      \label{fig:hist_vbs_xy}
\end{figure}

Finally, in Fig.~\ref{fig:hist_vbs_xy}, we depict the joint probability distribution, $\mathbf{P}(\mathbf{B}^x,\mathbf{B}^y)$, of the VBS order parameter along the x and y directions, evaluated at the OSM confinement transition. The visible circular symmetry supports the emergence of rotational $SO(2)$ symmetry at the OSM confinement transition.


\section{Dynamically imposed  constraint} 
\label{appC}

\begin{figure*}[ht]
\centering
   \includegraphics[width=0.8\textwidth]{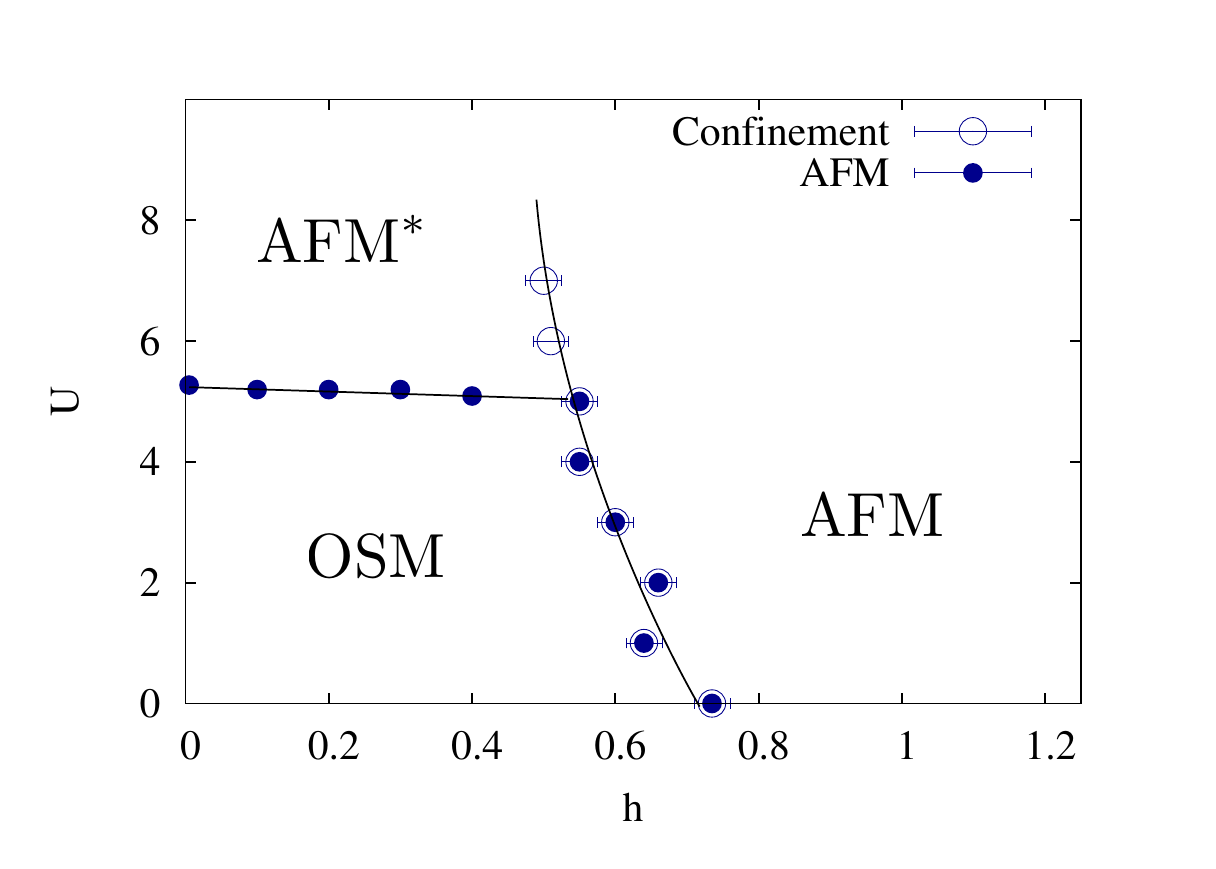}
    \caption{ \label{NC_Phase.fig}     Phase diagram  at    $J /t= -1$.  Data stems from simulations at  $L=10$ and $L=14$.  The AFM transition is  obtained by analyzing   the renormalization group invariant quantity $\lambda_{AFM}$.  The confinement transition  is obtained by monitoring $\chi_B  = \partial \langle \Phi \rangle/\partial h   $.  We have equally checked that at the  confinement transition, visons proliferate.  We have used values of $\Delta \tau = 0.4, 0.2, 0.1$  for growing values of $U/t$.  }
\end{figure*}

Imposing the constraint $Q_r = \pm{1}$    is necessary to  satisfy  local  $\mathbb{Z}_2$  gauge symmetry.  In fact, without  a constraint  the model is very  asymmetric in space and time:  $\langle f^{\dagger}_{r,\alpha}(\tau)  f^{\phantom\dagger}_{r',\alpha'} (\tau=0) \rangle = \delta_{r,r'}\delta_{\alpha, \alpha'} $.    Since $G_r$ commutes with the  Hamiltonian, the constraint will be dynamically imposed in the low temperature limit and for observables satisfying   $\left[ O,G_r\right] =0$,  we expect: 
\begin{equation}
	\lim_{L \rightarrow \infty } \lim_{T \rightarrow  0 }  \langle O(\tau) O \rangle_{NC} = \lim_{L \rightarrow \infty } \lim_{T \rightarrow  0}  \langle O(\tau) O \rangle_{C}.
\end{equation} 
Thus, provided that we first take the  zero temperature limit on a finite sized lattice, simulations with, $\langle \bullet \rangle_{C}$, or  without, $\langle \bullet \rangle_{NC}$,  constraint  should converge to the same result.  It is  very hard  to realize this  ordering of limits  numerically: as $h \rightarrow 0$  the relevant energy scale below which the constraint is dynamically imposed  vanishes. Above this energy scale, the  Ising fields  freeze. The model without constraint is amenable to  sign free QMC simulations at odd  flavors  and may be easier to simulate with   alternative methods such as the fermion bag approach \cite{Huffman17}.  It is hence certainly worth while comparing results with and without constraint.  
In this appendix, we briefly present QMC data  where the constraint is not explicitly taken into account, and show that consistent results are obtained.   We have used the ALF implementation of the auxiliary field QMC algorithm \cite{ALF_v1}.   In contrast to data presented in Ref.~\cite{Assaad2016},  we  have used parallel tempering schemes as well as global updates  to flip  blocks of spins along  the  imaginary  time.  These approaches aim at  reducing the long autocorrelation times  we encounter  in the vicinity of the 
OSM to AFM transtion. 
Fig.~\ref{NC_Phase.fig}  shows the phase diagram  at $J /t= -1$  in the $U$-$h$ plane.  We fix the temperature  to  $\beta t = 80$  such that, as mentioned above, and  in the low-h limit the Ising fields freeze  and we recover results of the $\pi$-flux Hubbard model \cite{Toldin14}.

\begin{figure*}
  \includegraphics[width=1.0\textwidth]{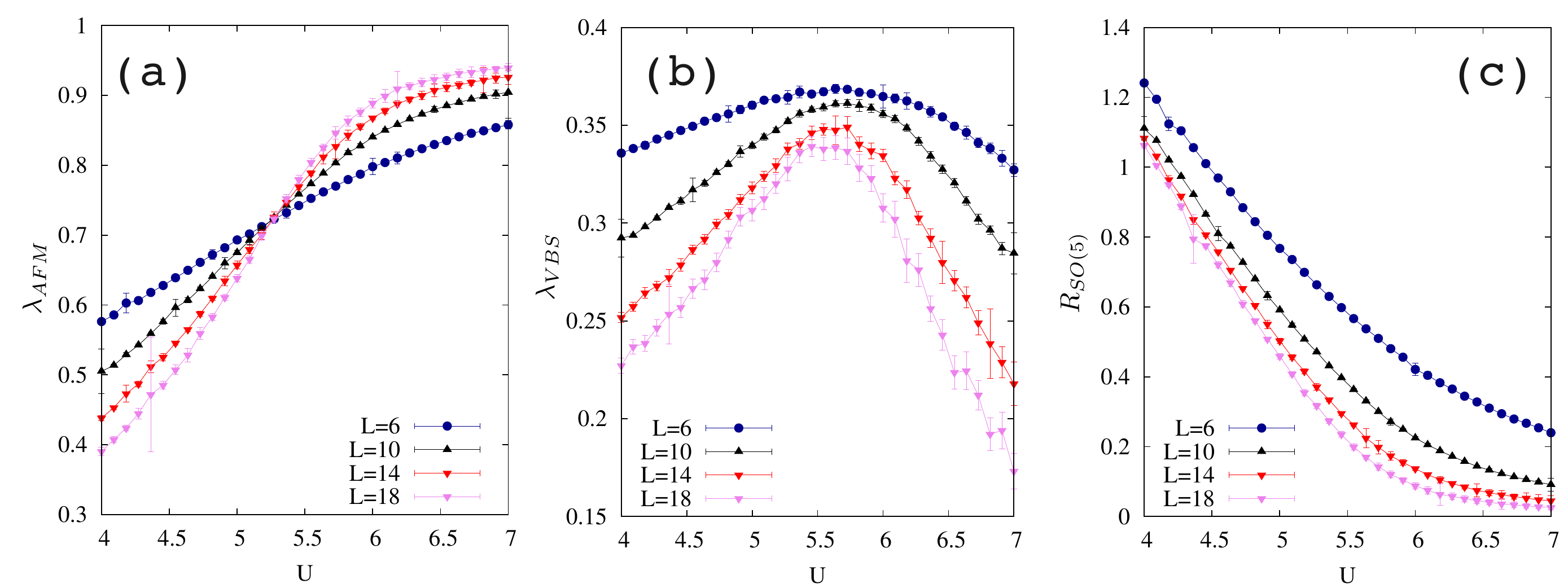}
   \caption{ \label{NC_GN.fig}    Simulations were carried out at $ L = \beta$,  $h/t = 0.05$, and  $\Delta \tau t  = 0.1$. For this  choice  of $h$ and $\beta t$ the Ising field are essentially frozen  and the flux per plaquette is very close to  -1. (a)  $\lambda_{\ttext{AFM}} $ as obtained from the susceptibilities. (b) $\lambda_{VBS} $ as obtained from  the susceptibilities.   (c)  Ratio  $R_{SO(5)} = \chi_{\ttext{VBS}}/\chi_{\ttext{AFM}}$.    }
\end{figure*}

\begin{figure*}
  \includegraphics[width=1.0\textwidth]{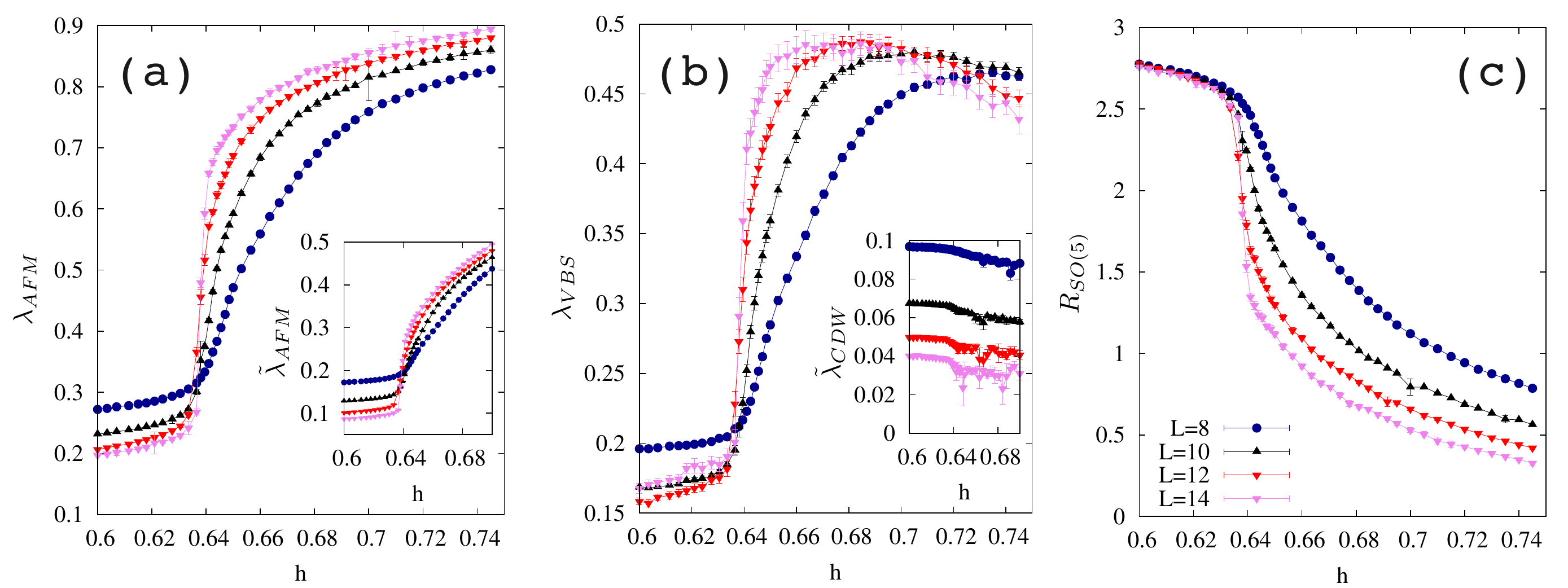}
   \caption{ \label{NC_QCD3.fig}   Simulations are carried out at  $ \beta t=80$,  $U/t=1$, $\Delta \tau = 0.4$.   The temperature was chosen as low as possible so as to attempt to satisfy the constraint.  (a)  $\lambda_{\ttext{AFM}} $ as obtained from the susceptibilities. Inset: $\tilde{\lambda}_{\ttext{AFM}} $ as  obtained from equal time correlation functions. (b)  $\lambda_{\ttext{VBS}} $ as obtained from  the susceptibilities. Inset:  Charge density wave  (CDW) correlation ratio at the antiferromgnetic wave vector, $\tilde{\lambda}_{\ttext{CDW}} $, as obtianed from equal time correlation functions. (c)  Ratio $R_{SO(5)} = \chi_{\ttext{VBS}}/\chi_{\ttext{AFM}}$. }
\end{figure*}


Figs.~\ref{NC_GN.fig}  and  \ref{NC_QCD3.fig}   plot the VBS and AFM  correlation ratios,  $\lambda_{\ttext{AFM}}$ and  $\lambda_{\ttext{VBS}}$   as obtained from the susceptibilities.   Both  $\lambda_{\ttext{AFM}}$ and  $\lambda_{\ttext{VBS}}$   are renormalization group invariant quantities and are expected  to cross at the critical point.  This  relies  on the assumption that the susceptibility is dominated by the singular part of the free energy  and  thereby requires $\eta < 2$. Note that the finite size scaling  form for susceptibilities reads $\chi \simeq L^{2-\eta}  G\left( L^z/\beta, L^{1/\nu} (g-g_c) \right) $.   Fig.~\ref{NC_GN.fig}(a) shows $\lambda_{\ttext{AFM}}$   for the O(3)-Gross-Neveu-Yukawa transition \cite{Assaad13,Toldin14,Sorella_GN} and clearly pins down the critical value of $U/t$.  The data of Fig.~\ref{NC_GN.fig}(b)  is consistent with a  constant value of $\lambda_{\ttext{VBS}}$   at the transition and in the thermodynamic limit thereby confirming  critical VBS fluctuations at the transition.  However, the quotient $\chi_{\ttext{VBS}}/\chi_{\ttext{AFM}}$ clearly  shows that at the O(3)-Gross-Neveu-Yukawa transition $ \eta_{\ttext{VBS}}  > \eta_{\ttext{AFM}}$ since at $U_c$, the curves fail to cross.    
Such a  statement does not hold  for the OSM  to AFM transition (see Fig.~\ref{NC_QCD3.fig}).  Here,  the data is consistent with 
 $ \eta_{\ttext{VBS}}  = \eta_{\ttext{AFM}}$ thereby supporting  an  emergent SO(5) symmetry.  The inset of Fig.~\ref{NC_QCD3.fig}(a) shows the AFM correlation ratio -- as obtained form equal time correlation functions.  Comparison with the equivalent data for the   charge density wave (CDW), inset of   Fig.~\ref{NC_QCD3.fig}(b),  shows that the SO(8) symmetry of Dirac fermions is  violated at the OSM to AFM transition.

\section{QED$_3$ and confinement transition of $\mathbb{Z}_4$ gauge theory}
\label{appD}
We consider the following conituum theory
\begin{equation}
\mathcal{L}=\sum_{i=1}^4\bar{\psi}_i\slashed{D}_a\psi_i+\frac{1}{2}|D_{4a}\phi|^2+r|\phi|^2+\lambda|\phi|^4+\frac{1}{4e^2}f_{\mu\nu}^2+\kappa V+\kappa^*V^*,
\end{equation}
where $a_{\mu}$ is a $U(1)$ gauge field, $\psi_i$ is a two-component Dirac fermion with gauge charge $q_g=1$, $\phi$ is a complex boson with gauge charge $q_g=4$, and $V$ schematically represents the monopole (instanton) operator of the $U(1)$ gauge field.

Let us first ignore the monopole term. When $r<r_c$ the Higgs condensate $\langle\phi\rangle\neq0$ will produce a $\mathbb{Z}_4$ gauge theory with $N_f=4$ gapless Dirac fermions. When $r>r_c$ the Higgs field can be ignored at low energy,  and we get a pure QED$_3$ with $N_f=4$, which appears to be a stable conformal field theory from numerical studies \cite{qedcft}. The critical point is expected to be also stable since the extra critical Higgs field $\phi$ should further control the gauge field fluctuation. Therefore in the absence of monopole the above theory describes a continuous transition from $N_f=4$ $\mathbb{Z}_4$ gauge theory to $N_f=4$ QED$_3$.

Now put the monopole terms back (assuming such a term is compatible with all global symmetries, as supported by previous study \cite{AliceaMonopole}). For the Higgs phase this has no effect. For the QED$_3$ phase, the monopole term is likely relevant \cite{KapustinQED, PufuQED} ($\Delta_V=0.265N_f-0.038+O(1/N_f)=1.022+O(1/N_f)<3$), and physically we expect confinement and spontaneous chiral symmetry breaking at low energy, producing a Neel-like state. At the critical point we expect the critical Higgs field to render the monopole irrelevant. This is because even without the gapless Dirac fermions, the monopole is known to be irrelevant because of the critical charge-$4$ Higgs field \cite{Anisotropy}, and physically we expect the gapless Dirac fermions to make the monopole even more irrelevant.  Therefore at the critical point the $U(1)$ gauge theory is effectively non-compact. Also notice that unlike the QCD$_3$ scenario for the transition studied in the main text, there is no cubic-like term for the Higgs field.  For these reasons we expect the QED$_3$-Higgs theory to describe a continuous transition is between a $\mathbb{Z}_4$ gauge theory with $N_f=4$ Dirac fermions and a Neel state.

\bibliography{so5_higgs}

\begin{thebibliography}{63}%
\makeatletter
\providecommand \@ifxundefined [1]{%
 \@ifx{#1\undefined}
}%
\providecommand \@ifnum [1]{%
 \ifnum #1\expandafter \@firstoftwo
 \else \expandafter \@secondoftwo
 \fi
}%
\providecommand \@ifx [1]{%
 \ifx #1\expandafter \@firstoftwo
 \else \expandafter \@secondoftwo
 \fi
}%
\providecommand \natexlab [1]{#1}%
\providecommand \enquote  [1]{``#1''}%
\providecommand \bibnamefont  [1]{#1}%
\providecommand \bibfnamefont [1]{#1}%
\providecommand \citenamefont [1]{#1}%
\providecommand \href@noop [0]{\@secondoftwo}%
\providecommand \href [0]{\begingroup \@sanitize@url \@href}%
\providecommand \@href[1]{\@@startlink{#1}\@@href}%
\providecommand \@@href[1]{\endgroup#1\@@endlink}%
\providecommand \@sanitize@url [0]{\catcode `\\12\catcode `\$12\catcode
  `\&12\catcode `\#12\catcode `\^12\catcode `\_12\catcode `\%12\relax}%
\providecommand \@@startlink[1]{}%
\providecommand \@@endlink[0]{}%
\providecommand \url  [0]{\begingroup\@sanitize@url \@url }%
\providecommand \@url [1]{\endgroup\@href {#1}{\urlprefix }}%
\providecommand \urlprefix  [0]{URL }%
\providecommand \Eprint [0]{\href }%
\providecommand \doibase [0]{http://dx.doi.org/}%
\providecommand \selectlanguage [0]{\@gobble}%
\providecommand \bibinfo  [0]{\@secondoftwo}%
\providecommand \bibfield  [0]{\@secondoftwo}%
\providecommand \translation [1]{[#1]}%
\providecommand \BibitemOpen [0]{}%
\providecommand \bibitemStop [0]{}%
\providecommand \bibitemNoStop [0]{.\EOS\space}%
\providecommand \EOS [0]{\spacefactor3000\relax}%
\providecommand \BibitemShut  [1]{\csname bibitem#1\endcsname}%
\let\auto@bib@innerbib\@empty
\bibitem [{\citenamefont {{Senthil}}\ \emph
  {et~al.}(2004{\natexlab{a}})\citenamefont {{Senthil}}, \citenamefont
  {{Vishwanath}}, \citenamefont {{Balents}}, \citenamefont {{Sachdev}},\ and\
  \citenamefont {{Fisher}}}]{Senthil_DC}%
  \BibitemOpen
  \bibfield  {author} {\bibinfo {author} {\bibfnamefont {T.}~\bibnamefont
  {{Senthil}}}, \bibinfo {author} {\bibfnamefont {A.}~\bibnamefont
  {{Vishwanath}}}, \bibinfo {author} {\bibfnamefont {L.}~\bibnamefont
  {{Balents}}}, \bibinfo {author} {\bibfnamefont {S.}~\bibnamefont
  {{Sachdev}}}, \ and\ \bibinfo {author} {\bibfnamefont {M.~P.~A.}\
  \bibnamefont {{Fisher}}},\ }\bibfield  {title} {\enquote {\bibinfo {title}
  {{Deconfined Quantum Critical Points}},}\ }\href {\doibase
  10.1126/science.1091806} {\bibfield  {journal} {\bibinfo  {journal}
  {Science}\ }\textbf {\bibinfo {volume} {303}},\ \bibinfo {pages} {1490}
  (\bibinfo {year} {2004}{\natexlab{a}})},\ \Eprint
  {http://arxiv.org/abs/cond-mat/0311326} {arXiv:cond-mat/0311326
  [cond-mat.str-el]} \BibitemShut {NoStop}%
\bibitem [{\citenamefont {Wegner}(1971)}]{WegnerILGT}%
  \BibitemOpen
  \bibfield  {author} {\bibinfo {author} {\bibfnamefont {F.~J.}\ \bibnamefont
  {Wegner}},\ }\bibfield  {title} {\enquote {\bibinfo {title} {{Duality in
  Generalized Ising Models and Phase Transitions without Local Order
  Parameters}},}\ }\href {\doibase 10.1063/1.1665530} {\bibfield  {journal}
  {\bibinfo  {journal} {Journal of Mathematical Physics}\ }\textbf {\bibinfo
  {volume} {12}},\ \bibinfo {pages} {2259} (\bibinfo {year}
  {1971})}\BibitemShut {NoStop}%
\bibitem [{\citenamefont {Wen}\ and\ \citenamefont {Wu}(1993)}]{WW93}%
  \BibitemOpen
  \bibfield  {author} {\bibinfo {author} {\bibfnamefont {X.-G.}\ \bibnamefont
  {Wen}}\ and\ \bibinfo {author} {\bibfnamefont {Y.-S.}\ \bibnamefont {Wu}},\
  }\bibfield  {title} {\enquote {\bibinfo {title} {{Transitions between the
  quantum Hall states and insulators induced by periodic potentials}},}\ }\href
  {\doibase 10.1103/PhysRevLett.70.1501} {\bibfield  {journal} {\bibinfo
  {journal} {Phys. Rev. Lett.}\ }\textbf {\bibinfo {volume} {70}},\ \bibinfo
  {pages} {1501} (\bibinfo {year} {1993})}\BibitemShut {NoStop}%
\bibitem [{\citenamefont {{Chen}}\ \emph {et~al.}(1993)\citenamefont {{Chen}},
  \citenamefont {{Fisher}},\ and\ \citenamefont {{Wu}}}]{CFW93}%
  \BibitemOpen
  \bibfield  {author} {\bibinfo {author} {\bibfnamefont {W.}~\bibnamefont
  {{Chen}}}, \bibinfo {author} {\bibfnamefont {M.~P.~A.}\ \bibnamefont
  {{Fisher}}}, \ and\ \bibinfo {author} {\bibfnamefont {Y.-S.}\ \bibnamefont
  {{Wu}}},\ }\bibfield  {title} {\enquote {\bibinfo {title} {{Mott transition
  in an anyon gas}},}\ }\href {\doibase 10.1103/PhysRevB.48.13749} {\bibfield
  {journal} {\bibinfo  {journal} {Phys. Rev. B}\ }\textbf {\bibinfo {volume}
  {48}},\ \bibinfo {pages} {13749} (\bibinfo {year} {1993})},\ \Eprint
  {http://arxiv.org/abs/cond-mat/9301037} {cond-mat/9301037} \BibitemShut
  {NoStop}%
\bibitem [{\citenamefont {Jalabert}\ and\ \citenamefont
  {Sachdev}(1991)}]{RJSS91}%
  \BibitemOpen
  \bibfield  {author} {\bibinfo {author} {\bibfnamefont {R.~A.}\ \bibnamefont
  {Jalabert}}\ and\ \bibinfo {author} {\bibfnamefont {S.}~\bibnamefont
  {Sachdev}},\ }\bibfield  {title} {\enquote {\bibinfo {title} {{Spontaneous
  alignment of frustrated bonds in an anisotropic, three-dimensional Ising
  model}},}\ }\href {\doibase 10.1103/PhysRevB.44.686} {\bibfield  {journal}
  {\bibinfo  {journal} {Phys. Rev. B}\ }\textbf {\bibinfo {volume} {44}},\
  \bibinfo {pages} {686} (\bibinfo {year} {1991})}\BibitemShut {NoStop}%
\bibitem [{\citenamefont {{Chubukov}}\ \emph {et~al.}(1994)\citenamefont
  {{Chubukov}}, \citenamefont {{Senthil}},\ and\ \citenamefont
  {{Sachdev}}}]{CSS94}%
  \BibitemOpen
  \bibfield  {author} {\bibinfo {author} {\bibfnamefont {A.~V.}\ \bibnamefont
  {{Chubukov}}}, \bibinfo {author} {\bibfnamefont {T.}~\bibnamefont
  {{Senthil}}}, \ and\ \bibinfo {author} {\bibfnamefont {S.}~\bibnamefont
  {{Sachdev}}},\ }\bibfield  {title} {\enquote {\bibinfo {title} {{Universal
  magnetic properties of frustrated quantum antiferromagnets in two
  dimensions}},}\ }\href {\doibase 10.1103/PhysRevLett.72.2089} {\bibfield
  {journal} {\bibinfo  {journal} {Phys. Rev. Lett.}\ }\textbf {\bibinfo
  {volume} {72}},\ \bibinfo {pages} {2089} (\bibinfo {year} {1994})},\ \Eprint
  {http://arxiv.org/abs/cond-mat/9311045} {cond-mat/9311045} \BibitemShut
  {NoStop}%
\bibitem [{\citenamefont {{Sachdev}}\ and\ \citenamefont
  {{Vojta}}(1999)}]{SSMV99}%
  \BibitemOpen
  \bibfield  {author} {\bibinfo {author} {\bibfnamefont {S.}~\bibnamefont
  {{Sachdev}}}\ and\ \bibinfo {author} {\bibfnamefont {M.}~\bibnamefont
  {{Vojta}}},\ }\bibfield  {title} {\enquote {\bibinfo {title} {{Translational
  symmetry breaking in two-dimensional antiferromagnets and
  superconductors}},}\ }\href@noop {} {\bibfield  {journal} {\bibinfo
  {journal} {J. Phys. Soc. Jpn {\bf 69}, Supp. B, 1}\ } (\bibinfo {year}
  {1999})},\ \Eprint {http://arxiv.org/abs/cond-mat/9910231} {cond-mat/9910231}
  \BibitemShut {NoStop}%
\bibitem [{\citenamefont {BenTov}(2015)}]{BenTov2015}%
  \BibitemOpen
  \bibfield  {author} {\bibinfo {author} {\bibfnamefont {Y.}~\bibnamefont
  {BenTov}},\ }\bibfield  {title} {\enquote {\bibinfo {title} {Fermion masses
  without symmetry breaking in two spacetime dimensions},}\ }\href {\doibase
  10.1007/JHEP07(2015)034} {\bibfield  {journal} {\bibinfo  {journal} {Journal
  of High Energy Physics}\ }\textbf {\bibinfo {volume} {2015}},\ \bibinfo
  {pages} {34} (\bibinfo {year} {2015})}\BibitemShut {NoStop}%
\bibitem [{\citenamefont {Ayyar}\ and\ \citenamefont
  {Chandrasekharan}(2016)}]{Chandrasekharan2016}%
  \BibitemOpen
  \bibfield  {author} {\bibinfo {author} {\bibfnamefont {V.}~\bibnamefont
  {Ayyar}}\ and\ \bibinfo {author} {\bibfnamefont {S.}~\bibnamefont
  {Chandrasekharan}},\ }\bibfield  {title} {\enquote {\bibinfo {title} {Origin
  of fermion masses without spontaneous symmetry breaking},}\ }\href {\doibase
  10.1103/PhysRevD.93.081701} {\bibfield  {journal} {\bibinfo  {journal} {Phys.
  Rev. D}\ }\textbf {\bibinfo {volume} {93}},\ \bibinfo {pages} {081701}
  (\bibinfo {year} {2016})}\BibitemShut {NoStop}%
\bibitem [{\citenamefont {Catterall}(2016)}]{Catterall2016}%
  \BibitemOpen
  \bibfield  {author} {\bibinfo {author} {\bibfnamefont {S.}~\bibnamefont
  {Catterall}},\ }\bibfield  {title} {\enquote {\bibinfo {title} {Fermion mass
  without symmetry breaking},}\ }\href {\doibase 10.1007/JHEP01(2016)121}
  {\bibfield  {journal} {\bibinfo  {journal} {Journal of High Energy Physics}\
  }\textbf {\bibinfo {volume} {2016}},\ \bibinfo {pages} {121} (\bibinfo {year}
  {2016})}\BibitemShut {NoStop}%
\bibitem [{\citenamefont {He}\ \emph {et~al.}(2016)\citenamefont {He},
  \citenamefont {Wu}, \citenamefont {You}, \citenamefont {Xu}, \citenamefont
  {Meng},\ and\ \citenamefont {Lu}}]{Meng2016}%
  \BibitemOpen
  \bibfield  {author} {\bibinfo {author} {\bibfnamefont {Y.-Y.}\ \bibnamefont
  {He}}, \bibinfo {author} {\bibfnamefont {H.-Q.}\ \bibnamefont {Wu}}, \bibinfo
  {author} {\bibfnamefont {Y.-Z.}\ \bibnamefont {You}}, \bibinfo {author}
  {\bibfnamefont {C.}~\bibnamefont {Xu}}, \bibinfo {author} {\bibfnamefont
  {Z.~Y.}\ \bibnamefont {Meng}}, \ and\ \bibinfo {author} {\bibfnamefont
  {Z.-Y.}\ \bibnamefont {Lu}},\ }\bibfield  {title} {\enquote {\bibinfo {title}
  {Quantum critical point of dirac fermion mass generation without spontaneous
  symmetry breaking},}\ }\href {\doibase 10.1103/PhysRevB.94.241111} {\bibfield
   {journal} {\bibinfo  {journal} {Phys. Rev. B}\ }\textbf {\bibinfo {volume}
  {94}},\ \bibinfo {pages} {241111} (\bibinfo {year} {2016})}\BibitemShut
  {NoStop}%
\bibitem [{\citenamefont {You}\ \emph {et~al.}(2018{\natexlab{a}})\citenamefont
  {You}, \citenamefont {He}, \citenamefont {Xu},\ and\ \citenamefont
  {Vishwanath}}]{You_2018}%
  \BibitemOpen
  \bibfield  {author} {\bibinfo {author} {\bibfnamefont {Y.-Z.}\ \bibnamefont
  {You}}, \bibinfo {author} {\bibfnamefont {Y.-C.}\ \bibnamefont {He}},
  \bibinfo {author} {\bibfnamefont {C.}~\bibnamefont {Xu}}, \ and\ \bibinfo
  {author} {\bibfnamefont {A.}~\bibnamefont {Vishwanath}},\ }\bibfield  {title}
  {\enquote {\bibinfo {title} {{Symmetric Fermion Mass Generation as Deconfined
  Quantum Criticality}},}\ }\href {\doibase 10.1103/PhysRevX.8.011026}
  {\bibfield  {journal} {\bibinfo  {journal} {Phys. Rev. X}\ }\textbf {\bibinfo
  {volume} {8}},\ \bibinfo {pages} {011026} (\bibinfo {year}
  {2018}{\natexlab{a}})}\BibitemShut {NoStop}%
\bibitem [{\citenamefont {You}\ \emph {et~al.}(2018{\natexlab{b}})\citenamefont
  {You}, \citenamefont {He}, \citenamefont {Vishwanath},\ and\ \citenamefont
  {Xu}}]{You_2018a}%
  \BibitemOpen
  \bibfield  {author} {\bibinfo {author} {\bibfnamefont {Y.-Z.}\ \bibnamefont
  {You}}, \bibinfo {author} {\bibfnamefont {Y.-C.}\ \bibnamefont {He}},
  \bibinfo {author} {\bibfnamefont {A.}~\bibnamefont {Vishwanath}}, \ and\
  \bibinfo {author} {\bibfnamefont {C.}~\bibnamefont {Xu}},\ }\bibfield
  {title} {\enquote {\bibinfo {title} {From bosonic topological transition to
  symmetric fermion mass generation},}\ }\href {\doibase
  10.1103/PhysRevB.97.125112} {\bibfield  {journal} {\bibinfo  {journal} {Phys.
  Rev. B}\ }\textbf {\bibinfo {volume} {97}},\ \bibinfo {pages} {125112}
  (\bibinfo {year} {2018}{\natexlab{b}})}\BibitemShut {NoStop}%
\bibitem [{\citenamefont {{Senthil}}\ and\ \citenamefont
  {{Fisher}}(2000)}]{Senthil_2000}%
  \BibitemOpen
  \bibfield  {author} {\bibinfo {author} {\bibfnamefont {T.}~\bibnamefont
  {{Senthil}}}\ and\ \bibinfo {author} {\bibfnamefont {M.~P.~A.}\ \bibnamefont
  {{Fisher}}},\ }\bibfield  {title} {\enquote {\bibinfo {title}
  {{$\mathbb{Z}_{2}$ gauge theory of electron fractionalization in strongly
  correlated systems}},}\ }\href {\doibase 10.1103/PhysRevB.62.7850} {\bibfield
   {journal} {\bibinfo  {journal} {Phys. Rev. B}\ }\textbf {\bibinfo {volume}
  {62}},\ \bibinfo {pages} {7850} (\bibinfo {year} {2000})},\ \Eprint
  {http://arxiv.org/abs/cond-mat/9910224} {cond-mat/9910224} \BibitemShut
  {NoStop}%
\bibitem [{\citenamefont {{Kitaev}}(2006)}]{Kitaev06}%
  \BibitemOpen
  \bibfield  {author} {\bibinfo {author} {\bibfnamefont {A.}~\bibnamefont
  {{Kitaev}}},\ }\bibfield  {title} {\enquote {\bibinfo {title} {{Anyons in an
  exactly solved model and beyond}},}\ }\href {\doibase
  10.1016/j.aop.2005.10.005} {\bibfield  {journal} {\bibinfo  {journal} {Annals
  of Physics}\ }\textbf {\bibinfo {volume} {321}},\ \bibinfo {pages} {2}
  (\bibinfo {year} {2006})},\ \Eprint {http://arxiv.org/abs/cond-mat/0506438}
  {cond-mat/0506438} \BibitemShut {NoStop}%
\bibitem [{\citenamefont {{Tanaka}}\ and\ \citenamefont
  {{Hu}}(2005)}]{Tanaka_2005}%
  \BibitemOpen
  \bibfield  {author} {\bibinfo {author} {\bibfnamefont {A.}~\bibnamefont
  {{Tanaka}}}\ and\ \bibinfo {author} {\bibfnamefont {X.}~\bibnamefont
  {{Hu}}},\ }\bibfield  {title} {\enquote {\bibinfo {title} {{Many-Body Spin
  Berry Phases Emerging from the {$\pi$}-Flux State: Competition between
  Antiferromagnetism and the Valence-Bond-Solid State}},}\ }\href {\doibase
  10.1103/PhysRevLett.95.036402} {\bibfield  {journal} {\bibinfo  {journal}
  {Phys. Rev. Lett.}\ }\textbf {\bibinfo {volume} {95}},\ \bibinfo {eid}
  {036402} (\bibinfo {year} {2005})},\ \Eprint
  {http://arxiv.org/abs/cond-mat/0501365} {cond-mat/0501365} \BibitemShut
  {NoStop}%
\bibitem [{\citenamefont {{Senthil}}\ and\ \citenamefont
  {{Fisher}}(2006)}]{Senthil_2006}%
  \BibitemOpen
  \bibfield  {author} {\bibinfo {author} {\bibfnamefont {T.}~\bibnamefont
  {{Senthil}}}\ and\ \bibinfo {author} {\bibfnamefont {M.~P.~A.}\ \bibnamefont
  {{Fisher}}},\ }\bibfield  {title} {\enquote {\bibinfo {title} {{Competing
  orders, nonlinear sigma models, and topological terms in quantum magnets}},}\
  }\href {\doibase 10.1103/PhysRevB.74.064405} {\bibfield  {journal} {\bibinfo
  {journal} {Phys. Rev. B}\ }\textbf {\bibinfo {volume} {74}},\ \bibinfo {eid}
  {064405} (\bibinfo {year} {2006})},\ \Eprint
  {http://arxiv.org/abs/cond-mat/0510459} {cond-mat/0510459} \BibitemShut
  {NoStop}%
\bibitem [{\citenamefont {Wang}\ \emph {et~al.}(2017)\citenamefont {Wang},
  \citenamefont {Nahum}, \citenamefont {Metlitski}, \citenamefont {Xu},\ and\
  \citenamefont {Senthil}}]{Wang_2017}%
  \BibitemOpen
  \bibfield  {author} {\bibinfo {author} {\bibfnamefont {C.}~\bibnamefont
  {Wang}}, \bibinfo {author} {\bibfnamefont {A.}~\bibnamefont {Nahum}},
  \bibinfo {author} {\bibfnamefont {M.~A.}\ \bibnamefont {Metlitski}}, \bibinfo
  {author} {\bibfnamefont {C.}~\bibnamefont {Xu}}, \ and\ \bibinfo {author}
  {\bibfnamefont {T.}~\bibnamefont {Senthil}},\ }\bibfield  {title} {\enquote
  {\bibinfo {title} {{Deconfined Quantum Critical Points: Symmetries and
  Dualities}},}\ }\href {\doibase 10.1103/PhysRevX.7.031051} {\bibfield
  {journal} {\bibinfo  {journal} {Phys. Rev. X}\ }\textbf {\bibinfo {volume}
  {7}},\ \bibinfo {pages} {031051} (\bibinfo {year} {2017})}\BibitemShut
  {NoStop}%
\bibitem [{\citenamefont {Nahum}\ \emph {et~al.}(2015)\citenamefont {Nahum},
  \citenamefont {Serna}, \citenamefont {Chalker}, \citenamefont {Ortu\~no},\
  and\ \citenamefont {Somoza}}]{Nahum_2015}%
  \BibitemOpen
  \bibfield  {author} {\bibinfo {author} {\bibfnamefont {A.}~\bibnamefont
  {Nahum}}, \bibinfo {author} {\bibfnamefont {P.}~\bibnamefont {Serna}},
  \bibinfo {author} {\bibfnamefont {J.~T.}\ \bibnamefont {Chalker}}, \bibinfo
  {author} {\bibfnamefont {M.}~\bibnamefont {Ortu\~no}}, \ and\ \bibinfo
  {author} {\bibfnamefont {A.~M.}\ \bibnamefont {Somoza}},\ }\bibfield  {title}
  {\enquote {\bibinfo {title} {{Emergent SO(5) Symmetry at the N\'eel to
  Valence-Bond-Solid Transition}},}\ }\href {\doibase
  10.1103/PhysRevLett.115.267203} {\bibfield  {journal} {\bibinfo  {journal}
  {Phys. Rev. Lett.}\ }\textbf {\bibinfo {volume} {115}},\ \bibinfo {pages}
  {267203} (\bibinfo {year} {2015})}\BibitemShut {NoStop}%
\bibitem [{\citenamefont {{Suwa}}\ \emph {et~al.}(2016)\citenamefont {{Suwa}},
  \citenamefont {{Sen}},\ and\ \citenamefont {{Sandvik}}}]{Suwa_2016}%
  \BibitemOpen
  \bibfield  {author} {\bibinfo {author} {\bibfnamefont {H.}~\bibnamefont
  {{Suwa}}}, \bibinfo {author} {\bibfnamefont {A.}~\bibnamefont {{Sen}}}, \
  and\ \bibinfo {author} {\bibfnamefont {A.~W.}\ \bibnamefont {{Sandvik}}},\
  }\bibfield  {title} {\enquote {\bibinfo {title} {{Level spectroscopy in a
  two-dimensional quantum magnet: Linearly dispersing spinons at the deconfined
  quantum critical point}},}\ }\href {\doibase 10.1103/PhysRevB.94.144416}
  {\bibfield  {journal} {\bibinfo  {journal} {Phys. Rev. B}\ }\textbf {\bibinfo
  {volume} {94}},\ \bibinfo {eid} {144416} (\bibinfo {year} {2016})},\ \Eprint
  {http://arxiv.org/abs/1607.05110} {arXiv:1607.05110 [cond-mat.str-el]}
  \BibitemShut {NoStop}%
\bibitem [{\citenamefont {Karthik}\ and\ \citenamefont
  {Narayanan}(2017)}]{Karthik_2017}%
  \BibitemOpen
  \bibfield  {author} {\bibinfo {author} {\bibfnamefont {N.}~\bibnamefont
  {Karthik}}\ and\ \bibinfo {author} {\bibfnamefont {R.}~\bibnamefont
  {Narayanan}},\ }\bibfield  {title} {\enquote {\bibinfo {title} {{Flavor and
  topological current correlators in parity-invariant three-dimensional
  QED}},}\ }\href {\doibase 10.1103/PhysRevD.96.054509} {\bibfield  {journal}
  {\bibinfo  {journal} {Phys. Rev. D}\ }\textbf {\bibinfo {volume} {96}},\
  \bibinfo {pages} {054509} (\bibinfo {year} {2017})}\BibitemShut {NoStop}%
\bibitem [{\citenamefont {Sato}\ \emph {et~al.}(2017)\citenamefont {Sato},
  \citenamefont {Hohenadler},\ and\ \citenamefont {Assaad}}]{Sato_2017}%
  \BibitemOpen
  \bibfield  {author} {\bibinfo {author} {\bibfnamefont {T.}~\bibnamefont
  {Sato}}, \bibinfo {author} {\bibfnamefont {M.}~\bibnamefont {Hohenadler}}, \
  and\ \bibinfo {author} {\bibfnamefont {F.~F.}\ \bibnamefont {Assaad}},\
  }\bibfield  {title} {\enquote {\bibinfo {title} {{Dirac Fermions with
  Competing Orders: Non-Landau Transition with Emergent Symmetry}},}\ }\href
  {\doibase 10.1103/PhysRevLett.119.197203} {\bibfield  {journal} {\bibinfo
  {journal} {Phys. Rev. Lett.}\ }\textbf {\bibinfo {volume} {119}},\ \bibinfo
  {pages} {197203} (\bibinfo {year} {2017})}\BibitemShut {NoStop}%
\bibitem [{\citenamefont {Sreejith}\ \emph {et~al.}(2018)\citenamefont
  {Sreejith}, \citenamefont {Powell},\ and\ \citenamefont
  {Nahum}}]{Powell_2018}%
  \BibitemOpen
  \bibfield  {author} {\bibinfo {author} {\bibfnamefont {G.~J.}\ \bibnamefont
  {Sreejith}}, \bibinfo {author} {\bibfnamefont {S.}~\bibnamefont {Powell}}, \
  and\ \bibinfo {author} {\bibfnamefont {A.}~\bibnamefont {Nahum}},\ }\bibfield
   {title} {\enquote {\bibinfo {title} {{Emergent SO(5) symmetry at the
  columnar ordering transition in the classical cubic dimer model}},}\
  }\href@noop {} {\  (\bibinfo {year} {2018})},\ \Eprint
  {http://arxiv.org/abs/1803.11218} {arXiv:1803.11218 [cond-mat.stat-mech]}
  \BibitemShut {NoStop}%
\bibitem [{\citenamefont {Assaad}\ and\ \citenamefont
  {Grover}(2016)}]{Assaad2016}%
  \BibitemOpen
  \bibfield  {author} {\bibinfo {author} {\bibfnamefont {F.~F.}\ \bibnamefont
  {Assaad}}\ and\ \bibinfo {author} {\bibfnamefont {T.}~\bibnamefont
  {Grover}},\ }\bibfield  {title} {\enquote {\bibinfo {title} {{Simple
  Fermionic Model of Deconfined Phases and Phase Transitions}},}\ }\href
  {\doibase 10.1103/PhysRevX.6.041049} {\bibfield  {journal} {\bibinfo
  {journal} {Phys. Rev. X}\ }\textbf {\bibinfo {volume} {6}},\ \bibinfo {pages}
  {041049} (\bibinfo {year} {2016})}\BibitemShut {NoStop}%
\bibitem [{\citenamefont {Gazit}\ \emph {et~al.}(2017)\citenamefont {Gazit},
  \citenamefont {Randeria},\ and\ \citenamefont {Vishwanath}}]{Gazit2017}%
  \BibitemOpen
  \bibfield  {author} {\bibinfo {author} {\bibfnamefont {S.}~\bibnamefont
  {Gazit}}, \bibinfo {author} {\bibfnamefont {M.}~\bibnamefont {Randeria}}, \
  and\ \bibinfo {author} {\bibfnamefont {A.}~\bibnamefont {Vishwanath}},\
  }\bibfield  {title} {\enquote {\bibinfo {title} {{Emergent Dirac fermions and
  broken symmetries in confined and deconfined phases of $Z_2$ gauge
  theories}},}\ }\href {http://dx.doi.org/10.1038/nphys4028} {\bibfield
  {journal} {\bibinfo  {journal} {Nature Physics}\ }\textbf {\bibinfo {volume}
  {13}},\ \bibinfo {pages} {484} (\bibinfo {year} {2017})},\ \Eprint
  {http://arxiv.org/abs/1607.03892} {arXiv:1607.03892 [cond-mat.str-el]}
  \BibitemShut {NoStop}%
\bibitem [{\citenamefont {Nandkishore}\ \emph {et~al.}(2012)\citenamefont
  {Nandkishore}, \citenamefont {Metlitski},\ and\ \citenamefont
  {Senthil}}]{Nandkishore_2012}%
  \BibitemOpen
  \bibfield  {author} {\bibinfo {author} {\bibfnamefont {R.}~\bibnamefont
  {Nandkishore}}, \bibinfo {author} {\bibfnamefont {M.~A.}\ \bibnamefont
  {Metlitski}}, \ and\ \bibinfo {author} {\bibfnamefont {T.}~\bibnamefont
  {Senthil}},\ }\bibfield  {title} {\enquote {\bibinfo {title} {{Orthogonal
  metals: The simplest non-Fermi liquids}},}\ }\href {\doibase
  10.1103/PhysRevB.86.045128} {\bibfield  {journal} {\bibinfo  {journal} {Phys.
  Rev. B}\ }\textbf {\bibinfo {volume} {86}},\ \bibinfo {pages} {045128}
  (\bibinfo {year} {2012})}\BibitemShut {NoStop}%
\bibitem [{\citenamefont {Zhang}(1990)}]{Zhang_PseudoSpin}%
  \BibitemOpen
  \bibfield  {author} {\bibinfo {author} {\bibfnamefont {S.}~\bibnamefont
  {Zhang}},\ }\bibfield  {title} {\enquote {\bibinfo {title} {{Pseudospin
  symmetry and new collective modes of the Hubbard model}},}\ }\href {\doibase
  10.1103/PhysRevLett.65.120} {\bibfield  {journal} {\bibinfo  {journal} {Phys.
  Rev. Lett.}\ }\textbf {\bibinfo {volume} {65}},\ \bibinfo {pages} {120}
  (\bibinfo {year} {1990})}\BibitemShut {NoStop}%
\bibitem [{\citenamefont {Auerbach}(2012)}]{auerbach_book}%
  \BibitemOpen
  \bibfield  {author} {\bibinfo {author} {\bibfnamefont {A.}~\bibnamefont
  {Auerbach}},\ }\href@noop {} {\emph {\bibinfo {title} {{Interacting electrons
  and quantum magnetism}}}}\ (\bibinfo  {publisher} {Springer Science \&
  Business Media},\ \bibinfo {year} {2012})\BibitemShut {NoStop}%
\bibitem [{\citenamefont {Kogut}(1979)}]{Kogut_RMP}%
  \BibitemOpen
  \bibfield  {author} {\bibinfo {author} {\bibfnamefont {J.~B.}\ \bibnamefont
  {Kogut}},\ }\bibfield  {title} {\enquote {\bibinfo {title} {{An introduction
  to lattice gauge theory and spin systems}},}\ }\href {\doibase
  10.1103/RevModPhys.51.659} {\bibfield  {journal} {\bibinfo  {journal} {Rev.
  Mod. Phys.}\ }\textbf {\bibinfo {volume} {51}},\ \bibinfo {pages} {659}
  (\bibinfo {year} {1979})}\BibitemShut {NoStop}%
\bibitem [{\citenamefont {Affleck}\ and\ \citenamefont
  {Marston}(1988)}]{Affleck_1988}%
  \BibitemOpen
  \bibfield  {author} {\bibinfo {author} {\bibfnamefont {I.}~\bibnamefont
  {Affleck}}\ and\ \bibinfo {author} {\bibfnamefont {J.~B.}\ \bibnamefont
  {Marston}},\ }\bibfield  {title} {\enquote {\bibinfo {title} {{Large-$n$
  limit of the Heisenberg-Hubbard model: Implications for high-${T}_{c}$
  superconductors}},}\ }\href {\doibase 10.1103/PhysRevB.37.3774} {\bibfield
  {journal} {\bibinfo  {journal} {Phys. Rev. B}\ }\textbf {\bibinfo {volume}
  {37}},\ \bibinfo {pages} {3774} (\bibinfo {year} {1988})}\BibitemShut
  {NoStop}%
\bibitem [{\citenamefont {Kitaev}(2003)}]{KITAEV_2003}%
  \BibitemOpen
  \bibfield  {author} {\bibinfo {author} {\bibfnamefont {A.}~\bibnamefont
  {Kitaev}},\ }\bibfield  {title} {\enquote {\bibinfo {title} {{Fault-tolerant
  quantum computation by anyons}},}\ }\href {\doibase
  https://doi.org/10.1016/S0003-4916(02)00018-0} {\bibfield  {journal}
  {\bibinfo  {journal} {Annals of Physics}\ }\textbf {\bibinfo {volume}
  {303}},\ \bibinfo {pages} {2 } (\bibinfo {year} {2003})}\BibitemShut
  {NoStop}%
\bibitem [{\citenamefont {Wen}(2002)}]{Wen_2002}%
  \BibitemOpen
  \bibfield  {author} {\bibinfo {author} {\bibfnamefont {X.-G.}\ \bibnamefont
  {Wen}},\ }\bibfield  {title} {\enquote {\bibinfo {title} {{Quantum orders and
  symmetric spin liquids}},}\ }\href {\doibase 10.1103/PhysRevB.65.165113}
  {\bibfield  {journal} {\bibinfo  {journal} {Phys. Rev. B}\ }\textbf {\bibinfo
  {volume} {65}},\ \bibinfo {pages} {165113} (\bibinfo {year}
  {2002})}\BibitemShut {NoStop}%
\bibitem [{\citenamefont {Herbut}\ \emph {et~al.}(2009)\citenamefont {Herbut},
  \citenamefont {Juri\ifmmode \check{c}\else \v{c}\fi{}i\ifmmode~\acute{c}\else
  \'{c}\fi{}},\ and\ \citenamefont {Vafek}}]{Herbut_GN}%
  \BibitemOpen
  \bibfield  {author} {\bibinfo {author} {\bibfnamefont {I.~F.}\ \bibnamefont
  {Herbut}}, \bibinfo {author} {\bibfnamefont {V.}~\bibnamefont {Juri\ifmmode
  \check{c}\else \v{c}\fi{}i\ifmmode~\acute{c}\else \'{c}\fi{}}}, \ and\
  \bibinfo {author} {\bibfnamefont {O.}~\bibnamefont {Vafek}},\ }\bibfield
  {title} {\enquote {\bibinfo {title} {{Relativistic Mott criticality in
  graphene}},}\ }\href {\doibase 10.1103/PhysRevB.80.075432} {\bibfield
  {journal} {\bibinfo  {journal} {Phys. Rev. B}\ }\textbf {\bibinfo {volume}
  {80}},\ \bibinfo {pages} {075432} (\bibinfo {year} {2009})}\BibitemShut
  {NoStop}%
\bibitem [{\citenamefont {Assaad}\ and\ \citenamefont
  {Herbut}(2013)}]{Assaad13}%
  \BibitemOpen
  \bibfield  {author} {\bibinfo {author} {\bibfnamefont {F.~F.}\ \bibnamefont
  {Assaad}}\ and\ \bibinfo {author} {\bibfnamefont {I.~F.}\ \bibnamefont
  {Herbut}},\ }\bibfield  {title} {\enquote {\bibinfo {title} {{Pinning the
  Order: The Nature of Quantum Criticality in the Hubbard Model on Honeycomb
  Lattice}},}\ }\href {\doibase 10.1103/PhysRevX.3.031010} {\bibfield
  {journal} {\bibinfo  {journal} {Phys. Rev. X}\ }\textbf {\bibinfo {volume}
  {3}},\ \bibinfo {pages} {031010} (\bibinfo {year} {2013})}\BibitemShut
  {NoStop}%
\bibitem [{\citenamefont {Parisen~Toldin}\ \emph {et~al.}(2015)\citenamefont
  {Parisen~Toldin}, \citenamefont {Hohenadler}, \citenamefont {Assaad},\ and\
  \citenamefont {Herbut}}]{Toldin14}%
  \BibitemOpen
  \bibfield  {author} {\bibinfo {author} {\bibfnamefont {F.}~\bibnamefont
  {Parisen~Toldin}}, \bibinfo {author} {\bibfnamefont {M.}~\bibnamefont
  {Hohenadler}}, \bibinfo {author} {\bibfnamefont {F.~F.}\ \bibnamefont
  {Assaad}}, \ and\ \bibinfo {author} {\bibfnamefont {I.~F.}\ \bibnamefont
  {Herbut}},\ }\bibfield  {title} {\enquote {\bibinfo {title} {{Fermionic
  quantum criticality in honeycomb and $\pi$-flux Hubbard models: Finite-size
  scaling of renormalization-group-invariant observables from quantum Monte
  Carlo}},}\ }\href {\doibase 10.1103/PhysRevB.91.165108} {\bibfield  {journal}
  {\bibinfo  {journal} {Phys. Rev. B}\ }\textbf {\bibinfo {volume} {91}},\
  \bibinfo {pages} {165108} (\bibinfo {year} {2015})}\BibitemShut {NoStop}%
\bibitem [{\citenamefont {Otsuka}\ \emph {et~al.}(2016)\citenamefont {Otsuka},
  \citenamefont {Yunoki},\ and\ \citenamefont {Sorella}}]{Sorella_GN}%
  \BibitemOpen
  \bibfield  {author} {\bibinfo {author} {\bibfnamefont {Y.}~\bibnamefont
  {Otsuka}}, \bibinfo {author} {\bibfnamefont {S.}~\bibnamefont {Yunoki}}, \
  and\ \bibinfo {author} {\bibfnamefont {S.}~\bibnamefont {Sorella}},\
  }\bibfield  {title} {\enquote {\bibinfo {title} {{Universal Quantum
  Criticality in the Metal-Insulator Transition of Two-Dimensional Interacting
  Dirac Electrons}},}\ }\href {\doibase 10.1103/PhysRevX.6.011029} {\bibfield
  {journal} {\bibinfo  {journal} {Phys. Rev. X}\ }\textbf {\bibinfo {volume}
  {6}},\ \bibinfo {pages} {011029} (\bibinfo {year} {2016})}\BibitemShut
  {NoStop}%
\bibitem [{\citenamefont {{Sachdev}}\ and\ \citenamefont
  {{Morinari}}(2002)}]{SSTM02}%
  \BibitemOpen
  \bibfield  {author} {\bibinfo {author} {\bibfnamefont {S.}~\bibnamefont
  {{Sachdev}}}\ and\ \bibinfo {author} {\bibfnamefont {T.}~\bibnamefont
  {{Morinari}}},\ }\bibfield  {title} {\enquote {\bibinfo {title} {{Strongly
  coupled quantum criticality with a Fermi surface in two dimensions:
  Fractionalization of spin and charge collective modes}},}\ }\href {\doibase
  10.1103/PhysRevB.66.235117} {\bibfield  {journal} {\bibinfo  {journal} {Phys.
  Rev. B}\ }\textbf {\bibinfo {volume} {66}},\ \bibinfo {eid} {235117}
  (\bibinfo {year} {2002})},\ \Eprint {http://arxiv.org/abs/cond-mat/0207167}
  {cond-mat/0207167} \BibitemShut {NoStop}%
\bibitem [{\citenamefont {Assaad}\ and\ \citenamefont
  {Evertz}(2008)}]{Assaad2008}%
  \BibitemOpen
  \bibfield  {author} {\bibinfo {author} {\bibfnamefont {F.}~\bibnamefont
  {Assaad}}\ and\ \bibinfo {author} {\bibfnamefont {H.}~\bibnamefont
  {Evertz}},\ }\enquote {\bibinfo {title} {{World-line and Determinantal
  Quantum Monte Carlo Methods for Spins, Phonons and Electrons}},}\ in\ \href
  {\doibase 10.1007/978-3-540-74686-7_10} {\emph {\bibinfo {booktitle}
  {{Computational Many-Particle Physics}}}},\ \bibinfo {editor} {edited by\
  \bibinfo {editor} {\bibfnamefont {H.}~\bibnamefont {Fehske}}, \bibinfo
  {editor} {\bibfnamefont {R.}~\bibnamefont {Schneider}}, \ and\ \bibinfo
  {editor} {\bibfnamefont {A.}~\bibnamefont {Wei{\ss}e}}}\ (\bibinfo
  {publisher} {Springer Berlin Heidelberg},\ \bibinfo {address} {Berlin,
  Heidelberg},\ \bibinfo {year} {2008})\ pp.\ \bibinfo {pages}
  {277--356}\BibitemShut {NoStop}%
\bibitem [{\citenamefont {Bercx}\ \emph {et~al.}(2017)\citenamefont {Bercx},
  \citenamefont {Goth}, \citenamefont {Hofmann},\ and\ \citenamefont
  {Assaad}}]{ALF_v1}%
  \BibitemOpen
  \bibfield  {author} {\bibinfo {author} {\bibfnamefont {M.}~\bibnamefont
  {Bercx}}, \bibinfo {author} {\bibfnamefont {F.}~\bibnamefont {Goth}},
  \bibinfo {author} {\bibfnamefont {J.~S.}\ \bibnamefont {Hofmann}}, \ and\
  \bibinfo {author} {\bibfnamefont {F.~F.}\ \bibnamefont {Assaad}},\ }\bibfield
   {title} {\enquote {\bibinfo {title} {{The ALF (Algorithms for Lattice
  Fermions) project release 1.0. Documentation for the auxiliary field quantum
  Monte Carlo code}},}\ }\href {\doibase 10.21468/SciPostPhys.3.2.013}
  {\bibfield  {journal} {\bibinfo  {journal} {SciPost Phys.}\ }\textbf
  {\bibinfo {volume} {3}},\ \bibinfo {pages} {013} (\bibinfo {year}
  {2017})}\BibitemShut {NoStop}%
\bibitem [{\citenamefont {Pujari}\ \emph {et~al.}(2016)\citenamefont {Pujari},
  \citenamefont {Lang}, \citenamefont {Murthy},\ and\ \citenamefont
  {Kaul}}]{Pujari_2016}%
  \BibitemOpen
  \bibfield  {author} {\bibinfo {author} {\bibfnamefont {S.}~\bibnamefont
  {Pujari}}, \bibinfo {author} {\bibfnamefont {T.~C.}\ \bibnamefont {Lang}},
  \bibinfo {author} {\bibfnamefont {G.}~\bibnamefont {Murthy}}, \ and\ \bibinfo
  {author} {\bibfnamefont {R.~K.}\ \bibnamefont {Kaul}},\ }\bibfield  {title}
  {\enquote {\bibinfo {title} {{Interaction-Induced Dirac Fermions from
  Quadratic Band Touching in Bilayer Graphene}},}\ }\href {\doibase
  10.1103/PhysRevLett.117.086404} {\bibfield  {journal} {\bibinfo  {journal}
  {Phys. Rev. Lett.}\ }\textbf {\bibinfo {volume} {117}},\ \bibinfo {pages}
  {086404} (\bibinfo {year} {2016})}\BibitemShut {NoStop}%
\bibitem [{\citenamefont {Polyakov}(1978)}]{Polyakov_1978}%
  \BibitemOpen
  \bibfield  {author} {\bibinfo {author} {\bibfnamefont {A.~M.}\ \bibnamefont
  {Polyakov}},\ }\bibfield  {title} {\enquote {\bibinfo {title} {{Thermal
  Properties of Gauge Fields and Quark Liberation}},}\ }\href {\doibase
  10.1016/0370-2693(78)90737-2} {\bibfield  {journal} {\bibinfo  {journal}
  {Phys. Lett.}\ }\textbf {\bibinfo {volume} {72B}},\ \bibinfo {pages} {477}
  (\bibinfo {year} {1978})}\BibitemShut {NoStop}%
\bibitem [{\citenamefont {{Isakov}}\ \emph {et~al.}(2011)\citenamefont
  {{Isakov}}, \citenamefont {{Hastings}},\ and\ \citenamefont
  {{Melko}}}]{Isakov2011}%
  \BibitemOpen
  \bibfield  {author} {\bibinfo {author} {\bibfnamefont {S.~V.}\ \bibnamefont
  {{Isakov}}}, \bibinfo {author} {\bibfnamefont {M.~B.}\ \bibnamefont
  {{Hastings}}}, \ and\ \bibinfo {author} {\bibfnamefont {R.~G.}\ \bibnamefont
  {{Melko}}},\ }\bibfield  {title} {\enquote {\bibinfo {title} {{Topological
  entanglement entropy of a Bose-Hubbard spin liquid}},}\ }\href {\doibase
  10.1038/nphys2036} {\bibfield  {journal} {\bibinfo  {journal} {Nature
  Physics}\ }\textbf {\bibinfo {volume} {7}},\ \bibinfo {pages} {772} (\bibinfo
  {year} {2011})},\ \Eprint {http://arxiv.org/abs/1102.1721} {arXiv:1102.1721
  [cond-mat.str-el]} \BibitemShut {NoStop}%
\bibitem [{\citenamefont {Grover}\ \emph {et~al.}(2013)\citenamefont {Grover},
  \citenamefont {Zhang},\ and\ \citenamefont {Vishwanath}}]{GroverEE_2013}%
  \BibitemOpen
  \bibfield  {author} {\bibinfo {author} {\bibfnamefont {T.}~\bibnamefont
  {Grover}}, \bibinfo {author} {\bibfnamefont {Y.}~\bibnamefont {Zhang}}, \
  and\ \bibinfo {author} {\bibfnamefont {A.}~\bibnamefont {Vishwanath}},\
  }\bibfield  {title} {\enquote {\bibinfo {title} {{Entanglement entropy as a
  portal to the physics of quantum spin liquids}},}\ }\href
  {http://stacks.iop.org/1367-2630/15/i=2/a=025002} {\bibfield  {journal}
  {\bibinfo  {journal} {New Journal of Physics}\ }\textbf {\bibinfo {volume}
  {15}},\ \bibinfo {pages} {025002} (\bibinfo {year} {2013})}\BibitemShut
  {NoStop}%
\bibitem [{\citenamefont {Fredenhagen}\ and\ \citenamefont
  {Marcu}(1986)}]{Fredenhagen_1986}%
  \BibitemOpen
  \bibfield  {author} {\bibinfo {author} {\bibfnamefont {K.}~\bibnamefont
  {Fredenhagen}}\ and\ \bibinfo {author} {\bibfnamefont {M.}~\bibnamefont
  {Marcu}},\ }\bibfield  {title} {\enquote {\bibinfo {title} {{Confinement
  criterion for QCD with dynamical quarks}},}\ }\href {\doibase
  10.1103/PhysRevLett.56.223} {\bibfield  {journal} {\bibinfo  {journal} {Phys.
  Rev. Lett.}\ }\textbf {\bibinfo {volume} {56}},\ \bibinfo {pages} {223}
  (\bibinfo {year} {1986})}\BibitemShut {NoStop}%
\bibitem [{\citenamefont {Gregor}\ \emph {et~al.}(2011)\citenamefont {Gregor},
  \citenamefont {Huse}, \citenamefont {Moessner},\ and\ \citenamefont
  {Sondhi}}]{Gregor_2010}%
  \BibitemOpen
  \bibfield  {author} {\bibinfo {author} {\bibfnamefont {K.}~\bibnamefont
  {Gregor}}, \bibinfo {author} {\bibfnamefont {D.~A.}\ \bibnamefont {Huse}},
  \bibinfo {author} {\bibfnamefont {R.}~\bibnamefont {Moessner}}, \ and\
  \bibinfo {author} {\bibfnamefont {S.~L.}\ \bibnamefont {Sondhi}},\ }\bibfield
   {title} {\enquote {\bibinfo {title} {{Diagnosing Deconfinement and
  Topological Order}},}\ }\href {\doibase 10.1088/1367-2630/13/2/025009}
  {\bibfield  {journal} {\bibinfo  {journal} {New J. Phys.}\ }\textbf {\bibinfo
  {volume} {13}},\ \bibinfo {pages} {025009} (\bibinfo {year} {2011})},\
  \Eprint {http://arxiv.org/abs/1011.4187} {arXiv:1011.4187 [cond-mat.str-el]}
  \BibitemShut {NoStop}%
\bibitem [{\citenamefont {{Sachdev}}(2018)}]{SSreview}%
  \BibitemOpen
  \bibfield  {author} {\bibinfo {author} {\bibfnamefont {S.}~\bibnamefont
  {{Sachdev}}},\ }\bibfield  {title} {\enquote {\bibinfo {title} {{Topological
  order and Fermi surface reconstruction}},}\ }\href@noop {} {\bibfield
  {journal} {\bibinfo  {journal} {ArXiv e-prints}\ } (\bibinfo {year}
  {2018})},\ \Eprint {http://arxiv.org/abs/1801.01125} {arXiv:1801.01125
  [cond-mat.str-el]} \BibitemShut {NoStop}%
\bibitem [{\citenamefont {{Senthil}}\ \emph
  {et~al.}(2004{\natexlab{b}})\citenamefont {{Senthil}}, \citenamefont
  {{Balents}}, \citenamefont {{Sachdev}}, \citenamefont {{Vishwanath}},\ and\
  \citenamefont {{Fisher}}}]{Senthil_DC2}%
  \BibitemOpen
  \bibfield  {author} {\bibinfo {author} {\bibfnamefont {T.}~\bibnamefont
  {{Senthil}}}, \bibinfo {author} {\bibfnamefont {L.}~\bibnamefont
  {{Balents}}}, \bibinfo {author} {\bibfnamefont {S.}~\bibnamefont
  {{Sachdev}}}, \bibinfo {author} {\bibfnamefont {A.}~\bibnamefont
  {{Vishwanath}}}, \ and\ \bibinfo {author} {\bibfnamefont {M.~P.~A.}\
  \bibnamefont {{Fisher}}},\ }\bibfield  {title} {\enquote {\bibinfo {title}
  {{Quantum criticality beyond the Landau-Ginzburg-Wilson paradigm}},}\ }\href
  {\doibase 10.1103/PhysRevB.70.144407} {\bibfield  {journal} {\bibinfo
  {journal} {Phys. Rev. B}\ }\textbf {\bibinfo {volume} {70}},\ \bibinfo {eid}
  {144407} (\bibinfo {year} {2004}{\natexlab{b}})},\ \Eprint
  {http://arxiv.org/abs/cond-mat/0312617} {cond-mat/0312617} \BibitemShut
  {NoStop}%
\bibitem [{\citenamefont {Fradkin}\ and\ \citenamefont
  {Shenker}(1979)}]{FradkinShenker}%
  \BibitemOpen
  \bibfield  {author} {\bibinfo {author} {\bibfnamefont {E.}~\bibnamefont
  {Fradkin}}\ and\ \bibinfo {author} {\bibfnamefont {S.~H.}\ \bibnamefont
  {Shenker}},\ }\bibfield  {title} {\enquote {\bibinfo {title} {{Phase diagrams
  of lattice gauge theories with Higgs fields}},}\ }\href {\doibase
  10.1103/PhysRevD.19.3682} {\bibfield  {journal} {\bibinfo  {journal} {Phys.
  Rev. D}\ }\textbf {\bibinfo {volume} {19}},\ \bibinfo {pages} {3682}
  (\bibinfo {year} {1979})}\BibitemShut {NoStop}%
\bibitem [{\citenamefont {{Sachdev}}\ and\ \citenamefont
  {{Chowdhury}}(2016)}]{SSNambu}%
  \BibitemOpen
  \bibfield  {author} {\bibinfo {author} {\bibfnamefont {S.}~\bibnamefont
  {{Sachdev}}}\ and\ \bibinfo {author} {\bibfnamefont {D.}~\bibnamefont
  {{Chowdhury}}},\ }\bibfield  {title} {\enquote {\bibinfo {title} {{The novel
  metallic states of the cuprates: Fermi liquids with topological order and
  strange metals}},}\ }\href {\doibase 10.1093/ptep/ptw110} {\bibfield
  {journal} {\bibinfo  {journal} {Progress of Theoretical and Experimental
  Physics}\ }\textbf {\bibinfo {volume} {2016}},\ \bibinfo {eid} {12C102}
  (\bibinfo {year} {2016})},\ \Eprint {http://arxiv.org/abs/1605.03579}
  {arXiv:1605.03579 [cond-mat.str-el]} \BibitemShut {NoStop}%
\bibitem [{\citenamefont {Chatterjee}\ \emph {et~al.}(2017)\citenamefont
  {Chatterjee}, \citenamefont {Sachdev},\ and\ \citenamefont
  {Scheurer}}]{CSS17}%
  \BibitemOpen
  \bibfield  {author} {\bibinfo {author} {\bibfnamefont {S.}~\bibnamefont
  {Chatterjee}}, \bibinfo {author} {\bibfnamefont {S.}~\bibnamefont {Sachdev}},
  \ and\ \bibinfo {author} {\bibfnamefont {M.}~\bibnamefont {Scheurer}},\
  }\bibfield  {title} {\enquote {\bibinfo {title} {{Intertwining topological
  order and broken symmetry in a theory of fluctuating spin density waves}},}\
  }\href {\doibase 10.1103/PhysRevLett.119.227002} {\bibfield  {journal}
  {\bibinfo  {journal} {Phys. Rev. Lett.}\ }\textbf {\bibinfo {volume} {119}},\
  \bibinfo {pages} {227002} (\bibinfo {year} {2017})},\ \Eprint
  {http://arxiv.org/abs/1705.06289} {arXiv:1705.06289 [cond-mat.str-el]}
  \BibitemShut {NoStop}%
\bibitem [{\citenamefont {{Sachdev}}\ \emph {et~al.}(2009)\citenamefont
  {{Sachdev}}, \citenamefont {{Metlitski}}, \citenamefont {{Qi}},\ and\
  \citenamefont {{Xu}}}]{SS09}%
  \BibitemOpen
  \bibfield  {author} {\bibinfo {author} {\bibfnamefont {S.}~\bibnamefont
  {{Sachdev}}}, \bibinfo {author} {\bibfnamefont {M.~A.}\ \bibnamefont
  {{Metlitski}}}, \bibinfo {author} {\bibfnamefont {Y.}~\bibnamefont {{Qi}}}, \
  and\ \bibinfo {author} {\bibfnamefont {C.}~\bibnamefont {{Xu}}},\ }\bibfield
  {title} {\enquote {\bibinfo {title} {{Fluctuating spin density waves in
  metals}},}\ }\href {\doibase 10.1103/PhysRevB.80.155129} {\bibfield
  {journal} {\bibinfo  {journal} {Phys. Rev. B}\ }\textbf {\bibinfo {volume}
  {80}},\ \bibinfo {eid} {155129} (\bibinfo {year} {2009})},\ \Eprint
  {http://arxiv.org/abs/0907.3732} {arXiv:0907.3732 [cond-mat.str-el]}
  \BibitemShut {NoStop}%
\bibitem [{\citenamefont {Lee}\ \emph {et~al.}(2006)\citenamefont {Lee},
  \citenamefont {Nagaosa},\ and\ \citenamefont {Wen}}]{LeeWenRMP}%
  \BibitemOpen
  \bibfield  {author} {\bibinfo {author} {\bibfnamefont {P.~A.}\ \bibnamefont
  {Lee}}, \bibinfo {author} {\bibfnamefont {N.}~\bibnamefont {Nagaosa}}, \ and\
  \bibinfo {author} {\bibfnamefont {X.-G.}\ \bibnamefont {Wen}},\ }\bibfield
  {title} {\enquote {\bibinfo {title} {{Doping a Mott insulator: Physics of
  high-temperature superconductivity}},}\ }\href {\doibase
  10.1103/RevModPhys.78.17} {\bibfield  {journal} {\bibinfo  {journal} {Rev.
  Mod. Phys.}\ }\textbf {\bibinfo {volume} {78}},\ \bibinfo {pages} {17}
  (\bibinfo {year} {2006})},\ \Eprint {http://arxiv.org/abs/cond-mat/0410445}
  {cond-mat/0410445} \BibitemShut {NoStop}%
\bibitem [{\citenamefont {{Xu}}\ and\ \citenamefont {{Sachdev}}(2010)}]{XS10}%
  \BibitemOpen
  \bibfield  {author} {\bibinfo {author} {\bibfnamefont {C.}~\bibnamefont
  {{Xu}}}\ and\ \bibinfo {author} {\bibfnamefont {S.}~\bibnamefont
  {{Sachdev}}},\ }\bibfield  {title} {\enquote {\bibinfo {title} {{Majorana
  Liquids: The Complete Fractionalization of the Electron}},}\ }\href {\doibase
  10.1103/PhysRevLett.105.057201} {\bibfield  {journal} {\bibinfo  {journal}
  {Phys. Rev. Lett.}\ }\textbf {\bibinfo {volume} {105}},\ \bibinfo {eid}
  {057201} (\bibinfo {year} {2010})},\ \Eprint {http://arxiv.org/abs/1004.5431}
  {arXiv:1004.5431 [cond-mat.str-el]} \BibitemShut {NoStop}%
\bibitem [{\citenamefont {{Thomson}}\ and\ \citenamefont
  {{Sachdev}}(2018)}]{Thomson_2018}%
  \BibitemOpen
  \bibfield  {author} {\bibinfo {author} {\bibfnamefont {A.}~\bibnamefont
  {{Thomson}}}\ and\ \bibinfo {author} {\bibfnamefont {S.}~\bibnamefont
  {{Sachdev}}},\ }\bibfield  {title} {\enquote {\bibinfo {title} {{Fermionic
  Spinon Theory of Square Lattice Spin Liquids near the N{\'e}el State}},}\
  }\href {\doibase 10.1103/PhysRevX.8.011012} {\bibfield  {journal} {\bibinfo
  {journal} {Phys. Rev. X}\ }\textbf {\bibinfo {volume} {8}},\ \bibinfo {eid}
  {011012} (\bibinfo {year} {2018})},\ \Eprint
  {http://arxiv.org/abs/1708.04626} {arXiv:1708.04626 [cond-mat.str-el]}
  \BibitemShut {NoStop}%
\bibitem [{\citenamefont {Karthik}\ and\ \citenamefont
  {Narayanan}(2018)}]{Rajamani_2018}%
  \BibitemOpen
  \bibfield  {author} {\bibinfo {author} {\bibfnamefont {N.}~\bibnamefont
  {Karthik}}\ and\ \bibinfo {author} {\bibfnamefont {R.}~\bibnamefont
  {Narayanan}},\ }\bibfield  {title} {\enquote {\bibinfo {title}
  {{Scale-invariance and scale-breaking in parity-invariant three-dimensional
  QCD}},}\ }\href {\doibase 10.1103/PhysRevD.97.054510} {\bibfield  {journal}
  {\bibinfo  {journal} {Phys. Rev. D}\ }\textbf {\bibinfo {volume} {97}},\
  \bibinfo {pages} {054510} (\bibinfo {year} {2018})},\ \Eprint
  {http://arxiv.org/abs/1801.02637} {arXiv:1801.02637 [hep-th]} \BibitemShut
  {NoStop}%
\bibitem [{Note1()}]{Note1}%
  \BibitemOpen
  \bibinfo {note} {Strictly speaking, the simulated QCD$_3$ at $N_f=2$ does not
  have the full $SO(5)$ symmetry on the lattice scale, because the full
  symmetry is anomalous. In principle, there is a more exotic scenario\cite
  {Wang_2017}, in which the QCD theory with full $SO(5)$ symmetry flows to the
  continuous Neel to VBS transition (the deconfined quantum critical point),
  and chiral symmetry breaking happens only when the full $SO(5)$ is explicitly
  broken (for example to $SO(3)\times SO(2)$). Our theory holds even if this
  scenario is correct, since the full $SO(5)$ is already broken in our
  microscopic model.}\BibitemShut {Stop}%
\bibitem [{\citenamefont {Huffman}(2016)}]{Huffman_2016}%
  \BibitemOpen
  \bibfield  {author} {\bibinfo {author} {\bibfnamefont {E.}~\bibnamefont
  {Huffman}},\ }\bibfield  {title} {\enquote {\bibinfo {title} {{Monte Carlo
  methods in continuous time for lattice Hamiltonians}},}\ }\bibfield
  {booktitle} {\emph {\bibinfo {booktitle} {{Proceedings, 34th International
  Symposium on Lattice Field Theory}}},\ }\href@noop {} {\  (\bibinfo {year}
  {2016})},\ \Eprint {http://arxiv.org/abs/1611.01680} {arXiv:1611.01680
  [hep-lat]} \BibitemShut {NoStop}%
\bibitem [{\citenamefont {Huffman}\ and\ \citenamefont
  {Chandrasekharan}(2017)}]{Huffman17}%
  \BibitemOpen
  \bibfield  {author} {\bibinfo {author} {\bibfnamefont {E.}~\bibnamefont
  {Huffman}}\ and\ \bibinfo {author} {\bibfnamefont {S.}~\bibnamefont
  {Chandrasekharan}},\ }\bibfield  {title} {\enquote {\bibinfo {title}
  {{Fermion bag approach to Hamiltonian lattice field theories in continuous
  time}},}\ }\href {\doibase 10.1103/PhysRevD.96.114502} {\bibfield  {journal}
  {\bibinfo  {journal} {Phys. Rev. D}\ }\textbf {\bibinfo {volume} {96}},\
  \bibinfo {pages} {114502} (\bibinfo {year} {2017})}\BibitemShut {NoStop}%
\bibitem [{\citenamefont {{Karthik}}\ and\ \citenamefont
  {{Narayanan}}(2016)}]{qedcft}%
  \BibitemOpen
  \bibfield  {author} {\bibinfo {author} {\bibfnamefont {N.}~\bibnamefont
  {{Karthik}}}\ and\ \bibinfo {author} {\bibfnamefont {R.}~\bibnamefont
  {{Narayanan}}},\ }\bibfield  {title} {\enquote {\bibinfo {title} {{Scale
  invariance of parity-invariant three-dimensional QED}},}\ }\href {\doibase
  10.1103/PhysRevD.94.065026} {\bibfield  {journal} {\bibinfo  {journal} {Phys.
  Rev. D}\ }\textbf {\bibinfo {volume} {94}},\ \bibinfo {eid} {065026}
  (\bibinfo {year} {2016})},\ \Eprint {http://arxiv.org/abs/1606.04109}
  {arXiv:1606.04109 [hep-th]} \BibitemShut {NoStop}%
\bibitem [{\citenamefont {{Alicea}}(2008)}]{AliceaMonopole}%
  \BibitemOpen
  \bibfield  {author} {\bibinfo {author} {\bibfnamefont {J.}~\bibnamefont
  {{Alicea}}},\ }\bibfield  {title} {\enquote {\bibinfo {title} {{Monopole
  quantum numbers in the staggered flux spin liquid}},}\ }\href {\doibase
  10.1103/PhysRevB.78.035126} {\bibfield  {journal} {\bibinfo  {journal} {Phys.
  Rev. B}\ }\textbf {\bibinfo {volume} {78}},\ \bibinfo {eid} {035126}
  (\bibinfo {year} {2008})},\ \Eprint {http://arxiv.org/abs/0804.0786}
  {arXiv:0804.0786 [cond-mat.str-el]} \BibitemShut {NoStop}%
\bibitem [{\citenamefont {{Borokhov}}\ \emph {et~al.}(2002)\citenamefont
  {{Borokhov}}, \citenamefont {{Kapustin}},\ and\ \citenamefont
  {{Wu}}}]{KapustinQED}%
  \BibitemOpen
  \bibfield  {author} {\bibinfo {author} {\bibfnamefont {V.}~\bibnamefont
  {{Borokhov}}}, \bibinfo {author} {\bibfnamefont {A.}~\bibnamefont
  {{Kapustin}}}, \ and\ \bibinfo {author} {\bibfnamefont {X.}~\bibnamefont
  {{Wu}}},\ }\bibfield  {title} {\enquote {\bibinfo {title} {{Topological
  Disorder Operators in Three-Dimensional Conformal Field Theory}},}\ }\href
  {\doibase 10.1088/1126-6708/2002/11/049} {\bibfield  {journal} {\bibinfo
  {journal} {Journal of High Energy Physics}\ }\textbf {\bibinfo {volume}
  {11}},\ \bibinfo {eid} {049} (\bibinfo {year} {2002})},\ \Eprint
  {http://arxiv.org/abs/hep-th/0206054} {hep-th/0206054} \BibitemShut {NoStop}%
\bibitem [{\citenamefont {{Dyer}}\ \emph {et~al.}(2013)\citenamefont {{Dyer}},
  \citenamefont {{Mezei}},\ and\ \citenamefont {{Pufu}}}]{PufuQED}%
  \BibitemOpen
  \bibfield  {author} {\bibinfo {author} {\bibfnamefont {E.}~\bibnamefont
  {{Dyer}}}, \bibinfo {author} {\bibfnamefont {M.}~\bibnamefont {{Mezei}}}, \
  and\ \bibinfo {author} {\bibfnamefont {S.~S.}\ \bibnamefont {{Pufu}}},\
  }\bibfield  {title} {\enquote {\bibinfo {title} {{Monopole Taxonomy in
  Three-Dimensional Conformal Field Theories}},}\ }\href@noop {} {\bibfield
  {journal} {\bibinfo  {journal} {ArXiv e-prints}\ } (\bibinfo {year}
  {2013})},\ \Eprint {http://arxiv.org/abs/1309.1160} {arXiv:1309.1160
  [hep-th]} \BibitemShut {NoStop}%
\bibitem [{\citenamefont {Manuel~Carmona}\ \emph {et~al.}(2000)\citenamefont
  {Manuel~Carmona}, \citenamefont {Pelissetto},\ and\ \citenamefont
  {Vicari}}]{Anisotropy}%
  \BibitemOpen
  \bibfield  {author} {\bibinfo {author} {\bibfnamefont {J.}~\bibnamefont
  {Manuel~Carmona}}, \bibinfo {author} {\bibfnamefont {A.}~\bibnamefont
  {Pelissetto}}, \ and\ \bibinfo {author} {\bibfnamefont {E.}~\bibnamefont
  {Vicari}},\ }\bibfield  {title} {\enquote {\bibinfo {title} {{$N$-component
  Ginzburg-Landau Hamiltonian with cubic anisotropy: A six-loop study}},}\
  }\href {\doibase 10.1103/PhysRevB.61.15136} {\bibfield  {journal} {\bibinfo
  {journal} {Phys. Rev. B}\ }\textbf {\bibinfo {volume} {61}},\ \bibinfo
  {pages} {15136} (\bibinfo {year} {2000})}\BibitemShut {NoStop}%
\end{thebibliography}%

\end{document}